\documentclass[twocolumn, aps, 10pt, floatfix]{revtex4-1}

\usepackage{lmodern}
\usepackage[T1]{fontenc}
\usepackage[utf8]{inputenc}
\usepackage[english]{babel}

\usepackage{amsmath}
\usepackage{amssymb}
\usepackage{mathtools}
\usepackage{bm}
\usepackage{graphicx,color}
\usepackage{hyperref}
\usepackage{booktabs}
\usepackage{epstopdf}

\hypersetup{
  colorlinks   = true, %Colours links instead of ugly boxes
  urlcolor     = blue, %Colour for external hyperlinks
  linkcolor    = blue, %Colour of internal links
  citecolor   = red, %Colour of citations
}

\def\CFT{\textsl{\tiny CFT}}
\def\TCI{\textsl{\tiny TCI}}
\def\APBC{\textsl{\tiny APBC}}
\def\PBC{\textsl{\tiny PBC}}
\def\Ising{\textsl{\tiny Ising}}
\def\PPP{\textsl{\tiny PPP}}
\def\APA{\textsl{\tiny APA}}
\def\AAP{\textsl{\tiny AAP}}
\def\QFT{\textsl{\tiny QFT}}

\begin{document}

%\title{Emergence of conformal and superconformal symmetry in quantum spin chains with twisted boundary conditions}
\title{Emergence of conformal symmetry in quantum spin chains:\\ anti-periodic boundary conditions and supersymmetry}
\author{Yijian Zou}
\email[]{yzou@pitp.ca}
\affiliation{Perimeter Institute for Theoretical Physics, Waterloo ON, N2L 2Y5, Canada}
\affiliation{University of Waterloo, Waterloo ON, N2L 3G1, Canada}
\author{Guifre Vidal}
\affiliation{Perimeter Institute for Theoretical Physics, Waterloo ON, N2L 2Y5, Canada}
\affiliation{Alphabet (Google) X, Mountain View, CA 94043, USA}
\date{\today}

\begin{abstract}
Universal properties of a critical quantum spin chain are encoded in the underlying conformal field theory (CFT). This underlying CFT is fully characterized by its conformal data. We propose a method to extract the conformal data from a critical quantum spin chain with both periodic and anti-periodic boundary conditions (PBC and APBC) based on low-energy eigenstates, generalizing previous work on spin chains with only PBC. First, scaling dimensions and conformal spins are extracted from the energies and momenta of the eigenstates. Second, the Koo-Saleur formula of lattice Virasoro generators is generalized to APBC and used to identify conformal towers. Third, local operators and string operators on the lattice are identified with CFT operators with PBC and APBC, respectively. Finally, operator product expansion coefficients are extracted by computing matrix elements of lattice primary operators in the low-energy subspaces with PBC and APBC. To go beyond exact diagonalization, tensor network methods based on periodic uniform matrix product states are used. We illustrate our approach with critical and tricritical Ising quantum spin chains. In the latter case, we propose lattice operators that correspond to supervirasoro generators and verify their action on low-energy eigenstates. In this way we explore the emergence of superconformal symmetry in the quantum spin chain.
\end{abstract}

\maketitle
\section{Introduction}
Understanding the universality class \cite{wilson_renormalization_1974} of a critical quantum system chain is an important problem in modern physics. At long distances, a critical quantum system is often described by a conformal field theory (CFT) \cite{belavin_infinite_1984,friedan_conformal_1984}, which is in turn specified by the conformal data.

Given a critical quantum spin Hamiltonian $H$ in 1+1 dimensions, an ambitious, yet challenging task is to extract the conformal data for the emergent CFT. One method is based on the \textit{operator-state correspondence} \cite{belavin_infinite_1984, friedan_conformal_1984, ginsparg_applied_1988, henkel_1999, francesco_2012}. This refers to the fact that in a 1+1 dimensional CFT, every state $|\psi^{\CFT}_\alpha\rangle$ on the circle is in one to one correspondence with a scaling operator $\psi^{\CFT}_\alpha$ (an operator that transforms covariantly under scale transformations and rotations of the plane). Each low energy eigenstate $|\psi_\alpha\rangle$ of the lattice Hamiltonian $H$ with periodic boundary conditions (PBC) is an approximate lattice realization of the CFT state $|\psi^{\CFT}_\alpha\rangle$. Conformal data can then be extracted from matrix elements of certain lattice operators $\langle \psi_\alpha|O|\psi_\beta\rangle$ \cite{cardy_conformal_1984, blote_1986, affleck_universal_1986, cardy_operator_1986, koo_representations_1994, milsted_extraction_2017, zou_conformal_2019} and extrapolation of the results to the thermodynamic limit. Moreover, there are recently proposed numerical methods \cite{zou_conformal_2018} to efficiently diagonalize all low energy eigenstates of a critical quantum spin chain with PBC up to several hundreds of spins. This significantly improves the accuracy of the extrapolation to large system sizes.

However, PBC only allows access to a subset of scaling operators \cite{cardy_finite_1984, cardy_1989}. For the rest, other boundary conditions are needed, such as antiperiodic boundary conditions (APBC) or, more generally, twisted boundary conditions. Furthermore, in case the CFT has an extended symmetry, the representations of the full symmetry group may involve mixing boundary conditions. In particular, when the conformal symmetry is enhanced by supersymmetry \cite{friedan_conformal_1984, friedan_superconformal_1985}, which transforms bosons into fermions, the fermionic excitations (appearing in the APBC sector) are needed to form a representation of the superconformal algebra. 

In this paper, we generalize the method of Refs. \cite{zou_conformal_2018, zou_conformal_2019} to critical spins chains with APBC (more general twisted boundary conditions can be addressed in a similar way). We propose a systematic way of extracting complete conformal data involving scaling operators in the APBC. The efficient diagonalization method of low energy eigenstates is also generalized to the APBC. As applications, we then extract accurate conformal data for two models with APBC, namely the Ising model and a tricritical Ising (TCI) model \cite{obrien_lattice_2018} due to O'Brien and Fendley. For the latter case the APBC excitations enable us to study the emergent superconformal symmetry. 

More specifically, given a critical quantum spin chain with PBC and APBC, we first diagonalize the low-energy eigenstates and identify each eigenstate on the lattice with a CFT scaling operator in the continuum. In particular, primary states can their conformal towers are identified for both boundary conditions. Then we find lattice operators that correspond to CFT primary operators in the continuum limit. All OPE coefficients involving primary operators can then be extracted. Furthermore, in the case where the conformal symmetry is enhanced by supersymmetry, such as the TCI model, we can identify generators of the extended algebra (supervirasoro algebra) on the lattice. By studying the action of supervirasoro generators on the low-energy subspace, we can identify supervirasoro primary states and supervirasoro conformal towers.

This paper is organized as following. In section II we review conformal data and operator-state correspondence of 1+1 dimensional CFT and previous work on extracting conformal data from a critical quantum spin chain. In section III we define APBC for spin chains and APBC for CFT and discuss their relations. In section IV we propose a systematic way of extracting conformal data based on the low energy spectrum of the spin chain with APBC. In section V we generalize the efficient diagonalization of low energy eigenstates of a critical quantum spin chain from PBC to APBC. In section VI, we extract conformal data from the Ising model. In section VII, we consider the more complicated TCI model and extract its conformal data. In section VIII, we construct lattice representations of superconformal generators that connect eigenstates in PBC and APBC and verify the superconformal structure of low energy excitations.

\section{Conformal data and its extraction}
In this section, we review some basic properties of conformal field theories \cite{belavin_infinite_1984, friedan_conformal_1984, ginsparg_applied_1988, henkel_1999, francesco_2012} and summarize the way of extracting conformal data based on the operator-state correspondence in PBC. These results are to be generalized to APBC in this paper.
\subsection{Conformal data} 
In 1+1 dimensions, a conformal field theory consists of a set of scaling operators $\psi^{\CFT}_\alpha(z,\bar{z})$, where $z=x^0+ix^1$ and $\bar{z}=x^0-ix^1$ are complex coordinates. Each scaling operator has a scaling dimension $\Delta^{\CFT}_\alpha$ and a conformal spin $s^{\CFT}_\alpha$, which describe how the scaling operators transform under dilations and rotations, respectively,
\begin{eqnarray}
z\rightarrow z'&=&\lambda z, ~~~~ \psi^{\CFT}_\alpha(z,\bar{z})\rightarrow \lambda^{-\Delta^{\CFT}_\alpha}\psi^{\CFT}_\alpha(z',\bar{z}') \\
z\rightarrow z'&=&e^{i\theta} z, ~~\psi^{\CFT}_\alpha(z,\bar{z})\rightarrow e^{-i\theta s^{\CFT}_\alpha}\psi^{\CFT}_\alpha(z',\bar{z}'),
\end{eqnarray}
where $\lambda>0$ and $0\leq\theta<2\pi$. All two point correlation functions in the CFT are determined by $(\Delta^{\CFT}_\alpha,s^{\CFT}_\alpha)$. 

Scaling operators are organized into conformal towers, with a \textit{primary field} $\phi^{\CFT}_\alpha$ in each conformal tower and the \textit{descendant fields} of this primary field. The descendant fields are related to the primary field by \textit{Virasoro generators} $L^{\CFT}_n$ and $\bar{L}^{\CFT}_n$ ($n\in \mathbb{Z}$) which act as ladder operators. They generate conformal transformations and satisfy the Virasoro algebra
\begin{eqnarray}
\label{VirasoroCFT}
&[&L^{\CFT}_n,L^{\CFT}_m]=(n-m) L^{\CFT}_{n+m} + \frac{c^{\CFT}}{12}n (n^2-1) \delta_{n+m,0} \\
&[&\bar{L}^{\CFT}_n,\bar{L}^{\CFT}_m]=(n-m) \bar{L}^{\CFT}_{n+m} + \frac{c^{\CFT}}{12}n (n^2-1) \delta_{n+m,0} \\
&[&L^{\CFT}_n,\bar{L}^{\CFT}_m]=0,
\end{eqnarray}
where $c^{\CFT}$ is the central charge that is a part of the conformal data. In particular, $D^{\CFT}=L^{\CFT}_0+\bar{L}^{\CFT}_0$ and $R^{\CFT}=L^{\CFT}_0-\bar{L}^{\CFT}_0$ generate dilations and rotations. Scaling operators $\psi^{\CFT}_\alpha(0,0)$ are eigenvectors of $D^{\CFT}$ and $R^{\CFT}$, with eigenvalues $\Delta^{\CFT}_{\alpha}$ and $s^{\CFT}_{\alpha}$,
\begin{align}
&[D^{\CFT},\psi^{\CFT}_\alpha(0,0)]=\Delta^{\CFT}_\alpha \psi^{\CFT}_\alpha(0,0) \\
&[R^{\CFT},\psi^{\CFT}_\alpha(0,0)]\, =s^{\CFT}_\alpha \psi^{\CFT}_\alpha(0,0).
\end{align}
Virasoro generators $L^{\CFT}_{n}$ and $\bar{L}^{\CFT}_{n}$ with $n<0$ are raising operators with respect to dilations -- they raise the scaling dimension by $|n|$, while those with $n>0$ lower the scaling dimension by $n$,
\begin{align}
&[D^{\CFT},[L^{\CFT}_n,\psi^{\CFT}_\alpha(0,0)]]=(\Delta^{\CFT}_\alpha-n) [L^{\CFT}_n,\psi^{\CFT}_\alpha(0,0)]\\
&[D^{\CFT},[\bar{L}^{\CFT}_n,\psi^{\CFT}_\alpha(0,0)]]=(\Delta^{\CFT}_\alpha-n) [\bar{L}^{\CFT}_n,\psi^{\CFT}_\alpha(0,0)].
\end{align}
 Primary fields are those whose scaling dimensions cannot be lowered
\begin{equation}
[L^{\CFT}_n, \phi^{\CFT}_\alpha(0,0)]=0,[\bar{L}^{\CFT}_n, \phi^{\CFT}_\alpha(0,0)]=0~~(n>0).
\end{equation}
Descendant fields in a conformal tower can be obtained by applying raising operators $L^{\CFT}_{n},\bar{L}^{\CFT}_n$ ($n<0$) to the primary field of that tower. 

The scaling operators form an associative algebra, the \textit{operator product expansion} (OPE), which relates the product of two scaling operators to a linear superposition of scaling operators. Three and higher point correlation functions are determined by the OPE. It turns out the OPE is completely characterized by the coefficients $C^{\CFT}_{\alpha\beta\gamma}$ involving primary fields \cite{belavin_infinite_1984, friedan_conformal_1984, cardy_operator_1986}. Therefore, we conclude that the \textit{conformal data} consists of $\{\Delta^{\CFT}_\alpha,s^{\CFT}_\alpha,C^{\CFT}_{\alpha\beta\gamma}\}$ for primary fields and the central charge $c^{\CFT}$.
 
\subsection{Operator-state correspondence and extraction of conformal data from a CFT on the cylinder}
The CFT can be put on a cylinder with a compactified spatial direction $x\in[0,L)$ and an uncompactified time direction $\tau\in(-\infty,\infty)$. The Hamiltonian and momentum are the integrals of the Hamiltonian and momentum density $h^{\CFT}(x)$ and $p^{\CFT}(x)$ on the time slice $\tau=0$,
\begin{eqnarray}
H^{\CFT}&=&\int_0^L dx\, h^{\CFT}(x)  \\
P^{\CFT}&=&\int_0^L dx\, p^{\CFT}(x).
\end{eqnarray}
They are related to the generators of dilations and rotations of the CFT on the complex plane by
\begin{eqnarray}
\label{HCFT}
H^{\CFT}&=&\frac{2\pi}{L}\left(L^{\CFT}_0+\bar{L}^{\CFT}_0-\frac{c}{12}\right)  \\
\label{PCFT}
P^{\CFT}&=& \frac{2\pi}{L}\left(L^{\CFT}_0-\bar{L}^{\CFT}_0\right).
\end{eqnarray}
The operator-state correspondence \cite{belavin_infinite_1984, friedan_conformal_1984, cardy_conformal_1984, blote_1986} says that all simultaneous eigenstates $|\psi^{\CFT}_\alpha\rangle$ of $H^{\CFT}$ and $P^{\CFT}$ are in one to one correspondence with scaling operators $\psi^{\CFT}_\alpha$. The energy and momentum are thus related to scaling dimensions and conformal spins by
\begin{eqnarray}
\label{EPCFT}
E^{\CFT}_\alpha&=& \frac{2\pi}{L}\left(\Delta^{\CFT}_\alpha-\frac{c}{12}\right)  \\
P^{\CFT}_\alpha&=& \frac{2\pi}{L}s^{\CFT}_\alpha.
\end{eqnarray}
Virasoro generators can be expressed as Fourier modes of $h(x)$ and $p(x)$. In particular, the Fourier modes of $h^{\CFT}(x)$,
\begin{equation}
H^{\CFT}_n\equiv\frac{L}{2\pi}\int_0^L dx\, h^{\CFT}(x) e^{inx 2\pi/L}
\end{equation}
equal a linear combination of Virasoro generators \cite{koo_representations_1994, milsted_extraction_2017},
\begin{equation}
\label{CFTLn}
H^{\CFT}_n=L^{\CFT}_{n}+\bar{L}^{\CFT}_{-n}-\frac{c^{\CFT}}{12}\delta_{n,0}.
\end{equation}
The central charge can be computed by
\begin{equation}
\label{cCFT}
c^{\CFT}=2|\langle T^{\CFT}|H^{\CFT}_{-2}|0^{\CFT}\rangle|^2,
\end{equation}
where $|0^{\CFT}\rangle$ is the ground state of the CFT and the state $|T^{\CFT}\rangle$ corresponds to the stress tensor of the CFT.

Primary fields $\phi^{\CFT}_\alpha$ correspond to \textit{primary states} $|\phi^{\CFT}_\alpha\rangle$, which are characterized by \cite{milsted_extraction_2017, zou_conformal_2018} 
\begin{equation}
L^{\CFT}_n|\phi^{\CFT}_\alpha\rangle=0, \bar{L}^{\CFT}_n|\phi^{\CFT}_\alpha\rangle=0 ~~(n>0).
\end{equation}
Note that the Virasoro algebra Eq.~\eqref{VirasoroCFT} implies that the above equalities hold for all $n>0$ if they hold for $n=1,2$, since other generators can be obtained by commutators of those with $n=1,2$.

The two Virasoro generators $L^{\CFT}_{-n}$ and $\bar{L}^{\CFT}_n$ in $H^{\CFT}_n$, Eq.~\eqref{CFTLn}, raise scaling dimensions and lower scaling dimensions respectively for $n\neq 0$. Therefore, primary states can be alternatively defined as 
\begin{equation}
\label{primaryCFT}
P^{\CFT}_{\phi_{\alpha}} H^{\CFT}_n|\phi^{\CFT}_\alpha\rangle=0 ~~(n=\pm 1, \pm 2),
\end{equation}
where $P^{\CFT}_{\phi_{\alpha}}$ is a projector onto the subspace spanned by states whose scaling dimension is smaller than that of $\phi^{\CFT}_\alpha$. 

Descendant states corresponding to descendant fields can be created by acting with $L^{\CFT}_n$,$\bar{L}^{\CFT}_n$ ($n<0$) on primary states.

OPE coefficients of primary fields are related to matrix elements of primary operators,
\begin{equation}
\label{OPECFT}
\langle \phi^{\CFT}_\alpha|\phi^{\CFT}_\beta(x=0)|\phi^{\CFT}_\gamma\rangle=\left(\frac{2\pi}{L}\right)^{\Delta^{\CFT}_\beta} C^{\CFT}_{\alpha\beta\gamma}.
\end{equation}
Translation invariance of the eigenstates allow us to express the OPE coefficients in terms of the Fourier mode of the CFT operator,
\begin{equation}
\label{OPECFT2}
\langle \phi^{\CFT}_\alpha|\phi^{\CFT,s_{\alpha}-s_{\gamma}}_\beta|\phi^{\CFT}_\gamma\rangle=\left(\frac{2\pi}{L}\right)^{\Delta^{\CFT}_\beta} C^{\CFT}_{\alpha\beta\gamma},
\end{equation}
where the Fourier mode of any CFT operaror is defined as
\begin{equation}
\label{CFTFourier}
\mathcal{O}^{\CFT,s}\equiv \frac{1}{L}\int_0^L dx\, \mathcal{O}^{\CFT}(x) e^{-isx2\pi/L}.
\end{equation}

\subsection{Extraction of conformal data from a spin chain}
Given a critical quantum spin chain $H=\sum_j h_j$ with PBC, it has been shown \cite{cardy_conformal_1984, blote_1986, affleck_universal_1986, cardy_operator_1986, koo_representations_1994, milsted_extraction_2017, zou_conformal_2019} that all conformal data can be extracted from appropriate matrix elements of lattice operators in the low energy eigenstates, which we now review. We will see that the formulas are analogous to Eqs.~\eqref{EPCFT}-\eqref{OPECFT}.

At low energies, each eigenstate of $H$ is in one to one correspondence with a CFT state,
\begin{equation}
|\psi_\alpha\rangle\sim|\psi^{\CFT}_\alpha\rangle.
\end{equation}
Denote the total number of sites as $N$. We can think of $N$ as analogous to the circumference $L$ of the cylinder. The energy and momentum $E_{\alpha}$ and $P_{\alpha}$ are related to $\Delta^{\CFT}_\alpha$ and $s^{\CFT}_\alpha$ by \cite{cardy_conformal_1984, blote_1986, affleck_universal_1986, cardy_operator_1986, cardy_logarithmic_1986}
\begin{eqnarray}
\label{EP}
E_{\alpha}&=&NA+\frac{B}{N}\left(\Delta^{\CFT}_\alpha-\frac{c}{12}\right)+o(N^{-1}) \\
\label{EP2}
P_{\alpha}&=&\frac{2\pi}{N}s^{\CFT}_\alpha,
\end{eqnarray}
where $A,B$ are non-universal constants related to the lattice model. At sufficiently large sizes, the subleading correction $o(N^{-1})$ is negligible. The constants $A$ and $B$ can be numerically estimated by using the fact that scaling dimensions of the ground state and the stress tensor state are always $\Delta_{I}=0$, $\Delta_T=2$ in a unitary CFT. $A$ is the ground state energy density in the thermodynamic limit and 
\begin{equation}
B\approx\frac{N}{2}(E_{T}-E_0).
\end{equation}
We can then extract the scaling dimensions and conformal spins as
\begin{eqnarray}
\label{DeltaPBC}
\Delta_\alpha&=&2\frac{E_\alpha-E_0}{E_T-E_0} \\
\label{sPBC}
s_\alpha &=& \frac{N}{2\pi}P_\alpha.
\end{eqnarray}
Notice that $\Delta_\alpha$ and $\Delta^{\CFT}_\alpha$ differ by finite-size corrections, but $s_\alpha=s^{\CFT}_\alpha$ (up to periodicity $N$) is exact because momenta are quantized. To obtain a more accurate approximation to $\Delta^{\CFT}_\alpha$, we can compute $\Delta_\alpha$ as a function of $N$ at finite sizes and extrapolate to $N\rightarrow \infty$ (the thermodynamic limit).

Furthermore, a local lattice operator $\mathcal{O}$ can be identified with a CFT operator $\mathcal{O}^{\CFT}$ \cite{cardy_operator_1986, zou_conformal_2019}
\begin{equation}
\label{operatorID}
\mathcal{O}\sim \mathcal{O}^{\CFT}=\sum_{\alpha} a_\alpha \psi^{\CFT}_\alpha,
\end{equation}
where $\psi^{\CFT}_\alpha$ are scaling operators. In general, the expansion Eq.~\eqref{operatorID} involves infinite scaling operators with arbitrarily large scaling dimensions. However, we will work with a truncated expansion to some maximal scaling dimension $\Delta^{\CFT}$. To compare a lattice operatr $\mathcal{O}$ with a CFT operator $\mathcal{O}^{\CFT}$, we first define Fourier modes of the lattice operator \footnote{For a multi-site operator $\mathcal{O}$, there is an ambiguity with the definition of the Fourier mode Eq.~\eqref{latticeFourier}, addressed in Ref.\cite{zou_conformal_2019} and further illustrated in appendix.},
\begin{equation}
\label{latticeFourier}
\mathcal{O}^s\equiv \frac{1}{N}\sum_{j=1}^N e^{-isj 2\pi/N} \mathcal{O}_j.
\end{equation}
They connect states with conformal spin $s_\alpha$ to states with conformal spin $s_\alpha+s$.  The coefficients $a_\alpha$ can be obtained by 
minimizing the cost function \cite{zou_conformal_2019}
\begin{equation}
\label{costf}
f^{\mathcal{O}}(\{a_\alpha\})=\sum_\beta |\langle \psi_\beta|\mathcal{O}^{s_\beta}|0\rangle-\langle \psi^{\CFT}_\beta|\mathcal{O}^{\CFT,{s_\beta}}|0^{\CFT}\rangle|^2
\end{equation}
for some subset of low energy eigenstates $\{|\psi_\beta\rangle\}$. Note that the CFT matrix elements $\langle \psi^{\CFT}_\beta|\mathcal{O}^{\CFT,{s_\beta}}|0^{\CFT}\rangle$ can be computed as a function of scaling dimensions, conformal spins and the central charge \cite{zou_conformal_2019}. Because there are finite-size corrections, the cost function is typically non-vanishing at finite $N$. An estimation of the coefficient $a_\alpha$ in the thermodynamic limit is obtained by an extrapolation of the finite-size data. 

In particular, the lattice Hamiltonian density $h_j$ corresponds to the CFT Hamiltonian density $h^{\CFT}(x)$ (up to a normalization factor and a constant shift), whose Fourier modes are linear combinations of Virasoro generators \cite{koo_representations_1994, read_associative-algebraic_2007, dubail_conformal_2010, vasseur_puzzle_2012,
gainutdinov_logarithmic_2013, gainutdinov_lattice_2013, bondesan_chiral_2015}. Therefore, 
\begin{equation}
\label{Ln}
H_n\equiv \frac{N}{B} \sum_{j=1}^N h_j e^{inj2\pi/N}\sim H^{\CFT}_n.
\end{equation}
Denote $P_{\phi_\alpha}$ the projector onto the subspace spanned by eigenstates whose energies are lower than that of $|\phi_\alpha\rangle$. Then, analogous to Eq.~\eqref{primaryCFT}, primary states can be identified by the condition \cite{milsted_extraction_2017, zou_conformal_2018} that
\begin{equation}
P_{\phi_\alpha}H_n|\phi_\alpha\rangle=0 ~~(n=\pm 1,\pm 2)
\end{equation}
in the thermodynamic limit. Other eigenstates can be approximately obtained by acting with $H_n~(n\neq 0)$ on the primary states, proceeding in analogy with the CFT.
 
An estimate $c$ of the central charge $c^{\CFT}$ is obtained from
\begin{equation}
\label{clat}
c=2|\langle T|H_{-2}|0\rangle|^2,
\end{equation}
analogous to Eq.~\eqref{cCFT}. Again, a suitable extrapolation to the thermodynamic limit wii be used.

The identification Eq.~\eqref{operatorID} allows us to find a lattice operator $\mathcal{O}_{\phi_\beta}$ corresponding to the primary operator $\phi^{\CFT}_\beta$ by inverting the linear expansion. The OPE coefficients $C^{\CFT}_{\alpha\beta\gamma}$ can be extracted approximately from \cite{cardy_operator_1986, zou_conformal_2019}
\begin{equation}
\label{OPE}
C_{\alpha\beta\gamma}=\left(\frac{2\pi}{N}\right)^{-\Delta_\beta} \langle \phi_\alpha|\mathcal{O}_{\phi_\beta,j=0}|\phi_\gamma\rangle.
\end{equation}

Due to translation invariance, the above matrix elements can also be computed by
\begin{equation}
C_{\alpha\beta\gamma}=\left(\frac{2\pi}{N}\right)^{-\Delta_\beta} \langle \phi_\alpha|\mathcal{O}^{s_\alpha-s_\gamma}_{\phi_\beta}|\phi_\gamma\rangle,
\end{equation}
analogous to Eq.~\eqref{OPECFT2}. Again, $C^{\CFT}_{\alpha\beta\gamma}$ can be obtained by extrapolating $C_{\alpha\beta\gamma}$ to the thermodynamic limit.

In order to extrapolate accurate conformal data in the thermodynamic limit, it is essential to obtain energy eigenstates for a series of system sizes $N$ that are sufficiently large, such that finite-size corrections are small. To go beyond exact diagonalization, we will use a tensor network method based on matrix product states \cite{white_1992, fannes_1992, rommer_1997, vidal_2004, vidal_2007a, degli_investigation_2004, tagliacozzo_2008, xavier_entanglement_2010, stojevic_2015, schollwock_2011}. It has been shown \cite{pirvu_exploiting_2011, pirvu_matrix_2012, zou_conformal_2018} that the tensor network method can be applied to obtain any energy eigenstate with sufficiently low energy for system sizes $N$ far beyond the reach of exact diagonalization, see section V.

We conclude this section with Table \ref{table} that summarizes the correspondence between lattice objects and CFT objects for the PBC.
\begin{table*}[htbp]
\begin{tabular}{|c|c|c|c|}
\hline 

CFT & Lattice (at low energies)   \\ \hline \hline
       Energy/momentum eigenstate $|\psi^{\CFT}_\alpha\rangle$ & Energy/momentum eigenstate $|\psi_\alpha\rangle$  \\ \hline
       Primary state $|\phi^{\CFT}_\alpha\rangle$ & Primary state $|\phi_\alpha\rangle$  \\ \hline
       Energy $E^{\CFT}_\alpha= \frac{2\pi}{L} (\Delta^{\CFT}_\alpha-{c^{\CFT}}/{12})$   & Energy $E_\alpha \approx NA+\frac{B}{2\pi}E^{\CFT}_\alpha$ (Eq.~\eqref{EP}) \\ \hline
       Momentum $P^{\CFT}_\alpha=\frac{2\pi}{L} s^{\CFT}_\alpha$          & Momentum $P_\alpha=P^{\CFT}_\alpha$ (Eq.~\eqref{EP2}) \\ \hline
       Local operator $\mathcal{O}^{\CFT}=\sum_\alpha a_\alpha \psi^{\CFT}_\alpha$     & Local operator $\mathcal{O}$ \\ \hline
        Fourier mode $\mathcal{O}^{\CFT,s}$ (Eq.~\eqref{CFTFourier}) & Fourier mode $\mathcal{O}^{s}$ (Eq.~\eqref{latticeFourier})\\ \hline
       $H^{\CFT}_n=L^{\CFT}_{n}+\bar{L}^{\CFT}_{-n}-{c^{\CFT}}/{12}\delta_{n,0}$     &  $H_n$ (Eq.~\eqref{Ln})  \\ \hline
      Primary operator $\phi^{\CFT}_\beta$   &  Operator $\mathcal{O}_{\phi_\beta}$ \\ \hline
      $C^{\CFT}_{\alpha\beta\gamma}= (2\pi/L)^{-\Delta^{\CFT}_\beta}\langle \phi^{\CFT}_\alpha|\phi^{\CFT, s_{\alpha}-s_{\gamma}}_{\beta}|\phi^{\CFT}_\gamma\rangle$  & $C_{\alpha\beta\gamma}= (2\pi/N)^{-\Delta_\beta}\langle \phi_\alpha|\mathcal{O}^{s_{\alpha}-s_{\gamma}}_{\phi_\beta}|\phi_\gamma\rangle \approx C^{\CFT}_{\alpha\beta\gamma} $  \\ \hline
\end{tabular}
\caption{Correspondence of lattice objects and CFT objects. The ''$\approx$'' means equal up to finite-size corrections.}
\label{table}
\end{table*}
\section{Antiperiodic boundary conditions}
In this section we review the definition and implications of antiperiodic boundary conditions (APBC) \cite{cardy_finite_1984, cardy_1989, henkel_1987, burkhardt_1985, aasen_2016, hauru_topological_2016} for spin chains and CFT. 
\subsection{APBC for spin chains}
In contrast with PBC, which can be defined for any spin chain, the APBC is only defined for a spin chain with an on-site $\mathbb{Z}_2$ symmetry. Denote the Hamiltonian density as $h_j$ and the lattice translation operator as $\mathcal{T}$,
\begin{equation}
\mathcal{T} h_j \mathcal{T}^{\dagger}=h_{j+1},
\end{equation}
where the site $j=N+1$ is identified with site $j=1$. The Hamiltonian with PBC is then
\begin{equation}
H^{\PBC}=\sum_{j=1}^N \mathcal{T}^{j-1} h_1 \mathcal{T}^{\dagger j-1}.
\end{equation}
To define the APBC, we first denote a representation of the $\mathbb{Z}_2$ generator as
\begin{equation}
\label{Z2latt}
\mathcal{Z}=\prod_{j=1}^N \mathcal{Z}_j,
\end{equation}
where 
\begin{equation}
\mathcal{Z}_j=\mathcal{Z}^{\dagger}_j=\mathcal{Z}^{-1}_j.
\end{equation}
The Hamiltonian density is invariant under the $\mathbb{Z}_2$ transformations,
\begin{equation}
h_j=\mathcal{Z}h_j\mathcal{Z}^{\dagger}.
\end{equation}
Define a twisted translation operator as
\begin{equation}
\tilde{\mathcal{T}}=\mathcal{Z}_1 \mathcal{T},
\end{equation}
then Hamiltonian for APBC is 
\begin{equation}
H^{\APBC}=\sum_{j=1}^N \tilde{\mathcal{T}}^{j-1} h_1 \tilde{\mathcal{T}}^{\dagger j-1},
\end{equation}
which is invariant under the twisted translation $\tilde{\mathcal{T}}$. The Hamiltonians with both boundary conditions are invariant under the $\mathbb{Z}_2$ transformation,
\begin{eqnarray}
H^{\PBC}&=&\mathcal{Z}H^{\PBC}\mathcal{Z}^{\dagger}. \\
H^{\APBC}&=&\mathcal{Z}H^{\APBC}\mathcal{Z}^{\dagger}.
\end{eqnarray}
Let us illustrate the APBC with the two models that are used in this paper. The first example is the critical Ising model,
\begin{equation}
h_{\Ising,j}=-X_j X_{j+1}-Z_j,
\end{equation}
where $X,Y,Z$ are Pauli operators,
\begin{equation}
X=\begin{bmatrix} 0 & 1 \\ 1 & 0 \end{bmatrix},
Y=\begin{bmatrix} 0 & -i \\ i & 0 \end{bmatrix},
Z=\begin{bmatrix} 1 & 0 \\ 0 & -1 \end{bmatrix}.
\end{equation}
 The $\mathbb{Z}_2$ symmetry is generated by $\mathcal{Z}_j=Z_j$. The Hamiltonian with APBC is
\begin{equation}
H^{\APBC}_{\Ising}=-\sum_{j=1}^{N-1} X_j X_{j+1}-\sum_{j=1}^N Z_j+X_NX_1.
\end{equation}
The other example is a tricritical Ising (TCI) model due to O'Brien and Fendley,
\begin{equation}
\label{hTCI}
h_{\TCI,j}=h_{\Ising,j}+\lambda^{*}(X_j X_{j+1} Z_{j+2}+Z_j X_{j+1} X_{j+2}),
\end{equation}
where $\lambda^{*}\approx 0.428$ is the tricritical point. It is also $\mathbb{Z}_2$ invariant under the global spin flip. The Hamiltonian with APBC is
\begin{eqnarray}
H^{\APBC}_{\TCI}&=&H^{\APBC}_{\Ising}+\lambda^{*}\sum_{j=1}^{N-2} (X_j X_{j+1}Z_{j+2}+Z_j X_{j+1} X_{j+2})  \nonumber \\
&+& \lambda^{*}(X_{N-1}X_NZ_1-X_NX_1Z_2) \label{TCI_b1} \\
&+&\lambda^{*}(-Z_{N-1}X_NX_1+Z_NX_1X_2). \label{TCI_b2}
\end{eqnarray}
We stress that the boundary term in the APBC Hamiltonian does not always have opposite sign to that of the PBC Hamiltonian, as is evident from Eqs.~\eqref{TCI_b1},\eqref{TCI_b2}. In a spin-1/2 chain with $\mathcal{Z}_j=Z_j$, the APBC is simply 
\begin{eqnarray}
\label{spinbc1}
X_{N+j}&=&-X_j \\
Y_{N+j}&=&-Y_j \\
\label{spinbc3}
Z_{N+j}&=&Z_j. 
\end{eqnarray}

Energy eigenstates can be labelled by eigenvalues of $H^{\PBC},\mathcal{T},\mathcal{Z}$ in the PBC, 
\begin{eqnarray}
H^{\PBC}|\psi^{\PBC}_\alpha\rangle&=&E^{\PBC}_\alpha |\psi^{\PBC}_\alpha\rangle \\
\mathcal{T}|\psi^{\PBC}_\alpha\rangle&=&e^{iP^{\PBC}_\alpha} |\psi^{\PBC}_\alpha\rangle \\
\mathcal{Z}|\psi^{\PBC}_\alpha\rangle&=&\mathcal{Z}^{\PBC}_\alpha |\psi^{\PBC}_\alpha\rangle,
\end{eqnarray}
where $\mathcal{Z}_\alpha=\pm 1$ corresponds to the even (odd) parity. In the APBC, the eigenstates are labelled by eigenvalues of $H^{\APBC},\tilde{\mathcal{T}},\mathcal{Z}$.
\begin{eqnarray}
H^{\APBC}|\psi^{\APBC}_\alpha\rangle&=&E^{\APBC}_\alpha |\psi^{\APBC}_\alpha\rangle \\
\tilde{\mathcal{T}}|\psi^{\APBC}_\alpha\rangle&=&e^{iP^{\APBC}_\alpha} |\psi^{\APBC}_\alpha\rangle \\
\mathcal{Z}|\psi^{\APBC}_\alpha\rangle&=&\mathcal{Z}^{\APBC}_\alpha |\psi^{\APBC}_\alpha\rangle.
\end{eqnarray}
The momenta are quantized in both cases. Denote
\begin{eqnarray}
P^{\PBC}_\alpha &=& \frac{2\pi}{N} s^{\PBC}_\alpha \\
P^{\APBC}_\alpha &=& \frac{2\pi}{N} s^{\APBC}_\alpha.
\end{eqnarray}
Since $\mathcal{T}^N=\mathbf{1}$ (a translation by $N$ sites in PBC is trivial) and $\tilde{\mathcal{T}}^N=\mathcal{Z}\mathcal{T}^N=\mathcal{Z}$ (a translation by $N$ sites in APBC amounts to a global spin flip), 
\begin{eqnarray}
\label{slatt1}
s^{\PBC}_\alpha &\in& \mathbb{Z}\\
\label{slatt2}
s^{\APBC}_\alpha &\in& \mathbb{Z} ~~~(\mathcal{Z}^{\APBC}_\alpha=1) \\
\label{slatt3}
s^{\APBC}_\alpha &\in& \mathbb{Z}+\frac{1}{2} ~~~(\mathcal{Z}^{\APBC}_\alpha=-1),
\end{eqnarray}
For critical theories, it means that the conformal spin of a $\mathbb{Z}_2$ even operator in APBC is integer but that of a $\mathbb{Z}_2$ odd operator is half integer. 

\subsection{APBC for CFT}
The PBC and APBC for a conformal field theory on the cylinder are defined as
\begin{equation}
\Psi^{\CFT}(x+L)=\pm\Psi^{\CFT}(x)
\end{equation}
for some fundamental field $\Psi^{\CFT}(x)$, where $+$ is for PBC and $-$ is for APBC. Any field can be expanded into Fourier modes,
\begin{equation}
\psi^{\CFT, s}_\alpha\equiv  \frac{1}{L}\int_0^L dx\, \psi^{\CFT}_\alpha(x) e^{isx 2\pi/L}.
\end{equation} 
In the path integral formalism, the choice of boundary condition affects the Fourier mode of fields that enter into the partition function \cite{cardy_1989}. Therefore the boundary condition affects the operator content of the CFT. 

For concretness, we will consider the case where the fundamental field $\Psi^{\CFT}(x)$ is a fermionic operator. The boundary conditions $\pm$ are usually referred to as Ramond (R) \cite{ramond_dual_1971} and Neveu-Schwarz (NS) \cite{neveu_tachyon_1971}, respectively. 

If the field $\psi^{\CFT}_\alpha(x)$ is a fermionic field, then the only nonvanishing Fourier modes have $s\in \mathbb{Z}+1/2$ in the NS sector, and $s\in \mathbb{Z}$ in the R sector. If the field $\psi^{\CFT}_\alpha(x)$ is a bosonic field (e.g. product of two fermionic fields), then the only nonvanishing Fourier modes have $s\in \mathbb{Z}$ in both NS and R sectors. As a result, the conformal spins of the scaling operators satisfy
\begin{eqnarray}
\label{scft1}
s^{\CFT}_\alpha &\in& \mathbb{Z}+\frac{1}{2} ~~\mathrm{(NS, fermion)} \\
\label{scft2}
s^{\CFT}_\alpha &\in& \mathbb{Z} ~~\mathrm{(NS, boson)} \\
\label{scft3}
s^{\CFT}_\alpha &\in& \mathbb{Z} ~~(\mathrm{R}) 
\end{eqnarray}
Eqs.~\eqref{slatt1}-\eqref{slatt3} are somewhat similar to Eqs.~\eqref{scft1}-\eqref{scft3}. They are related by the well known Jordan-Wigner transformation \cite{jordan_1928} that relates a spin-1/2 chain in 1+1 dimensions to a fermion chain with specific choices of boundary conditions. We will review this below.

\subsection{Jordan-Wigner transformation}
Given a spin-1/2 chain with the above $\mathbb{Z}_2$ symmetry, the Jordan-Wigner transformation 
\begin{eqnarray}
\gamma_{2j-1}&=&\left(\prod_{n=1}^{j-1} Z_n\right) X_j  \\
\gamma_{2j}&=&\left(\prod_{n=1}^{j-1} Z_n\right) Y_j 
\end{eqnarray}
makes it a Majorana fermion chain \cite{obrien_lattice_2018, rahmani_phase_2015-1, rahmani_emergent_2015} with $2N$ Majorana modes $\gamma_j$. It can be verified that the Majorana operators $\gamma_j$s are Hermitian, and satisfy anticommutation relations
\begin{equation}
\{\gamma_j,\gamma_l\}=2\delta_{jl}.
\end{equation}
They are fermionic operators because they anticommute on different sites (as opposed to spin operators which commute on different sites). Also notice that the Majorana operators are nonlocal in terms of spin operators. They are string operators in a spin chain (see the following section). However, local products of an even number of Majorana operators are local in spins operators, e.g.
\begin{eqnarray}
Z_j &=& i\gamma_{2j-1}\gamma_{2j} \\
X_jX_{j+1} &=& i\gamma_{2j}\gamma_{2j+1}.
\end{eqnarray}

Majorana operators are odd under $\mathbb{Z}_2$,
\begin{equation}
\mathcal{Z} \gamma_j \mathcal{Z}^{\dagger}=-\gamma_j.
\end{equation}
Then the R (NS) boundary conditions for a Majorana fermion chain are
\begin{equation}
\label{fermionbc}
\gamma_{2N+j}=\pm \gamma_j.
\end{equation}
In the underlying field theory, they correspond to R (NS) boundary conditions, respectively. Comparing Eq.~\eqref{fermionbc} with Eqs.~\eqref{spinbc1}-\eqref{spinbc3}, it is straightforward to see that under the Jordan-Wigner transformation, the R boundary condition for the fermion chain corresponds to the $\mathbb{Z}_2$ odd sector of PBC of the spin chain, and the $\mathbb{Z}_2$ even sector of APBC of the spin chain. The NS boundary condition for the fermion chain corresponds to the $\mathbb{Z}_2$ even sector of PBC of the spin chain and the $\mathbb{Z}_2$ odd sector of APBC of the spin chain, i.e.,
\begin{eqnarray}
\mathrm{NS}=\mathrm{PBC}(\mathcal{Z}=1)~+~\mathrm{APBC}(\mathcal{Z}=-1) \\
\mathrm{R}=\mathrm{PBC}(\mathcal{Z}=-1)~+~\mathrm{APBC}(\mathcal{Z}=1).
\end{eqnarray}
The conformal spins of scaling operators in each boundary condition, i.e., Eqs.~\eqref{slatt1}-\eqref{slatt3} for spin chains and Eqs.~\eqref{scft1}-\eqref{scft3} for fermion chains are consistent with this assignment.

The Ising model and the TCI model studied in this paper can be transformed into Majorana models \cite{obrien_lattice_2018}. The Hamiltonian densities are 
\begin{eqnarray}
\label{IsingFermion}
h_{\Ising,j}&=&-i(\gamma_{2j-1}\gamma_{2j}+\gamma_{2j}\gamma_{2j+1}) \\
\label{TCIFermion}
h_{\TCI,j}&=&-i(\gamma_{2j-1}\gamma_{2j}+\gamma_{2j}\gamma_{2j+1}) \\
&-&\lambda^{*}(\gamma_{2j-1}\gamma_{2j}\gamma_{2j+2}\gamma_{2j+3}+\gamma_{2j}\gamma_{2j+1}\gamma_{2j+3}\gamma_{2j+4}) \nonumber.
\end{eqnarray} 
It can be seen that the Ising model is quadratic in terms of Majorana operators. Therefore it is a free fermion model and can be solved exactly. On the other hand, the TCI model is not free (it contains quartic terms), nor integrable. Its low energy eigenstates can only be computed numerically. However, the underlying CFT of the TCI model is a unitary minimal model, which can be solved exactly.

Both the Ising model and the TCI model are invariant under the translation of the Majorana modes:
\begin{equation}
\gamma_{2j}\rightarrow \gamma_{2j+1},\,\,\gamma_{2j+1}\rightarrow \gamma_{2(j+1)}.
\end{equation}
In spin variables, the transformation exchanges $Z$ and $XX$, 
\begin{equation}
Z_j\rightarrow X_j X_{j+1}, \,\, X_j X_{j+1}\rightarrow Z_{j+1}.
\end{equation}
This nontrivial transformation is known as the Kramers-Wannier duality. Notice that acting with the duality twice amounts to the translation operator $\mathcal{T}$. We will later see its role in classifying scaling operators.
\section{Conformal data from low energy eigenstates in APBC}
In this section, we generalize the methods in section II.C to extract conformal data from critical spin chains with APBC. 
\subsection{The Hilbert space}
We shall first make some comments on the Hilbert space of the spin chain with PBC and APBC. It is clear that low energy states of PBC and APBC reside in the \textit{same} Hilbert space on the lattice. However, the low energy states with respect to $H^{\PBC}$ are not low energy eigenstates with respect to $H^{\APBC}$ (and vice versa). Indeed the boundary defect costs $O(1)$ energy as opposed to $O(1/N)$ (Eq.~\eqref{EP}). Therefore, the low energy subspaces of PBC and APBC are orthogonal in the thermodynamic limit. This is to be expected because they correspond to different primary operators, and states in different conformal towers are orthogonal in the CFT. Later, we will call a CFT operator a (A)PBC operator if the corresponding state appears in the low-energy spectrum of the spin chain with (A)PBC.

Fourier modes of local operators can only connect low-energy states within the same boundary condition. To connect low-energy PBC states and low-energy APBC states, we need nonlocal string operators (introduced below). As a result, lattice operators that correspond to APBC scaling operators are string operators, in contrast to local operators which correspond to PBC scaling operators. 
\subsection{Scaling dimensions and conformal spins}
Since Eqs.~\eqref{HCFT},\eqref{PCFT} is still valid for the CFT \cite{cardy_finite_1984}, we expect that Eqs.~\eqref{EP},\eqref{EP2} are also valid for excitations with APBC. Therefore, scaling dimensions and conformal spins can still be extracted by
\begin{eqnarray}
\label{DeltaAPBC}
\Delta^{\APBC}_\alpha&=&2\frac{E^{\APBC}_\alpha-E^{\PBC}_0}{E^{\PBC}_T-E^{\PBC}_0} \\
\label{sAPBC}
s^{\APBC}_\alpha &=& \frac{N}{2\pi}P^{\APBC}_\alpha.
\end{eqnarray}
The only difference from the PBC is that now $s^{\APBC}_\alpha$ can be integer or half integer depending on the $\mathbb{Z}_2$ sector. 
\subsection{Virasoro generators}
Eq.~\eqref{CFTLn} is still valid for CFT with both boundary conditions. However, on the lattice, the boundary term of $H_n$ should be chosen such that it respects the boundary condition. For PBC, Eq.~\eqref{Ln} can be rewritten as
\begin{equation}
\label{LnPBC}
H^{\PBC}_n=\frac{N}{B}\sum_{j=1}^N \mathcal{T}^{j-1} h_1 \mathcal{T}^{\dagger j-1} e^{inj2\pi/N}.
\end{equation} 
$H^{\PBC}_n$ is covariant under the translation operator of PBC,
\begin{equation}
\mathcal{T}H^{\PBC}_n \mathcal{T}^{\dagger}=e^{-in2\pi/N}H^{\PBC}_n.
\end{equation}

For the APBC, the translation operator is $\tilde{\mathcal{T}}$. Therefore the definition of Fourier modes is changed accordingly. We can define
\begin{equation}
\label{LnAPBC}
H^{\APBC}_n=\frac{N}{B}\sum_{j=1}^N \tilde{\mathcal{T}}^{j-1} h_1 \tilde{\mathcal{T}}^{\dagger j-1} e^{inj2\pi/N},
\end{equation} 
which is covariant under $\tilde{\mathcal{T}}$,
\begin{equation}
\tilde{\mathcal{T}}H^{\APBC}_n \tilde{\mathcal{T}}^{\dagger}=e^{-in2\pi/N}H^{\APBC}_n.
\end{equation}
It is expected that
\begin{equation}
H^{\APBC}_n\sim H^{\CFT}_n = L^{\CFT}_{-n}+\bar{L}^{\CFT}_n-\frac{c^{\CFT}}{12}\delta_{n,0}
\end{equation}
for low energy states in APBC.

Note that Eq.~\eqref{LnAPBC} is a sum of local operators, and therefore it only connects low energy states within APBC in the scaling limit. This is consistent with the action of Virasoro generators, since scaling operators in different boundary conditions necessarily belong to different conformal towers.  

Once we have the lattice Virasoro generators, we can proceed as in the case of PBC \cite{koo_representations_1994, milsted_extraction_2017, zou_conformal_2018} to identify primary and descendant states.
\subsection{Local operators and OPE coefficients}
For a general local lattice operator $\mathcal{O}_j$, we can define its Fourier modes with respect to the  translation operator $\mathcal{T}$ for PBC and $\tilde{\mathcal{T}}$ for APBC as
\begin{equation}
\label{FourierPBC}
\mathcal{O}^{s}=\frac{1}{N}\sum_{j=1}^N \mathcal{T}^{j-1} \mathcal{O}_1 \mathcal{T}^{\dagger j-1} e^{-isj2\pi/N},
\end{equation} 
which is just a rewritten form of the Eq.~\eqref{latticeFourier}, and
\begin{equation}
\label{FourierAPBC}
\tilde{\mathcal{O}}^{s}=\frac{1}{N}\sum_{j=1}^N \tilde{\mathcal{T}}^{j-1} \mathcal{O}_1 \tilde{\mathcal{T}}^{\dagger j-1} e^{-isj2\pi/N}.
\end{equation} 
For the latter case, $s$ is integer if $\mathcal{O}$ is $\mathbb{Z}_2$ even, and $s$ is half integer if $\mathcal{O}$ is $\mathbb{Z}_2$ odd. 

Given a local operator $\mathcal{O}_{\phi_\beta}$ corresponding to a PBC primary field $\phi^{\CFT,\PBC}_\beta$, OPE coefficients can be extracted by 
\begin{eqnarray}
\label{OPEPPP}
C^{\PPP}_{\alpha\beta\gamma} &=& \left(\frac{2\pi}{N}\right)^{-\Delta^{\PBC}_{\beta}}\langle \psi^{\PBC}_\alpha|\mathcal{O}^{s_{\alpha}-s_\gamma}_{\phi_\beta}|\psi^{\PBC}_\gamma\rangle. \\
\label{OPEAPBC}
C^{\APA}_{\alpha\beta\gamma} &=& \left(\frac{2\pi}{N}\right)^{-\Delta^{\PBC}_{\beta}}\langle \psi^{\APBC}_\alpha|\tilde{\mathcal{O}}^{s_{\alpha}-s_\gamma}_{\phi_\beta}|\psi^{\APBC}_\gamma\rangle,
\end{eqnarray}
where we have divided nonzero OPE coefficients into two classes, the first one involving only PBC primary operators  (with superscript $\PPP$), and the second one involving one PBC primary operator and two APBC primary operators (with superscript $\APA$). The rest of combinations of operators must have vanishing OPE coefficients. 

It is worth noting that in Eq.~\eqref{OPEAPBC} momentum conservation automatically forces $\mathbb{Z}_2$ parity conservation. Consider, for example, the case where $\mathcal{O}$ is $\mathbb{Z}_2$ even, then $s_\alpha-s_\gamma\in \mathbb{Z}$. Then Eqs.~\eqref{slatt2},\eqref{slatt3} imply that $|\psi^{\APBC}_\alpha\rangle$ and $|\psi^{\APBC}_\gamma\rangle$ neccessarily have the same parity.

For Hermitian primary operators, OPE coefficients transform in a simple way when the indices get permuted. Under an even permutation, the OPE coefficient does not change, such as $C^{\CFT,\PPP}_{\alpha\beta\gamma}=C^{\CFT,\PPP}_{\beta\gamma\alpha}$. Under an odd permutation, the OPE coefficient becomes complex conjugated, such as $C^{\CFT,\AAP}_{\beta\alpha\gamma}=C^{\CFT,\APA *}_{\alpha\beta\gamma}$. It is a nontrivial check of our method if we can also directly compute $C^{\AAP}_{\beta\alpha\gamma}$ (to be defined in Eq.~\eqref{stringOPE}) with a lattice representation of $\phi^{\CFT,\APBC}_\alpha$, which we demonstrate below.

\subsection{String operators and OPE coefficients} 
A lattice representation of scaling operators in the APBC should connect the ground state (in PBC) with states in APBC. To connect states in different boundary conditions, we introduce \textit{string operators},
\begin{equation}
\label{Stringopdef}
\mathcal{S}_{\mathcal{O},j}=\left(\prod_{n=1}^{j-1} \mathcal{Z}_n\right) \mathcal{O}_j,
\end{equation}
where $\mathcal{O}$ is a local operator. Its Fourier modes are defined by
\begin{equation}
\label{stringFourier}
\mathcal{S}^s_{\mathcal{O}}=\sum_{j=1}^N \tilde{\mathcal{T}}^{j-1} \mathcal{O}_1 \mathcal{T}^{\dagger j-1} e^{-isj 2\pi/N},
\end{equation}
where $s$ can either be integer or half integer. Notice that we use both $\mathcal{T}$ and $\tilde{\mathcal{T}}$. If $\mathcal{O}$ is a one-site operator, then it is manifest that
\begin{equation}
\label{stringFourier1site}
\mathcal{S}^s_{\mathcal{O}}=\sum_{j=1}^N \mathcal{S}_{\mathcal{O},j} e^{-isj2\pi/N}.
\end{equation}
For a multi-site operator $\mathcal{O}$, Eq.~\eqref{stringFourier1site} is no longer valid. However, it is still not hard to compute all the terms in $\mathcal{S}^s_\mathcal{O}$ from the definition Eq.~\eqref{stringFourier}. For example,
  \begin{equation}
  \mathcal{S}^s_{YZ}=\sum_{j=1}^{N-1}e^{-isj 2\pi/N} \left(\prod_{l=1}^{j-1} Z_l\right)Y_{j}Z_{j+1}+\mathcal{B}^{s}_{YZ},
  \end{equation}
 where 
 \begin{equation}
 \mathcal{B}^{s}_{YZ}=e^{-is2\pi} I_1 \left(\prod_{l=2}^{N-1} Z_{l}\right) Y_N
\end{equation}
as discussed in the appendix.

The covariance of Eq.~\eqref{stringFourier} under translations is more involved. Recall that it maps PBC states to APBC states. First, we can always decompose $\mathcal{S}_\mathcal{O}$ into $\mathbb{Z}_2$ even and $\mathbb{Z}_2$ odd parts. Therefore we will only consider $\mathcal{S}_\mathcal{O}$ with definite parity,
\begin{equation}
\mathcal{Z} \mathcal{S}_{\mathcal{O},j} \mathcal{Z}^{\dagger}=\mathcal{Z}_{\mathcal{O}} \mathcal{S}_{\mathcal{O},j},
\end{equation}
where $\mathcal{Z}_{\mathcal{O}}=\pm 1$ is the parity of both $\mathcal{O}$ and $\mathcal{S}_\mathcal{O}$. We will show that, when acting with $\mathcal{S}^s_\mathcal{O}$ on a PBC eigenstate $|\psi^{\PBC}_\alpha\rangle$ with conformal spin $s^{\PBC}_\alpha$, the result is a linear combination of APBC eigenstates $|\psi^{\APBC}_\beta\rangle$ with conformal spin $s^{\APBC}_\beta=s+s^{\PBC}_\alpha$ if and only if
\begin{equation}
\label{stringTI}
e^{-is2\pi}=\mathcal{Z}_{\mathcal{O}}\mathcal{Z}^{\PBC}_{\alpha},
\end{equation}
which constrains whether $s$ is integer or half integer. 

In order to map $|\psi^{\PBC}_\alpha\rangle$ to the eigenstate of $\tilde{\mathcal{T}}$, it is enough that the operator $\mathcal{S}^s_\mathcal{O}$ satisfies
\begin{equation}
\label{TstringT}
\tilde{\mathcal{T}}\mathcal{S}^s_{\mathcal{O}}\mathcal{T}^{\dagger}|\psi^{\PBC}_\alpha\rangle=e^{is2\pi/N} \mathcal{S}^s_{\mathcal{O}} |\psi^{\PBC}_\alpha\rangle.
\end{equation}
However, this is not manifestly true for all states $|\psi^{\PBC}_\alpha\rangle$. We can explicitly compute that
\begin{equation}
\tilde{\mathcal{T}}\mathcal{S}^s_{\mathcal{O}}\mathcal{T}^{\dagger}-e^{is2\pi/N} \mathcal{S}^s_{\mathcal{O}}=e^{-is2\pi}\mathcal{Z}\mathcal{O}_1-\mathcal{O}_1.
\end{equation}
Then Eq.~\eqref{TstringT} is satisfied if and only if
\begin{equation}
e^{-is2\pi}\mathcal{Z}\mathcal{O}_1|\psi^{\PBC}_\alpha\rangle=\mathcal{O}_1|\psi^{\PBC}_\alpha\rangle,
\end{equation}
which is equivalent to Eq.~\eqref{stringTI}.

Similar to the case of PBC, we can relate a lattice string operator $\mathcal{S}_\mathcal{O}$ to a linear combination of APBC scaling operators in the CFT,
\begin{equation}
\label{stringCFT}
\mathcal{S}_\mathcal{O}\sim \mathcal{O}^{\CFT}=\sum_{\alpha}a_\alpha \psi^{\CFT,\APBC}_\alpha
\end{equation}
The variational parameters can be obtained by minimizing a similar cost function as Eq.~\eqref{costf},
\begin{equation}
\label{costf2}
f^{\mathcal{O}}(\{a_\alpha\})=\sum_\beta |\langle \psi^{\APBC}_\beta|\mathcal{S}^{s_\beta}_\mathcal{O}|0\rangle-\langle \psi^{\CFT}_\beta|\mathcal{O}^{\CFT,{s_\beta}}|0^{\CFT}\rangle|^2.
\end{equation}
 Note that Eq.~\eqref{stringTI} is automatically satisfied if we choose all $|\psi^{\APBC}_\beta\rangle$ to be in the correct parity sector, i.e., $\mathcal{Z}^{\APBC}_{\beta}=\mathcal{Z}_{\mathcal{O}}$, since the ground state is always $\mathbb{Z}_2$ even and $e^{is^{\APBC}_\beta 2\pi}=\mathcal{Z}^{\APBC}_{\beta}$.

Again, Eq.~\eqref{stringCFT} allows us to construct a lattice representation of primary fields $\phi^{\CFT,\APBC}_\beta$ as a string operator, denoted as $\mathcal{S}_{\phi_\beta}$. The OPE coefficients can be extracted by
\begin{equation}
\label{stringOPE}
C^{\AAP}_{\alpha\beta\gamma} = \left(\frac{2\pi}{N}\right)^{-\Delta^{\APBC}_{\beta}}\langle \psi^{\APBC}_\alpha|\mathcal{S}^{s_{\alpha}-s_\gamma}_{\phi_\beta}|\psi^{\PBC}_\gamma\rangle.
\end{equation}
We conclude with a table that summarizes the correspondence between lattice objects and CFT objects for the APBC.
\begin{table*}[htbp]
\begin{tabular}{|c|c|c|c|}
\hline
CFT & Lattice (at low energies)   \\ \hline \hline
       Energy/momentum eigenstate $|\psi^{\CFT,\APBC}_\alpha\rangle$ & Energy/momentum eigenstate $|\psi^{\APBC}_\alpha\rangle$  \\ \hline
       Primary state $|\phi^{\CFT,\APBC}_\alpha\rangle$ & Primary state $|\phi^{\APBC}_\alpha\rangle$  \\ \hline
      Energy $E^{\CFT}_\alpha=\frac{2\pi}{L} (\Delta^{\CFT,\APBC}_\alpha-c^{\CFT}/12)$     & Energy $E^{\APBC}_\alpha \approx NA+\frac{B}{2\pi}E^{\CFT}_\alpha$ (Eq.~\eqref{EP}) \\ \hline
       Momentum $P^{\CFT}_\alpha=\frac{2\pi}{L} s^{\CFT,\APBC}_\alpha$          & Momentum $P^{\APBC}_\alpha=P^{\CFT}_\alpha$ (Eq.~\eqref{EP2}) \\ \hline
       Fourier mode of a PBC operator $\mathcal{O}^{\CFT,s}$   &  Fourier mode $\tilde{\mathcal{O}}^{s}$ (acting on eigenstates of $H^{\APBC}$) (Eq.~\eqref{FourierAPBC}). \\ \hline
      $H^{\CFT}_n=L^{\CFT}_{n}+\bar{L}^{\CFT}_{-n}-{c^{\CFT}}/{12}\delta_{n,0}$                                   & $H^{\APBC}_n$ (acting on eigenstates of $H^{\APBC}$) (Eq.~\eqref{LnAPBC}) \\ \hline
      
    $C^{\APA,\CFT}_{\alpha\beta\gamma}= (2\pi/L)^{-\Delta^{\CFT}_\beta}\langle \phi^{\CFT,\APBC}_\alpha|\phi^{\CFT,s_{\alpha}-s_{\gamma}}_{\beta}|\phi^{\CFT,\APBC}_\gamma\rangle$    & $C^{\APA}_{\alpha\beta\gamma}= (2\pi/N)^{-\Delta_\beta}\langle \phi^{\APBC}_\alpha|\tilde{\mathcal{O}}^{s_{\alpha}-s_{\gamma}}_{\phi_\beta}|\phi^{\APBC}_\gamma\rangle\approx C^{\APA,\CFT}_{\alpha\beta\gamma}$   \\ \hline
       
      APBC operator $\mathcal{O}^{\CFT}=\sum_\alpha a_{\alpha}\psi^{\CFT,\APBC}_\alpha$    & String operator $\mathcal{S}_\mathcal{O}$ (Eq.~\eqref{Stringopdef})\\ \hline
         Fourier mode of an APBC operator $\mathcal{O}^{\CFT,s}$ & Fourier mode of a string operator $\mathcal{S}^s_\mathcal{O}$ (Eq.~\eqref{stringFourier}) \\ \hline
      APBC primary operator $\phi^{\CFT,\APBC}_\beta$    & String operator $\mathcal{S}_{\phi_\beta}$  \\ \hline
      $C^{\AAP,\CFT}_{\alpha\beta\gamma}=(2\pi/L)^{-\Delta^{\CFT}_\beta} \langle \phi^{\CFT,\APBC}_\alpha|\phi^{\CFT,s_{\alpha}-s_{\gamma}}_{\beta}|\phi^{\CFT,\PBC}_\gamma\rangle$    &  $C^{\AAP}_{\alpha\beta\gamma}=(2\pi/N)^{-\Delta_\beta}\langle \phi^{\APBC}_\alpha|\mathcal{S}^{s_{\alpha}-s_{\gamma}}_{\phi_\beta}|\phi^{\PBC}_\gamma\rangle\approx C^{\AAP,\CFT}_{\alpha\beta\gamma} $  \\ \hline
\end{tabular}
\caption{Correspondence of lattice objects and CFT objects for spin chains with APBC}
\label{table2}
\end{table*}
\section{Diagonalization of low energy eigenstates in APBC}
As explained previously, in this work we extract conformal data by (i) computing matrix elements on a finite (anti-)periodic spin chain of size $N$, (ii) repeating the computation for increasing sizes $N$, and (iii) extrapolating the results to large $N$. In order to reduce finite-size errors in the extrapolation of conformal data, we should diagonalize the low energy eigenstates for a series of sizes $N$ that are sufficiently large. The use of periodic uniform matrix product states (puMPS) has enabled us to efficiently obtain low energy eigenstates for a critical spin chain with PBC up to several hundreds of spins. In this section we first review the puMPS techniques \cite{pirvu_exploiting_2011, pirvu_matrix_2012, zou_conformal_2018} for PBC, and then generalize it to APBC. We note that the generalization is in a sense similar to \cite{Stauber_topological_2018} where matrix product states are used to represent spinon excitations on an infinite line, although here we work on the circle. The generalization allows us to compute all low-energy eigenstates in APBC, as well as matrix elements of both local operators and string operators, with a computational cost that scales as in the case of PBC.  However, the extraction of conformal data is independent of how the low-energy states are diagonalized. Therefore, unless specifically interested in the use of puMPS, the reader may skip this section.
\subsection{puMPS for eigenstates in PBC}
For a lattice model with $N$ sites, where each site hosts a local Hilbert space of dimension $d$, a puMPS \cite{rommer_1997}
\begin{equation}
\label{puMPS}
|\Psi(A)\rangle=\sum_{s_1,s_2\cdots s_N=1}^d\mathrm{Tr}(A^{s_1}A^{s_2}\cdots A^{s_N})|\vec{s}\rangle
\end{equation}
 is an ansatz with variational parameters inside matrices $A^{s},~s=1,2,\cdots,d$, where each $A^s$ is a $D\times D$ matrix, and $|\vec{s}\rangle\equiv |s_1 s_2\cdots s_N\rangle$ is a basis of the Hilbert space. $D$ is referred to as bond dimension, and determines the maximal amount of entanglement the state can have. Note that the puMPS is manifestly translation invariant. The variational parameters are determined by minimizing the energy functional
 \begin{equation}
 E(A,\bar{A})=\frac{\langle \Psi(\bar{A})|H|\Psi(A)\rangle}{\langle \Psi(\bar{A})|\Psi(A)\rangle}.
 \end{equation}
In practice it is done by an iterative approach where each step costs $\mathcal{O}(ND^5)$.
 
It has been shown \cite{verstraete_matrix_2006} that the puMPS can faithfully represent the ground state of a critical quantum Hamiltonian, with bond dimension $D$ that grows polynomially with $N$. This growth is in accordance with the logarithmic scaling of entanglement entropy in CFT.

To represent low energy excited states, we use a Bloch-wave like ansatz \cite{pirvu_exploiting_2011, pirvu_matrix_2012, zou_conformal_2018} with momentum $p$,
\begin{equation}
|\Phi_p (B;A)\rangle=\sum_{j=1}^{N}e^{-ipj} \mathcal{T}^j \sum_{\vec{s}=1}^d\mathrm{Tr}(B^{s_1}A^{s_2}\cdots A^{s_N})|\vec{s}\rangle,
\end{equation}
where $B$ is a variational tensor that has size $d\times D\times D$. It is an eigenstate of the translation operator $\mathcal{T}$,
\begin{equation}
\mathcal{T}|\Phi_p (B;A)\rangle=e^{ip}|\Phi_p (B;A)\rangle,
\end{equation}
because
\begin{eqnarray}
&&\mathcal{T}|\Phi_p (B;A)\rangle \nonumber \\
&=&\sum_{j=1}^{N}e^{-ipj} \mathcal{T}^{j+1} \sum_{\vec{s}=1}^d\mathrm{Tr}(B^{s_1}A^{s_2}\cdots A^{s_N})|\vec{s}\rangle \nonumber \\
&=&\sum_{j=2}^{N+1}e^{-ip(j-1)} \mathcal{T}^{j} \sum_{\vec{s}=1}^d\mathrm{Tr}(B^{s_1}A^{s_2}\cdots A^{s_N})|\vec{s}\rangle \nonumber \\
&=&e^{ip}\sum_{j=2}^{N+1}e^{-ipj} \mathcal{T}^{j} \sum_{\vec{s}=1}^d\mathrm{Tr}(B^{s_1}A^{s_2}\cdots A^{s_N})|\vec{s}\rangle \nonumber \\
&=&e^{ip}\sum_{j=1}^{N}e^{-ipj} \mathcal{T}^{j} \sum_{\vec{s}=1}^d\mathrm{Tr}(B^{s_1}A^{s_2}\cdots A^{s_N})|\vec{s}\rangle \nonumber \\
&=&e^{ip}|\Phi_p (B;A)\rangle, \nonumber
\end{eqnarray}
where the fourth equality makes use of the fact that $e^{-ip(N+1)}\mathcal{T}^{N+1}=e^{-ip}\mathcal{T}$, as a result of $e^{-ipN}=1$ and $\mathcal{T}^N=\mathbf{1}$.

The tensor $B$ can be determined by requiring the state to be at a saddle point of the energy functional, which translates into solving a generalized eigenvalue equation with cost $\mathcal{O}(ND^6)$ \cite{pirvu_exploiting_2011, pirvu_matrix_2012, zou_conformal_2018}.

It has been demonstrated \cite{pirvu_exploiting_2011, pirvu_matrix_2012, zou_conformal_2018} that this ansatz can capture any excited state at sufficiently low energies for a critical quantum spin chain. With the cost growing polynomially with $N$, it can be applied to critical quantum spin chains with several hundreds of spins.

\subsection{Symmetric tensors}
To proceed to APBC, we first need the notion of \textit{symmetric tensors} \cite{singh_tensor_2010, singh_tensor_2011, singh_global_2013}, which we review below.

For the spin-1/2 chain in this paper, we will use $\mathbb{Z}_2$ symmetric tensors $A$ of the form
\begin{equation}
A^{1}=\begin{bmatrix}
A^{1}_{11} & 0 \\
0 & A^{1}_{22}
\end{bmatrix}, ~~
A^{2}=\begin{bmatrix}
0 & A^{2}_{12} \\
A^{2}_{21} & 0
\end{bmatrix},
\end{equation}
where each block is a $D/2\times D/2$ matrix (more generally the two blocks can have different dimensions which add up to $D$). The tensor is invariant under the on-site $\mathbb{Z}_2$ symmetry up to a gauge transformation,
\begin{equation}
\label{symT}
\sum_{s'}\mathcal{Z}_{ss'} A^{s'} = U_B(\mathcal{Z})A^s U^{\dagger}_B(\mathcal{Z}),
\end{equation}
where $\mathcal{Z}_{ss'}=Z_{ss'}$ (that is the $ss'$ component of the Pauli matrix $Z$) is the representation of the $\mathbb{Z}_2$ generator on one site of the lattice ($\mathcal{Z}_j$ in Eq.~\eqref{Z2latt}), and 
\begin{equation}
\label{UBZ}
U_B(\mathcal{Z})=
\begin{bmatrix}
I_{D/2} & 0 \\
0 & -I_{D/2}
\end{bmatrix}
\end{equation}
is a $D\times D$ dimensional representation of the $\mathbb{Z}_2$ generator on the bond, where $I_{D/2}$ is a $D/2$ dimensional identity matrix.  Importantly, the use of $\mathbb{Z}_2$ symmetric tensors forces the puMPS to be invariant under $\mathbb{Z}_2$,
\begin{equation}
\mathcal{Z}|\Psi(A)\rangle=|\Psi(A)\rangle.
\end{equation}
This can be seen by
\begin{eqnarray}
&&\mathcal{Z}|\Psi(A) \nonumber \rangle \\
&=&\sum_{\vec{s}=1}^d\mathrm{Tr}(A^{s_1}A^{s_2}\cdots A^{s_N})\mathcal{Z}|\vec{s}\rangle \nonumber \\
&=&\sum_{\vec{s}=1}^d\mathrm{Tr}(\mathcal{Z}_{s_1 s'_1}A^{s'_1}\mathcal{Z}_{s_2 s'_2}A^{s_2}\cdots \mathcal{Z}_{s_N s'_N}A^{s_N})|\vec{s}\rangle \nonumber \\
&=&\sum_{\vec{s}=1}^d\mathrm{Tr}(U_B(\mathcal{Z})A^{s_1}U^{\dagger}_B(\mathcal{Z})\cdots U_B(\mathcal{Z})A^{s_N}U^{\dagger}_B(\mathcal{Z}))|\vec{s}\rangle \nonumber \\
&=&\sum_{\vec{s}=1}^d\mathrm{Tr}(A^{s_1}A^{s_2}\cdots A^{s_N})|\vec{s}\rangle \nonumber \\
&=&|\Psi(A)\rangle, \nonumber
\end{eqnarray}
where in the third line we change the order of tensor contraction, in the fourth line we use the definition of symmetric tensors, Eq.~\eqref{symT}, and in the fifth line we use the cyclic property of trace and unitarity of $U_B(\mathcal{Z})$.

The excitation ansatz can also be forced to be $\mathbb{Z}_2$ invariant by requiring that
\begin{equation}
\label{symB}
\sum_{s'}\mathcal{Z}_{ss'} B^{s'} = \pm U_B(\mathcal{Z})B^s U^{\dagger}_B(\mathcal{Z}),
\end{equation}
where the $\pm$ represents $\mathbb{Z}_2$ even or odd excitations,
\begin{equation}
\mathcal{Z}|\Phi_p(B;A)\rangle=\pm |\Phi_p(B;A)\rangle.
\end{equation}

The use of symmetric tensors has three advantages. First, it reduces the number of variational parameters by one half, leading to more efficient algorithms. Second, it enables us to diagonalize states separately in each symmetry sector, with the symmetry forced exactly. Third and most importantly, it allows us to write down a simple generalization of the excitation ansatz to APBC, and more generally, twisted boundary conditions.

\subsection{puMPS for eigenstates in APBC}
We propose that low energy eigenstates of critical quantum spin chains with APBC can be represented by
\begin{equation}
\label{phiapbc}
|\Phi^{\APBC}_p(B;A)\rangle= \sum_{j=1}^{N}e^{-ipj} \tilde{\mathcal{T}}^j \sum_{\vec{s}=1}^d\mathrm{Tr}(B^{s_1}A^{s_2}\cdots A^{s_N})|\vec{s}\rangle,
\end{equation}
where $A$ is the same tensor appearing the ground state ansatz Eq.~\eqref{puMPS} for the spin chain with \textit{periodic boundary conditions} (PBC), where Eq.~\eqref{symT} can be enforced. The tensor $B$ satisfies Eq.~\eqref{symB}, where the $\pm$ determines the $\mathbb{Z}_2$ sector $\mathcal{Z}=\pm 1$ of the state. The momentum $p$ is restricted to 
\begin{eqnarray}
\label{pevenMPS}
p &\in& \frac{2\pi}{N}\mathbb{Z} ~(\mathcal{Z}=1). \\
\label{poddMPS}
p &\in& \frac{2\pi}{N}\left(\mathbb{Z}+\frac{1}{2}\right) ~(\mathcal{Z}=-1).
\end{eqnarray}
Importantly, the explicit enforcement of $\mathbb{Z}_2$ symmetry ensures the ansatz to be translation invariant under $\tilde{\mathcal{T}}$,
\begin{equation}
\tilde{\mathcal{T}}|\Phi^{\APBC}_p(B;A)\rangle= e^{ip}|\Phi^{\APBC}_p(B;A)\rangle.
\end{equation}
This can be readily seen by
\begin{eqnarray}
&&\tilde{\mathcal{T}}|\Phi^{\APBC}_p (B;A)\rangle \nonumber \\
&=&\sum_{j=1}^{N}e^{-ipj} \tilde{\mathcal{T}}^{j+1} \sum_{\vec{s}=1}^d\mathrm{Tr}(B^{s_1}A^{s_2}\cdots A^{s_N})|\vec{s}\rangle \nonumber \\
&=&\sum_{j=2}^{N+1}e^{-ip(j-1)} \tilde{\mathcal{T}}^{j} \sum_{\vec{s}=1}^d\mathrm{Tr}(B^{s_1}A^{s_2}\cdots A^{s_N})|\vec{s}\rangle \nonumber \\
&=&e^{ip}\sum_{j=2}^{N+1}e^{-ipj} \tilde{\mathcal{T}}^{j} \sum_{\vec{s}=1}^d\mathrm{Tr}(B^{s_1}A^{s_2}\cdots A^{s_N})|\vec{s}\rangle \nonumber \\
&=&e^{ip}\left[\sum_{j=2}^{N}e^{-ipj} \tilde{\mathcal{T}}^{j} \sum_{\vec{s}=1}^d\mathrm{Tr}(B^{s_1}A^{s_2}\cdots A^{s_N})|\vec{s}\rangle\right.  \nonumber \\
&+& \left. e^{-ip(N+1)}\tilde{T}^{N+1} \sum_{\vec{s}=1}^d\mathrm{Tr}(B^{s_1}A^{s_2}\cdots A^{s_N})|\vec{s}\rangle\right] \nonumber \\
&=&e^{ip}|\Phi_p (B;A)\rangle. \nonumber
\end{eqnarray}
In the last equality we have used 
\begin{eqnarray}
&&e^{-ip(N+1)}\tilde{\mathcal{T}}^{N+1}\sum_{\vec{s}=1}^d\mathrm{Tr}(B^{s_1}A^{s_2}\cdots A^{s_N})|\vec{s}\rangle \nonumber \\
&=&e^{-ip}\tilde{\mathcal{T}} \left(e^{-ipN}\mathcal{Z}\sum_{\vec{s}=1}^d\mathrm{Tr}(B^{s_1}A^{s_2}\cdots A^{s_N})|\vec{s}\rangle\right) \nonumber \\
&=&\pm e^{-ipN} \left(e^{-ip}\tilde{\mathcal{T}} \sum_{\vec{s}=1}^d\mathrm{Tr}(B^{s_1}A^{s_2}\cdots A^{s_N})|\vec{s}\rangle\right) \nonumber \\
&=&e^{-ip}\tilde{\mathcal{T}}\sum_{\vec{s}=1}^d\mathrm{Tr}(B^{s_1}A^{s_2}\cdots A^{s_N})|\vec{s}\rangle,
\end{eqnarray}
where in the third line $e^{-ipN}=\pm 1$ depending on which of Eqs.~\eqref{pevenMPS},\eqref{poddMPS} is satisfied.

We reiterate that, given the $A$ tensor for the PBC ground state, we use the ansatz Eq.~\eqref{phiapbc} to represent \textit{any} low-energy eigenstate of APBC with sufficiently low energy, including the APBC ground state. The tensor $B$ can be determined by requiring the state to be at a saddle point of the energy functional with respect to $H^{\APBC}$. The algorithm is quite similar to the case of PBC and has the same numerical cost $\mathcal{O}({ND^6})$. We leave the details of the algorithm to the Appendix.
\section{The Ising model}
In this section, we use our methods in previous sections to extract the conformal data from the Ising model.
\subsection{The Ising CFT}
The Ising CFT belongs to the unitary minimal models \cite{belavin_infinite_1984, friedan_conformal_1984} which can be exactly solved, meaning that all conformal data can be solved exactly. It has central charge $c^{\CFT}=1/2$. The Ising CFT has 3 primary operators in the PBC sector, denoted as $\mathbf{1},\epsilon,\sigma$ and 3 primary operators in the APBC sector, denoted as $\mu,\psi,\bar{\psi}$ \cite{cardy_finite_1984}. As noted in Eq.~\eqref{IsingFermion}, the Ising model can be mapped to a free Majorana fermion chain via the Jordan-Wigner transformation. Primary operators can also be classified by the boundary condition of the fermion, see Table \ref{table3}. Under the Kramers-Wannier duality, $|\mathbf{1}\rangle,|\bar{\psi}\rangle$ are even, and $|\epsilon\rangle,|\psi\rangle$ are odd. $|\sigma\rangle$ and $|\mu\rangle$ are not eigenstates of the duality.
\begin{table}[htbp]
\begin{tabular}{|c|c|c|c|c|c|}
\hline
$\phi^{\CFT}_\alpha$ & $\Delta^{\CFT}_\alpha$ & $s^{\CFT}_\alpha$ &$\mathcal{Z}_\alpha$& spin chain B.C. & fermion B.C.  \\ \hline
  $\mathbf{1}$ & 0 & 0 & +  & PBC & NS \\ \hline
  $\epsilon$ & 1 & 0 & +  & PBC & NS \\ \hline
  $\sigma$ & 1/8 & 0 & -- & PBC & R \\ \hline
  $\psi$ & 1/2 & 1/2 & -- & APBC & NS \\ \hline
  $\bar{\psi}$ & 1/2 & --1/2 & -- & APBC & NS \\ \hline
  $\mu$ & 1/8 & 0 & + & APBC & R \\ \hline
\end{tabular}
\caption{Primary fields of the Ising CFT.}
\label{table3}
\end{table}

 There are 5 nonzero OPE coefficients \cite{francesco_critical_1987, ginsparg_applied_1988} (up to permutation of indices) that do not involve the identity operator,
\begin{eqnarray}
C^{\CFT}_{\sigma\sigma\epsilon}&=&\frac{1}{2} \\
C^{\CFT}_{\mu\mu\epsilon}&=&-\frac{1}{2},~~~  C^{\CFT}_{\bar{\psi}\psi\epsilon}=i \\
C^{\CFT}_{\psi\mu\sigma}&=&\frac{1}{2}(1-i), ~~~ C^{\CFT}_{\bar{\psi}\mu\sigma}=\frac{1}{2}(1+i).
\end{eqnarray}
OPE coefficients involving the identity operator are trivial $C^{\CFT}_{\alpha\beta \mathbf{1}}=\delta_{\alpha\beta}$. Notice that other OPE coefficients (such as $C^{\CFT}_{\sigma\sigma\sigma}, C^{\CFT}_{\epsilon\epsilon\epsilon}$) vanish because otherwise they are incompatible with either $\mathbb{Z}_2$ symmetry or the Kramers-Wannier duality.
  
The fact that $\sigma^{\CFT}$ and $\mu^{\CFT}$ have the same scaling dimension can be explained by the free fermion picture \cite{boyanovsky_field_1989}. They are related by
\begin{align}
|\mu^{\CFT}\rangle&=b^{\CFT}_0|\sigma^{\CFT}\rangle \\
|\sigma^{\CFT}\rangle&=b^{\CFT}_0|\mu^{\CFT}\rangle,
\end{align}
where 
\begin{equation}
b^{\CFT}_0\propto \psi^{\CFT,s=0} = \frac{1}{L}\int_0^L dx\, \psi^{\CFT}(x)
\end{equation}
 is the fermionic zero mode. In the free fermion CFT, the fermionic zero mode commutes with $L^{\CFT}_0$. Therefore the action of $b^{\CFT}_0$ leaves the scaling dimension invariant. The fermionic zero mode is present only in the R sector of the free fermion CFT as a result of Eqs.~\eqref{scft1},\eqref{scft3}. Actually there is a general theorem saying that the double degeneracy is a robust feature for all Majorana chains (whether free or interacting) with Kramers-Wannier self duality \cite{hsieh_all_2016}. Later, we will see another example, namely the tricritical Ising CFT.
 
\subsection{Scaling dimensions, conformal spins and central charge from the Ising model}
\begin{figure}
\includegraphics[width=0.49\linewidth]{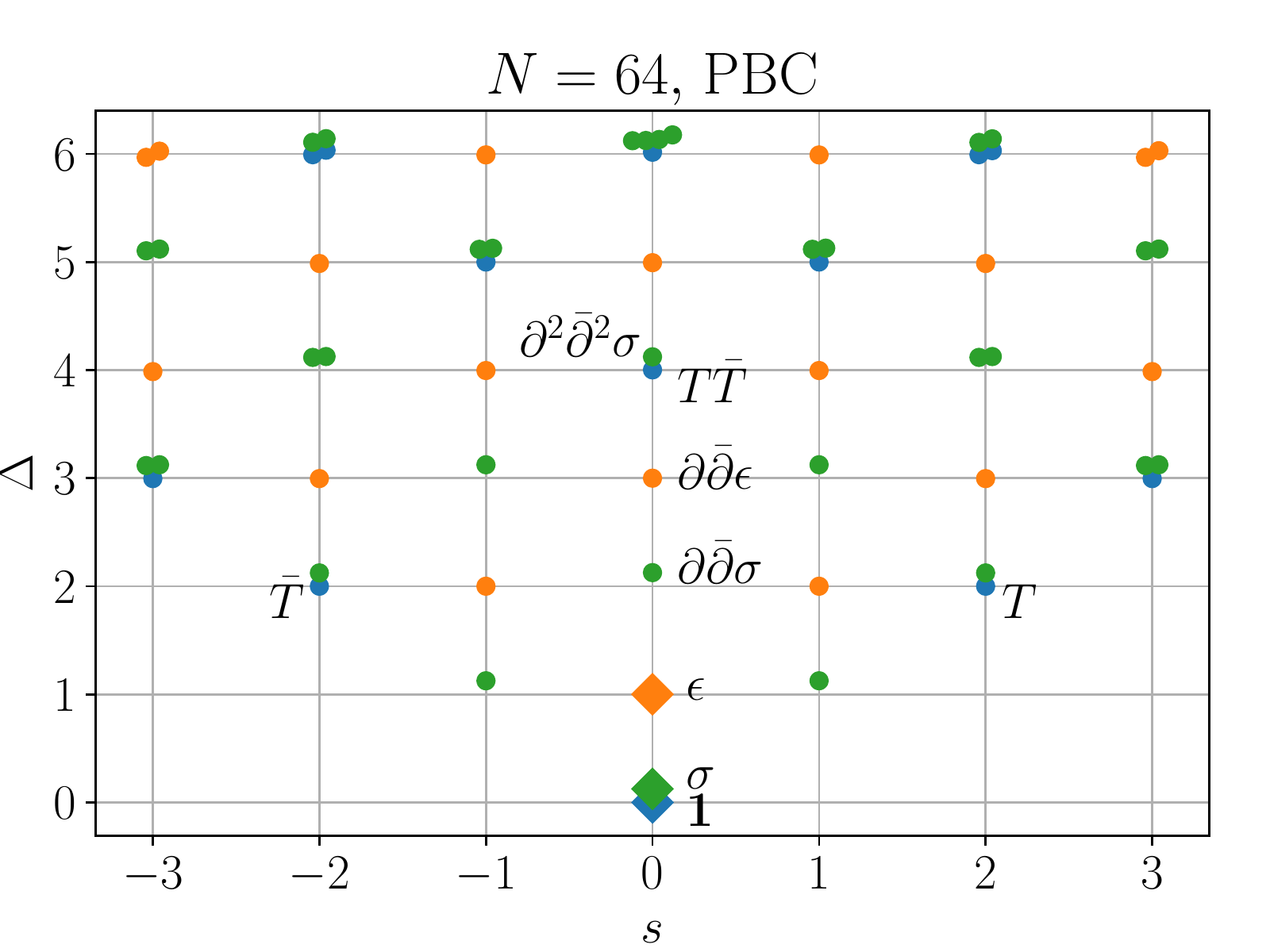} 
\includegraphics[width=0.49\linewidth]{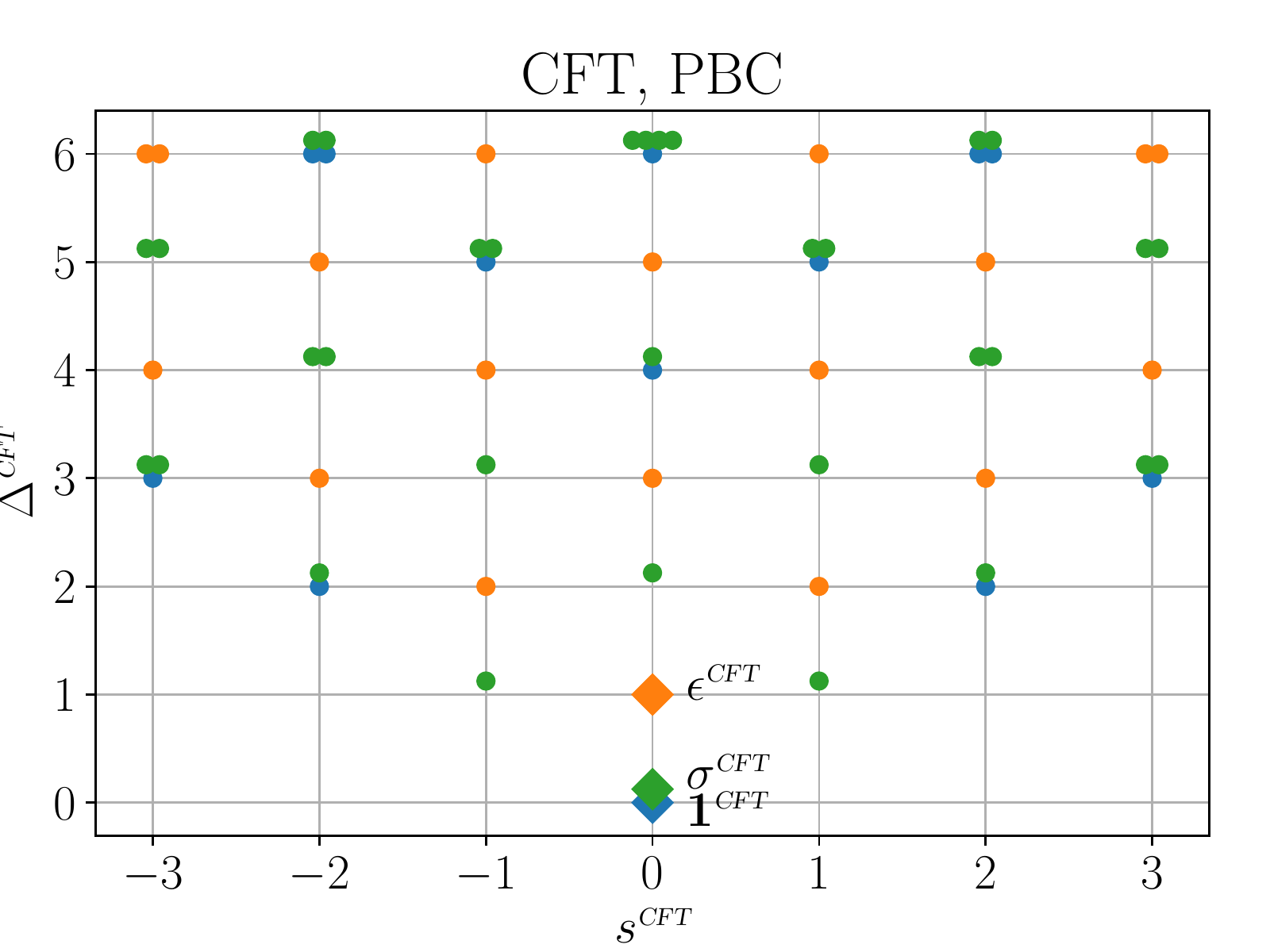} 
\includegraphics[width=0.49\linewidth]{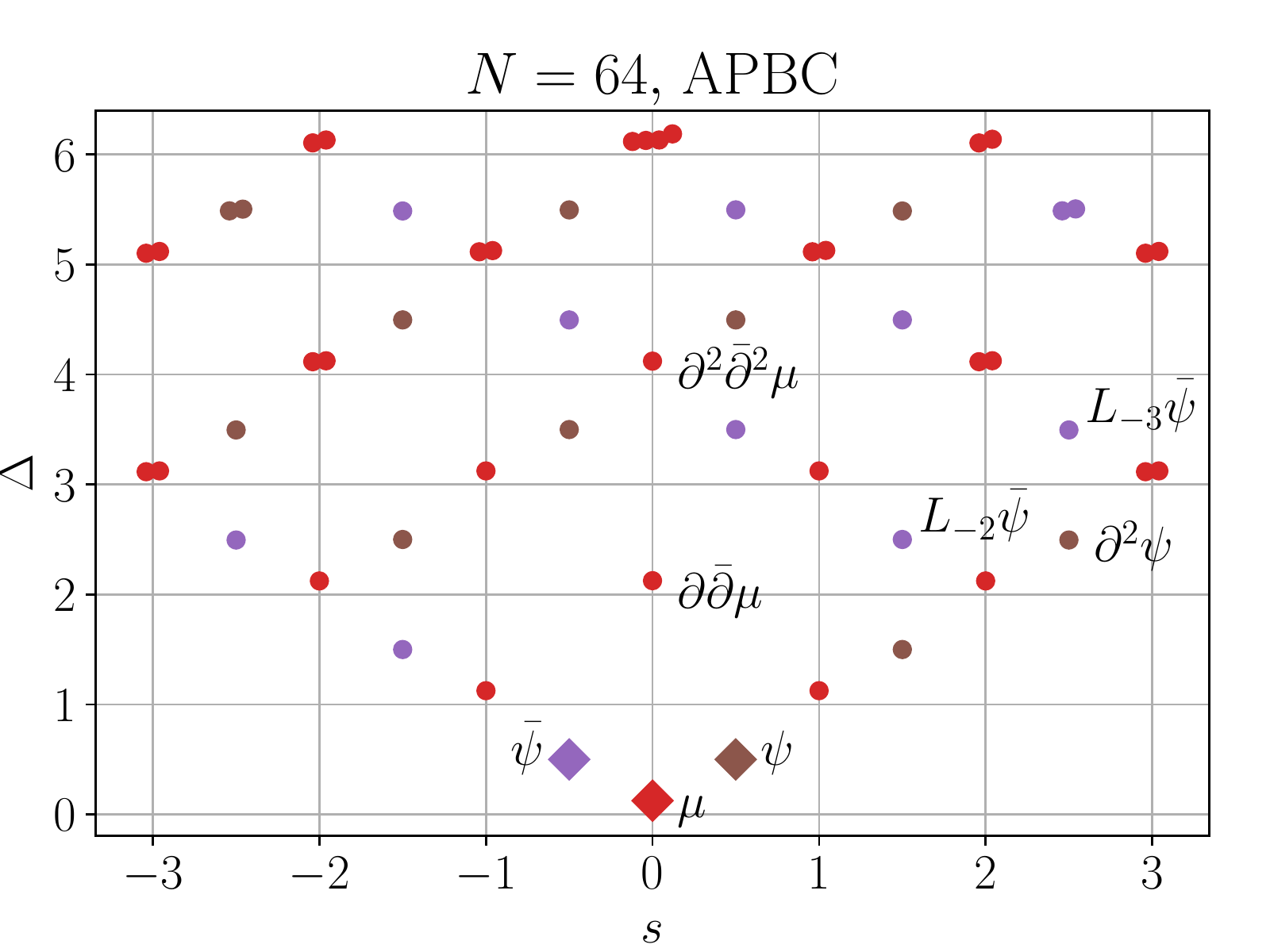} 
\includegraphics[width=0.49\linewidth]{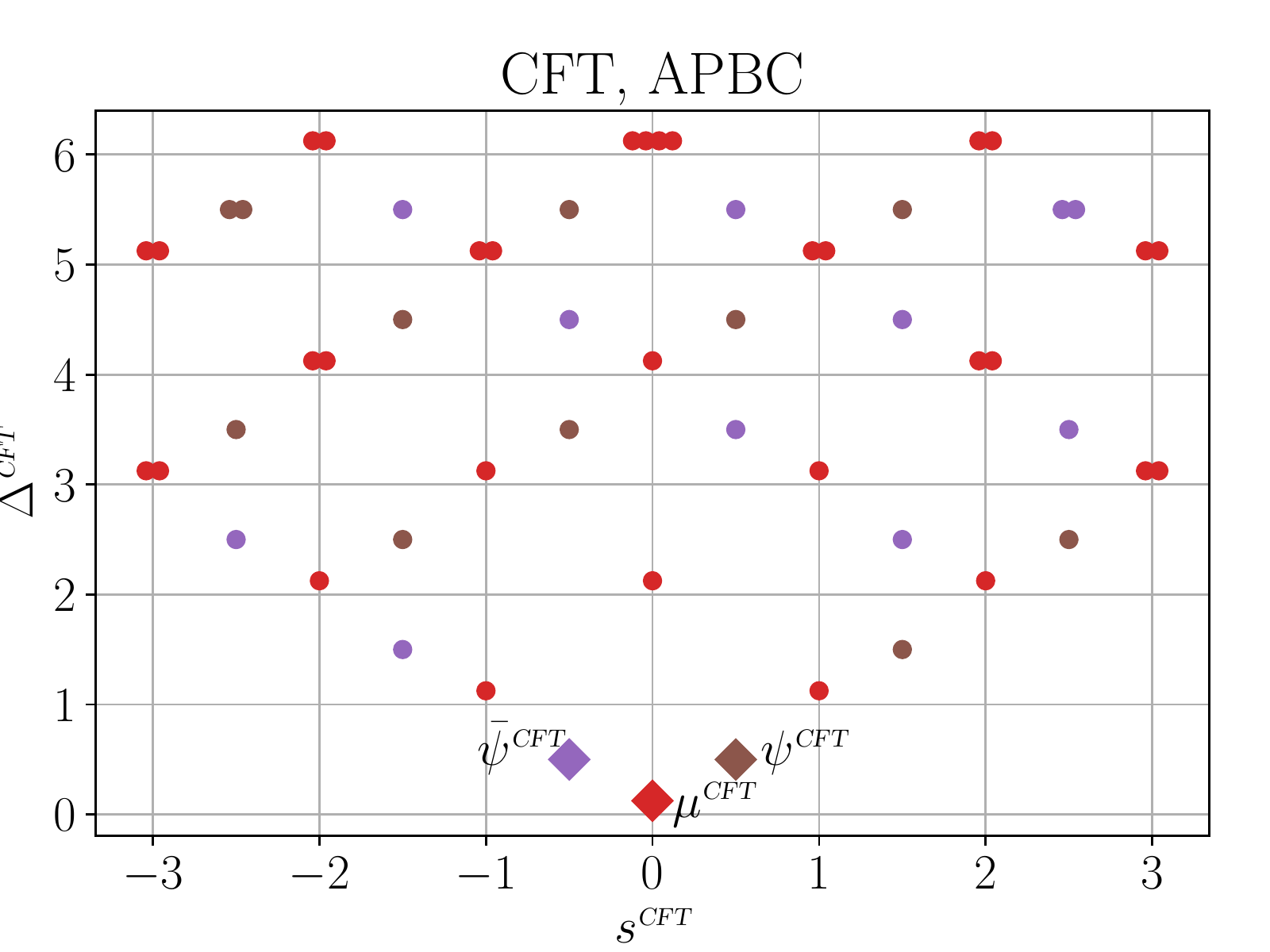} 
\caption{\label{Fig:Isingspec} (Left) Low-energy spectrum of the Ising model with the PBC and the APBC at $N=64$, diagonalized using puMPS with bond dimension $D=28$. Different colors indicate different conformal towers, with diamonds labeling the primary states. We have only shown the states with conformal spins $|s|\leq 3$. (Right) The spectrum of the Ising CFT up to $\Delta^{\CFT}\leq 6+1/8$ and $-3\leq s^{\CFT}\leq 3$. (Note: degenerate states are plotted with a slightly shifted conformal spin to show degeneracy. For example, there are 4 descendant states of $\sigma^{\CFT}$ at scaling dimension $6+1/8$ and conformal spin $0$.)}
\end{figure}

We use puMPS with bond dimension $18\leq D\leq 44$ to diagonalize the low-energy spectrum of the Ising model with both PBC and APBC for $32\leq N\leq 160$. For example, at $N=64$, we use puMPS with bond dimension $D=28$ to compute eigenstates with both boundary conditions up to $\Delta^{\CFT}\leq 6+1/8$. The results are shown in Fig. (\ref{Fig:Isingspec}). Comparing with the CFT spectrum, we see that all low energy eigenstates in both boundary conditions are captured. Ref.\cite{zou_conformal_2018} has considered the same puMPS ansatz for the PBC. The difference is that here we have employed $\mathbb{Z}_2$ symmetric tensors. 

Primary states and conformal towers are identified using the matrix elements of $H^{\PBC}_n$ and $H^{\APBC}_n$ in Eqs.~\eqref{LnPBC},\eqref{LnAPBC}, with eigenstates in the PBC and APBC sectors, respectively. We found that the identification of conformal towers is correct for all eigenstates in the figure.  

\begin{figure}
\includegraphics[width=0.49\linewidth]{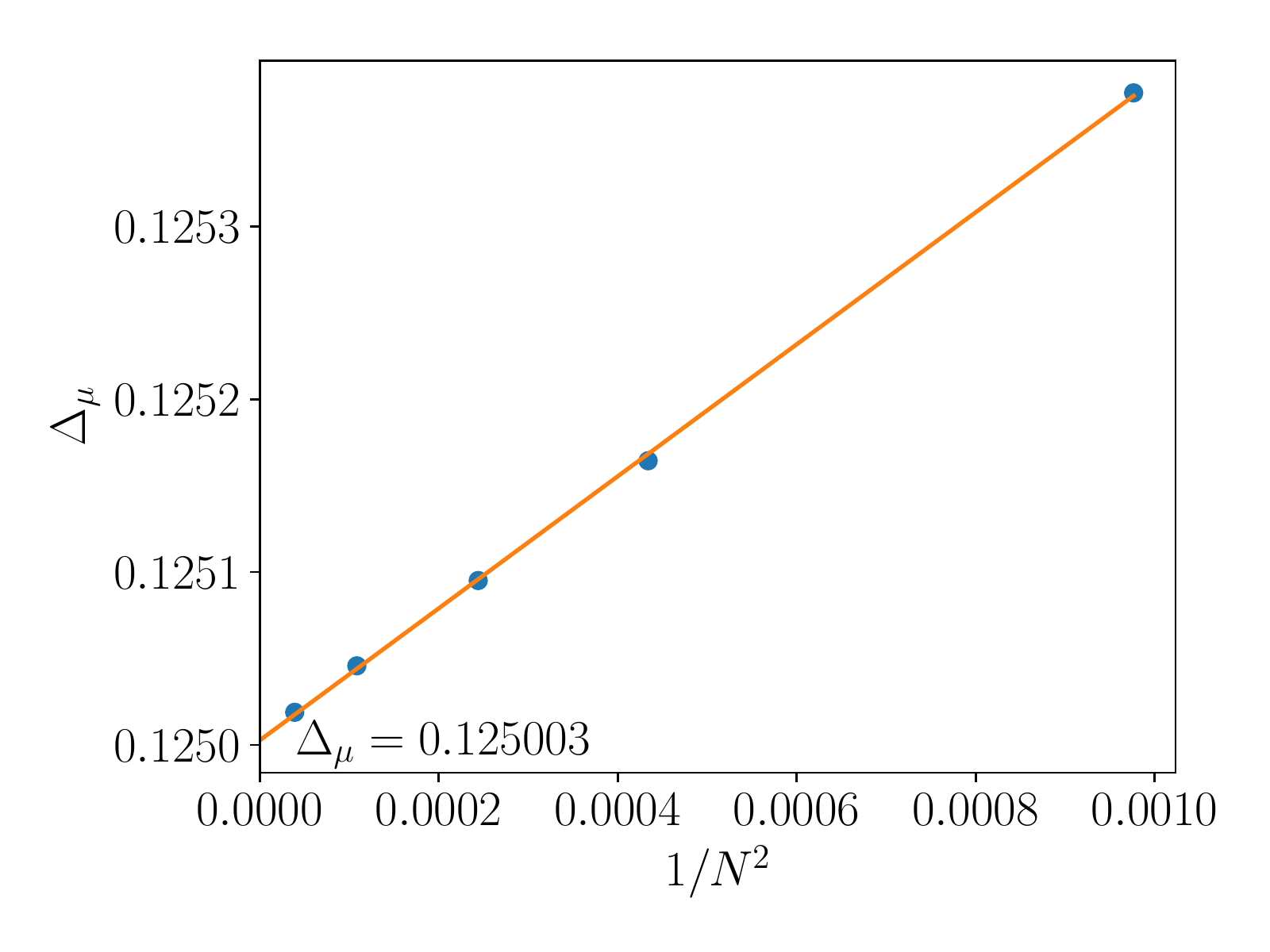} 
\includegraphics[width=0.49\linewidth]{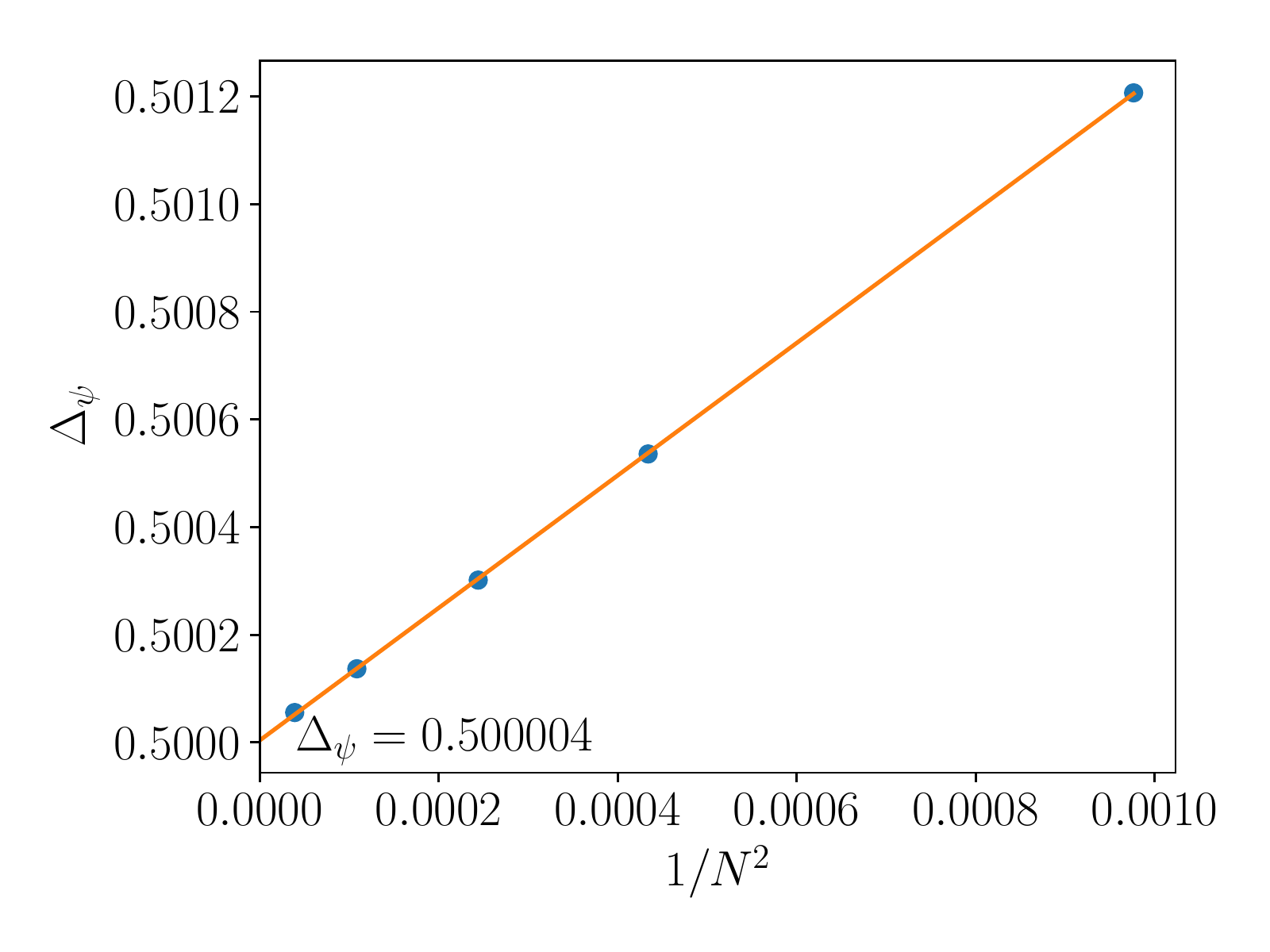} 
\caption{\label{Fig:IsingDelta} Extrapolation of the scaling dimensions $\Delta_{\mu}$ and $\Delta_{\psi}$ for the Ising model with $32 \leq N \leq 160$.}
\end{figure}

We can compute dimensions and conformal spins using Eqs.\eqref{DeltaPBC},\eqref{sPBC},\eqref{DeltaAPBC},\eqref{sAPBC} with different $N$, and then extrapolate the scaling dimensions to the thermodynamic limit. For example, the extrapolation for $\Delta_\mu$ and $\Delta_\psi$ is shown in Fig. (\ref{Fig:IsingDelta}). Table \ref{table4} shows the comparison between the numerical estimations of scaling dimensions with the exact values for all primary states and several descendant states. For completeness, we also extract the central charge with Eq.~\eqref{clat}.
\begin{table}[htbp]
\begin{tabular}{|c|c|c|c|}
\hline
        & Exact  & puMPS & error \\ \hline
  $c$ & $0.5$ & 0.499996 & $4\times 10^{-6}$  \\ \hline
  $\Delta_\epsilon$ & 1 & 1.000001 & $10^{-6}$   \\ \hline
  $\Delta_\sigma$ & 0.125 & 0.125005 & $5\times 10^{-6}$  \\ \hline
  $\Delta_\psi$ & 0.5 & 0.500004 &  $4\times 10^{-6}$ \\ \hline
  $\Delta_{\bar{\psi}}$ &0.5 & 0.500004 & $4\times 10^{-6}$  \\ \hline
  $\Delta_{\mu}$ & 0.125 & 0.125003 & $3\times 10^{-6}$   \\ \hline
  $\Delta_{\partial\bar{\partial}\epsilon}$ & 3 &3.00002 & $2\times 10^{-5}$  \\ \hline
  $\Delta_{T\bar{T}}$ & 4 & 4.001 & $10^{-3}$  \\ \hline
  $\Delta_{\partial\bar{\partial}\sigma}$  & 2.125 & 2.12501& $10^{-5}$ \\ \hline
  $\Delta_{\partial^2\bar{\partial}^2\sigma}$  & 4.125 & 4.126 & $10^{-3}$ \\ \hline
  $\Delta_{\partial^2\psi}$  &  2.5 & 2.499995 & $5\times 10^{-6}$\\ \hline
  $\Delta_{L_{-2}\bar{\psi}}$ & 2.5 &2.50001 & $10^{-5}$ \\ \hline
  $\Delta_{L_{-3}\bar{\psi}}$ & 3.5 & 3.50002 & $2\times 10^{-5}$ \\ \hline
  $\Delta_{\partial\bar{\partial}\mu}$  & 2.125 & 2.125003 & $3\times 10^{-6}$\\ \hline
  $\Delta_{\partial^2\bar{\partial}^2 \mu}$  & 4.125 &4.1256 & $6\times 10^{-4}$\\ \hline
\end{tabular}
\caption{Scaling dimensions from the Ising model with $32\leq N \leq 160$.}
\label{table4}
\end{table}

We see that the accuracy of the scaling dimensions is better for lower-lying excited states. The errors come from two sources. First, the diagonalization with puMPS is approximate. This error can be reduced by using a larger $D$. Second, the finite-size corrections also increase with energy. This error can be reduced by simulating larger sizes $N$. In order to make a reliable extrapolation to the thermodynamic limit, we should choose sufficiently large bond dimensions $D$ such that the finite-$D$ error is much smaller than the finite-size errors. However, for a fixed bond dimension, the finite-$D$ error also increases with the energy. In the above example, the finite-$D$ error is much smaller than the finite-size error for the $|\mu\rangle$ and $|\psi\rangle$ states, as is evident from Fig. (\ref{Fig:IsingDelta}). However, for higher excited states such as $|T\bar{T}\rangle$, the finite-$D$ error is comparable to the finite-size error, which makes the extrapolation not as accurate.

\subsection{OPE coefficients from the Ising model}
In order to extract the OPE coefficients, we first associate each lattice operator with a truncated expansion of scaling operators in the CFT, as Eqs.~\eqref{operatorID},\eqref{stringCFT} for local operators and string operators, respectively. For the purpose of finding the lattice primary operators for the Ising model, it is sufficient to limit the scaling operators in the expansion to the primary operators, i.e., $\mathbf{1}^{\CFT},\sigma^{\CFT},\epsilon^{\CFT}$ in the PBC, and $\mu^{\CFT},\psi^{\CFT},\bar{\psi}^{\CFT}$ in the APBC. The coefficients $a_{\alpha}$ can be obtained by minimizing the cost functions Eqs.~\eqref{costf},\eqref{costf2} for local operators and string operators, respectively. The cost function is specified by a set of low-energy eigenstates $|\psi_{\beta}\rangle$, which we choose to be the set of primary states $|\phi_\beta\rangle$ for simplicity. The CFT matrix elements in the cost function Eq.~\eqref{costf} are
\begin{equation}
\langle \phi^{\CFT}_\beta|\phi^{\CFT,s_{\beta}}_\alpha|0^{\CFT}\rangle=\left(\frac{2\pi}{L}\right)^{\Delta^{\CFT}_{\alpha}}\delta_{\alpha\beta}.
\end{equation}
The cost function Eq.~\eqref{costf} becomes
\begin{equation}
f^{\mathcal{O}}(\{a_\alpha\})=\sum_\alpha \left|\langle \phi_\alpha|\mathcal{O}^{s_\alpha}|0\rangle-\left(\frac{2\pi}{N}\right)^{\Delta^{\CFT}_{\alpha}}a_\alpha\right|^2.
\end{equation}
Its minimum is achieved by
\begin{equation}
\label{aalphasimp}
a_{\alpha}=\left(\frac{2\pi}{N}\right)^{-\Delta^{\CFT}_{\alpha}}\langle \phi_\alpha|\mathcal{O}^{s_\alpha}|0\rangle. 
\end{equation}
Similar expressions hold for string operators, with $\phi_\alpha$ APBC operators and $\mathcal{O}^{s}$ substituted with $\mathcal{S}^s_{\mathcal{O}}$. We will apply Eq.~\eqref{aalphasimp} to several lattice operators and extrapolate the $a_\alpha$ coefficients to the thermodynamic limit. The resulting expansions
\begin{equation}
\label{exppri}
\mathcal{O}\sim \sum_\alpha a_\alpha\phi^{\CFT}_\alpha
\end{equation}
for several operarors $\mathcal{O}$'s are shown in Table \ref{table:Isingops}. We start by considering single site lattice operators $\mathcal{O}$ and add operators with larger support (two-site, etc) if they are needed to invert the expansion Eq.~\eqref{exppri} to obtain a lattice representation of the primary fields.

For the Ising model, we have computed $a_\alpha$'s for local operators $\mathcal{O}_j=X_j,Y_j,Z_j,X_jX_{j+1}$ and string operators $\mathcal{S}_{\mathcal{O},j}=\mathcal{S}_{I,j}, \mathcal{S}_{X,j},\mathcal{S}_{Y,j}$ with $20\leq N \leq 96$. Note that the $Y$ operator does not correspond to any primary field, but to the descendant $\partial_\tau\sigma^{\CFT}$ in the CFT \cite{zou_conformal_2019}. We can then find lattice operators that correspond to CFT primary operators, listed in Table \ref{table:Isingops}. Notice that the result is consistent with the $\mathbb{Z}_2$ symmetry and the Kramers-Wannier duality. For example, the lattice operator $\mathcal{O}_{\epsilon}=1.5708(XX-Z)$ is even under $\mathbb{Z}_2$ and odd under the duality. 
\begin{table}[htbp]
\begin{tabular}{|c|c|}
\hline
      Lattice  & CFT \\ \hline
       $X$     &  $0.80312\sigma^{\CFT}$      \\ \hline 
       $Y$     &  $0.0000\sigma^{\CFT}$      \\ \hline 
       $Z$     &  $0.63662\mathbf{1}^{\CFT}-0.31831\epsilon^{\CFT}$      \\ \hline 
       $XX$   &  $0.63662\mathbf{1}^{\CFT}+0.31831\epsilon^{\CFT}$      \\ \hline 
       $\mathcal{S}_I$     &  $0.80312\mu^{\CFT}$      \\ \hline 
       $\mathcal{S}_X$     &  $0.39894\psi^{\CFT}+0.39894\bar{\psi}^{\CFT}$      \\ \hline
       $\mathcal{S}_Y$     &  $-0.39894\psi^{\CFT}+0.39894\bar{\psi}^{\CFT}$      \\ \hline
\end{tabular}
\\

\begin{tabular}{|c|c|}
\hline
      CFT  & Lattice \\ \hline
       $\sigma^{\CFT}$     &  $1.2451X$      \\ \hline 
       $\epsilon^{\CFT}$     &  $1.5708XX-1.5708Z$      \\ \hline 
       $\mu^{\CFT}$     &  $1.2451 \mathcal{S}_X$      \\ \hline 
       $\psi^{\CFT}$   &  $1.2533\mathcal{S}_X-1.2533\mathcal{S}_Y$      \\ \hline 
       $\bar{\psi}^{\CFT}$     &  $1.2533\mathcal{S}_X+1.2533\mathcal{S}_Y$      \\ \hline 
\end{tabular}
\caption{Correspondence between lattice operators and CFT operators for the Ising model. (Top) Lattice operators expressed as a linear combination of a truncated set of CFT operators. (Bottom) CFT  primary operators expressed as a linear combination of lattice operators by inverting the top table.}
\label{table:Isingops}
\end{table}

\begin{figure}
\includegraphics[width=0.49\linewidth]{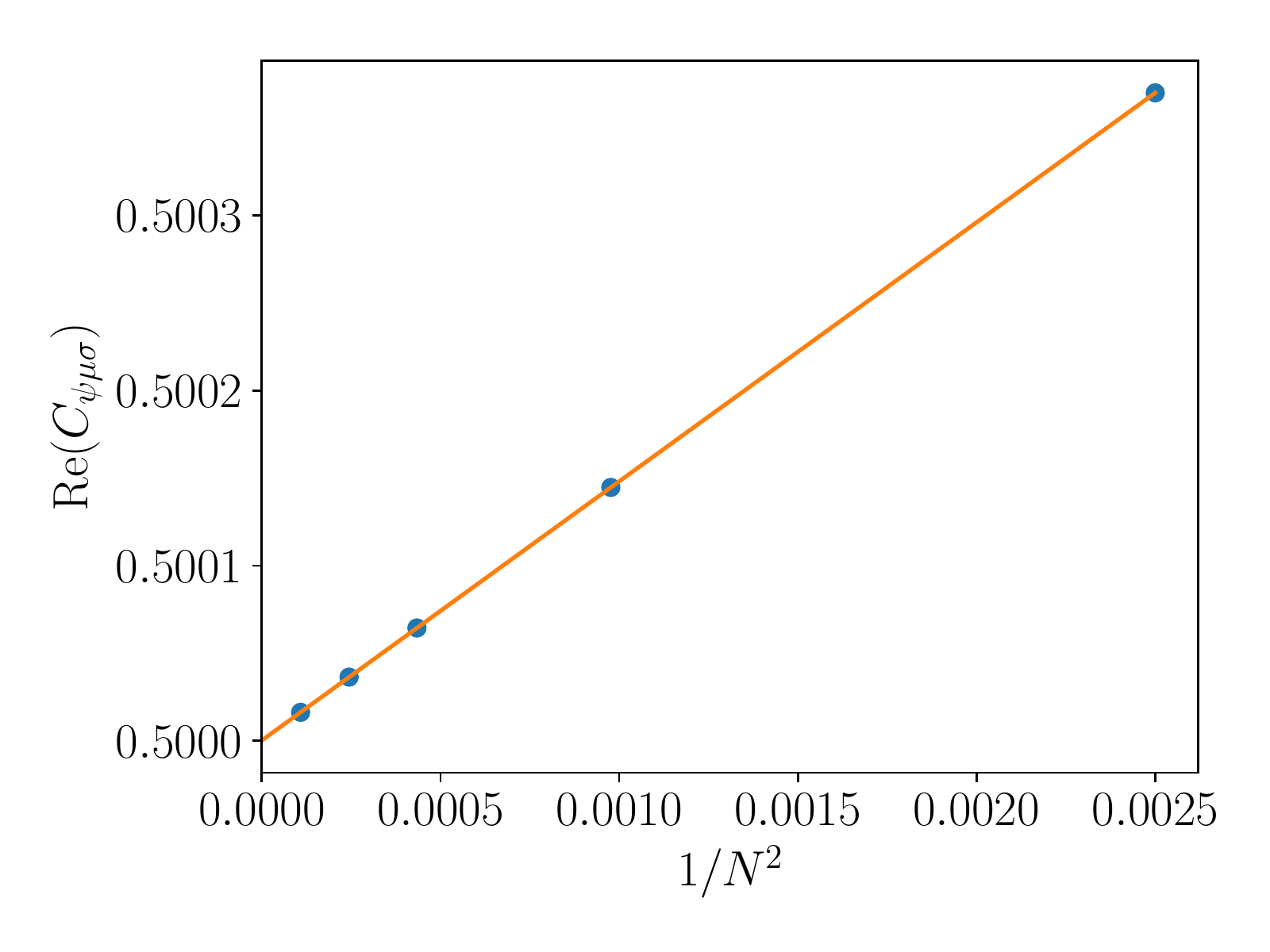}
\includegraphics[width=0.49\linewidth]{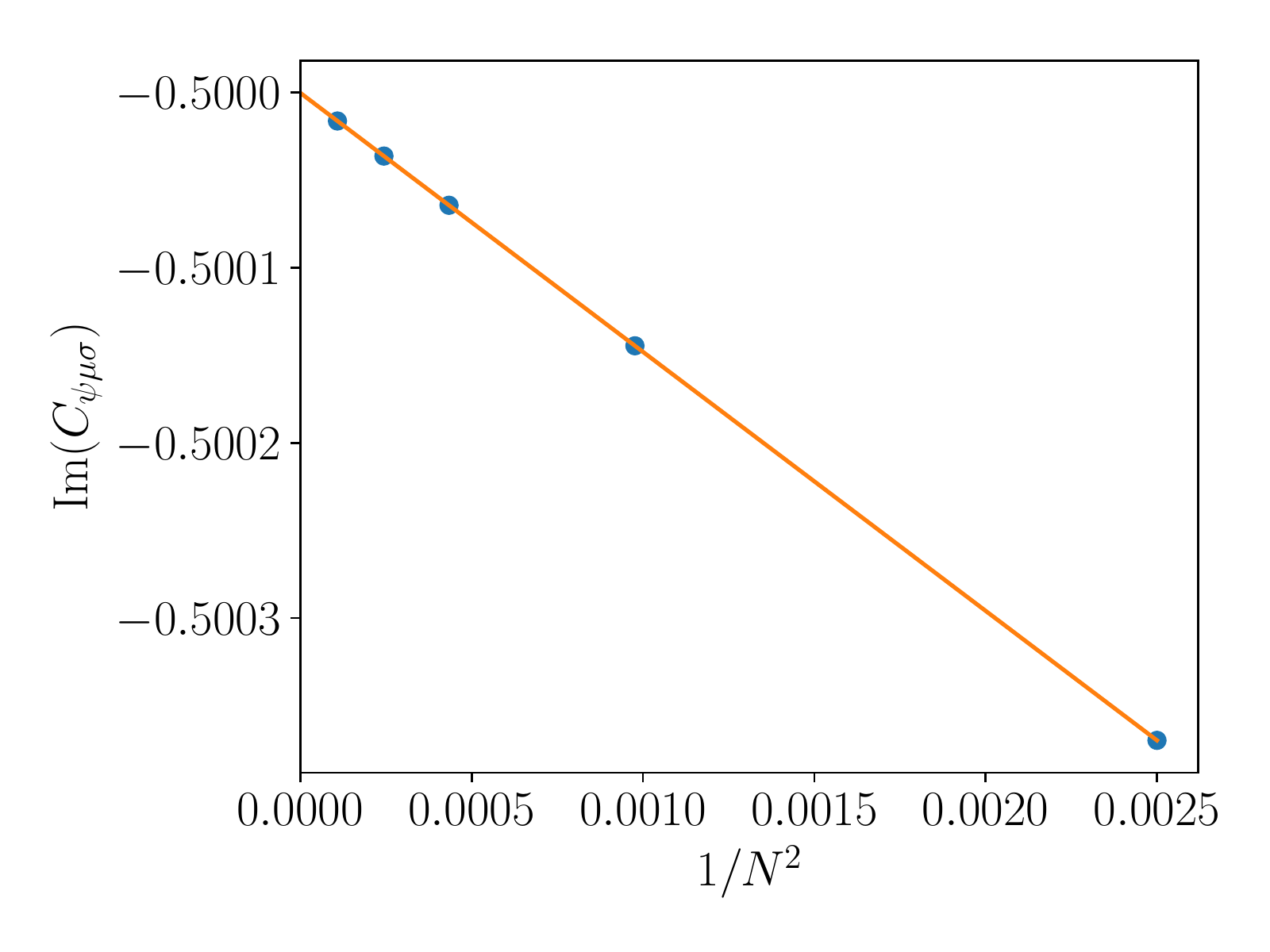}
\caption{\label{Fig:Cpsimusigma} Extrapolation of the real and imaginary part of the OPE coefficient $C_{\psi\mu\sigma}$.}
\end{figure}

We can then use these lattice operators to compute OPE coefficients from Eqs.~\eqref{OPEPPP},\eqref{OPEAPBC}. Again, an extrapolation to the thermodynamic limit is performed. For complex OPE coefficients, the real part and the imaginary part are extrapolated independently. The extrapolation of $C_{\psi\mu\sigma}$ is shown in Fig. (\ref{Fig:Cpsimusigma}) as an example.

Recall that each OPE coefficient can be computed in different ways by permuting the indices, where the second index labels the lattice primary operators and the other indices label the eigenstates. We tried all possible permutation of the indices for each nontrivial OPE coefficient of the Ising model, and the results are listed in Table \ref{table:IsingOPE}. All numerical results agree with the exact results in at least 5 digits.
\begin{table}
\begin{tabular}{|c|c|c|}
\hline
      OPE coefficient & Exact  & Numerical \\ \hline
       $C_{\sigma\sigma\epsilon}$     &  $0.5$   & $0.500001$   \\ \hline 
       $C_{\sigma\epsilon\sigma}$     &  $0.5$   & $0.500000$ \\ \hline 
       $C_{\mu\mu\epsilon}$     &  $-0.5$    & $-0.500000$ \\ \hline 
       $C_{\mu\epsilon\mu}$   &  $-0.5$   &  $-0.500000$ \\ \hline 
       $C_{\psi\mu\sigma}$     &  $0.5-0.5i$ &  $0.500000-0.500000i$    \\ \hline 
       $C_{\psi\sigma\mu}$     &  $0.5+0.5i$ &  $0.500000+0.500000i$    \\ \hline
       $C_{\mu\psi\sigma}$     &  $0.5+0.5i$ &   $0.500003+0.500003i$   \\ \hline
       $C_{\bar{\psi}\mu\sigma}$     &  $0.5+0.5i$ &  $0.500000+0.500000i$    \\ \hline 
       $C_{\bar{\psi}\sigma\mu}$     &  $0.5-0.5i$ &   $0.500000-0.500000i$   \\ \hline
       $C_{\mu\bar{\psi}\sigma}$     &  $0.5-0.5i$ &   $0.500003-0.500003i$   \\ \hline 
       $C_{\psi\epsilon\bar{\psi}}$    &  $i$           &   $1.00000i$   \\ \hline
       $C_{\bar{\psi}\psi\epsilon}$    &  $i$           &    $1.00001i$  \\ \hline
       $C_{\psi\bar{\psi}\epsilon}$    &  $-i$           &    $-1.00001i$  \\ \hline
\end{tabular}
\caption{OPE coefficients computed from the Ising model with $20\leq N\leq 96$. Numerical data are kept up to 6 digits. }
\label{table:IsingOPE}
\end{table}

We note that the expansion Eqs.~\eqref{operatorID},\eqref{stringCFT} may also involve descendant fields. This has been done in \cite{zou_conformal_2019}, where the extra coefficients in the expansion are used to obtain improved lattice representations of primary fields. The finite-size corrections of OPE coefficients are reduced to $N^{-4}$. However, in this paper, we will only find lattice operators which  correspond to the primary operators in the leading order. This is enough for extracting OPE coefficients.
\section{The tricritical Ising model}
In this section we apply our methods to a more complicated model, the tricritical Ising (TCI) model. It famously has emergent supersymmetry combined with the conformal symmmetry \cite{friedan_superconformal_1985}. In this section we will proceed as if we did not know about the emergent supersymmetry and extract the conformal data. In the next section we will analyze the emergent supersymmetry in more detail.

\subsection{The TCI CFT}
The TCI CFT also belongs to the unitary minimal models, whose conformal data can be solved exactly. It has central charge $c^{\CFT}=7/10$. The TCI CFT has 6 primary fields in the PBC sector, $\mathbf{1},\epsilon,\epsilon',\epsilon'',\sigma,\sigma$, and 6 primary fields in the APBC sector, $\psi,\bar{\psi},T_F,\bar{T}_F,\mu,\mu'$ \cite{lassig_scaling_1991}. Similar to the Ising CFT, we can also classify the above primary fields by the fermionic boundary conditions. The primary fields are summarized in Table \ref{table:TCICFTdeltas}. Again, operators in the NS sector can be classified with the eigenvalue under the Kramers-Wannier duality. $|\mathbf{1}\rangle, |\epsilon'\rangle,|\bar{\psi}\rangle, |T_F\rangle$ are even under the duality and $|\epsilon\rangle,|\epsilon''\rangle,|\psi\rangle,|\bar{T}_F\rangle$ are odd \cite{lassig_scaling_1991}.
\begin{table}[htbp]
\begin{tabular}{|c|c|c|c|c|c|}
\hline
$\phi^{\CFT}_\alpha$ & $\Delta^{\CFT}_\alpha$ & $s^{\CFT}_\alpha$ &$\mathcal{Z}_\alpha$& spin chain B.C. & fermion B.C.  \\ \hline
  $\mathbf{1}$ & 0 & 0 & +  & PBC & NS \\ \hline
  $\epsilon$ & 1/5 & 0 & +  & PBC & NS \\ \hline
  $\epsilon'$ & 6/5 & 0 & +  & PBC & NS \\ \hline
  $\epsilon''$ & 3 & 0 & +  & PBC & NS \\ \hline
  $\sigma$ & 3/40 & 0 & -- & PBC & R \\ \hline
  $\sigma'$ & 7/8 & 0 & --  & PBC & R \\ \hline
  $\psi$ & 7/10 & 1/2 & --  & APBC & NS \\ \hline
  $\bar{\psi}$ & 7/10 & --1/2 & --  & APBC & NS \\ \hline
  $T_F$ & 3/2 & 3/2 & --  & APBC & NS \\ \hline
  $\bar{T}_F$ & 3/2 & --3/2 & --  & APBC & NS \\ \hline
  $\mu$ & 3/40 & 0 & + & APBC & R \\ \hline
  $\mu'$ & 7/8 & 0 & + & APBC & R \\ \hline
\end{tabular}
\caption{Virasoro primary fields of the TCI CFT.}
\label{table:TCICFTdeltas}
\end{table}

As in the Ising CFT, we see a double degeneracy in the R sector. The degenerate states are related to each other by a supersymmetry transformation, which we will discuss in detail in the next section. 

As in the Ising model, the OPE must be consistent with the $\mathbb{Z}_2$ symmetry and the duality. The nonzero nontrivial ones (up to permutation of indices) are
\begin{eqnarray}
&\mathrm{PBC-PBC-PBC} \nonumber \\
&C^{\CFT}_{\epsilon\epsilon\epsilon'}=c_1, ~~ C^{\CFT}_{\epsilon'\epsilon'\epsilon'}=c_1, ~~ C^{\CFT}_{\epsilon\epsilon'\epsilon''}=3/7 ~~  \\
&C^{\CFT}_{\sigma\sigma\epsilon}=3c_1/2, ~~ C^{\CFT}_{\sigma\sigma\epsilon'}=c_1/4, ~~ C^{\CFT}_{\sigma\sigma\epsilon''}=1/56 ~~ \\
\label{BestOPE}
&C^{\CFT}_{\sigma\sigma'\epsilon}=1/2, ~~ C^{\CFT}_{\sigma\sigma'\epsilon'}=3/4, ~~  C^{\CFT}_{\sigma'\sigma'\epsilon''}=7/8 \\
& \nonumber \\
&\mathrm{APBC-APBC-PBC}, ~~ \mathcal{Z}=-1,-1,+1 \nonumber \\
&C^{\CFT}_{\psi\psi\epsilon}=C^{\CFT}_{\bar{\psi}\bar{\psi}\epsilon'}=-c_1 \\
&C^{\CFT}_{\bar{\psi}\psi\epsilon'}=-ic_1, ~~ C^{\CFT}_{\bar{\psi}\psi\epsilon''}=3i/7\\
&C^{\CFT}_{\psi T_F\epsilon}=C^{\CFT*}_{\bar{\psi}\bar{T}_F\epsilon}=-i \sqrt{3/7}, ~~ C^{\CFT}_{\bar{T}_F T_F\epsilon''}=-i ~~ \\
&C^{\CFT}_{\bar{\psi} T_F\epsilon'}=C^{\CFT}_{\psi\bar{T}_F\epsilon'}=\sqrt{3/7} \\
& \nonumber \\
&\mathrm{APBC-APBC-PBC}, ~~\mathcal{Z}=+1,-1,-1 \nonumber \\
&C^{\CFT}_{\mu T_F\sigma}=C^{\CFT*}_{\mu\bar{T}_F\sigma}=\sqrt{1/56}e^{-i\pi/4} \\
&C^{\CFT}_{\mu' T_F\sigma'}=C^{\CFT*}_{\mu'\bar{T}_F\sigma'}=\sqrt{7/8}e^{-3i\pi/4} \\
&C^{\CFT}_{\mu \psi\sigma}=C^{\CFT*}_{\mu\bar{\psi}\sigma}=\sqrt{3/8}c_1e^{-i\pi/4} \\
&C^{\CFT}_{\mu' \psi\sigma}=C^{\CFT*}_{\mu'\bar{\psi}\sigma}=\sqrt{3/8}e^{-3i\pi/4} \\
&C^{\CFT}_{\mu \psi\sigma'}=C^{\CFT*}_{\mu\bar{\psi}\sigma'}=\sqrt{3/8}e^{-3i\pi/4} \\
& \nonumber \\
&\mathrm{APBC-APBC-PBC}, ~~ \mathcal{Z}=+1,+1,+1 \nonumber \\
&C^{\CFT}_{\mu\mu\epsilon}=-3c_1/2, ~ C^{\CFT}_{\mu\mu\epsilon'}=c_1/4,  ~ C^{\CFT}_{\mu\mu\epsilon''}=-1/56 ~~ \\
&C^{\CFT}_{\mu'\mu\epsilon}=-1/2, ~ C^{\CFT}_{\mu'\mu\epsilon'}=3/4, ~ C^{\CFT}_{\mu'\mu'\epsilon''}=-7/8, ~~~~
\end{eqnarray}
where
\begin{equation}
c_1=\frac{2}{3}\sqrt{\frac{\Gamma(4/5)\Gamma^3(2/5)}{\Gamma(1/5)\Gamma^3(3/5)}}\approx 0.61030.
\end{equation}
The OPE coefficients of the PBC primary operators are taken from \cite{Mossa_2008}. We note that the OPE coefficient $C^{\CFT}_{\sigma'\sigma'\epsilon''}$ was written in \cite{lassig_scaling_1991} as $7c_1/8$, instead of $7/8$ in Eq.~\eqref{BestOPE}. Our numerical result below agrees with \cite{Mossa_2008}. OPE coefficients involving APBC operators are generally complex. Their phases depend on the convention of the normalization of scaling operators. Here we have chosen the convention such that they match the lattice calculations below. 
\subsection{Scaling dimensions, conformal spins and central charge from the TCI model}
\begin{figure}
\includegraphics[width=0.49\linewidth]{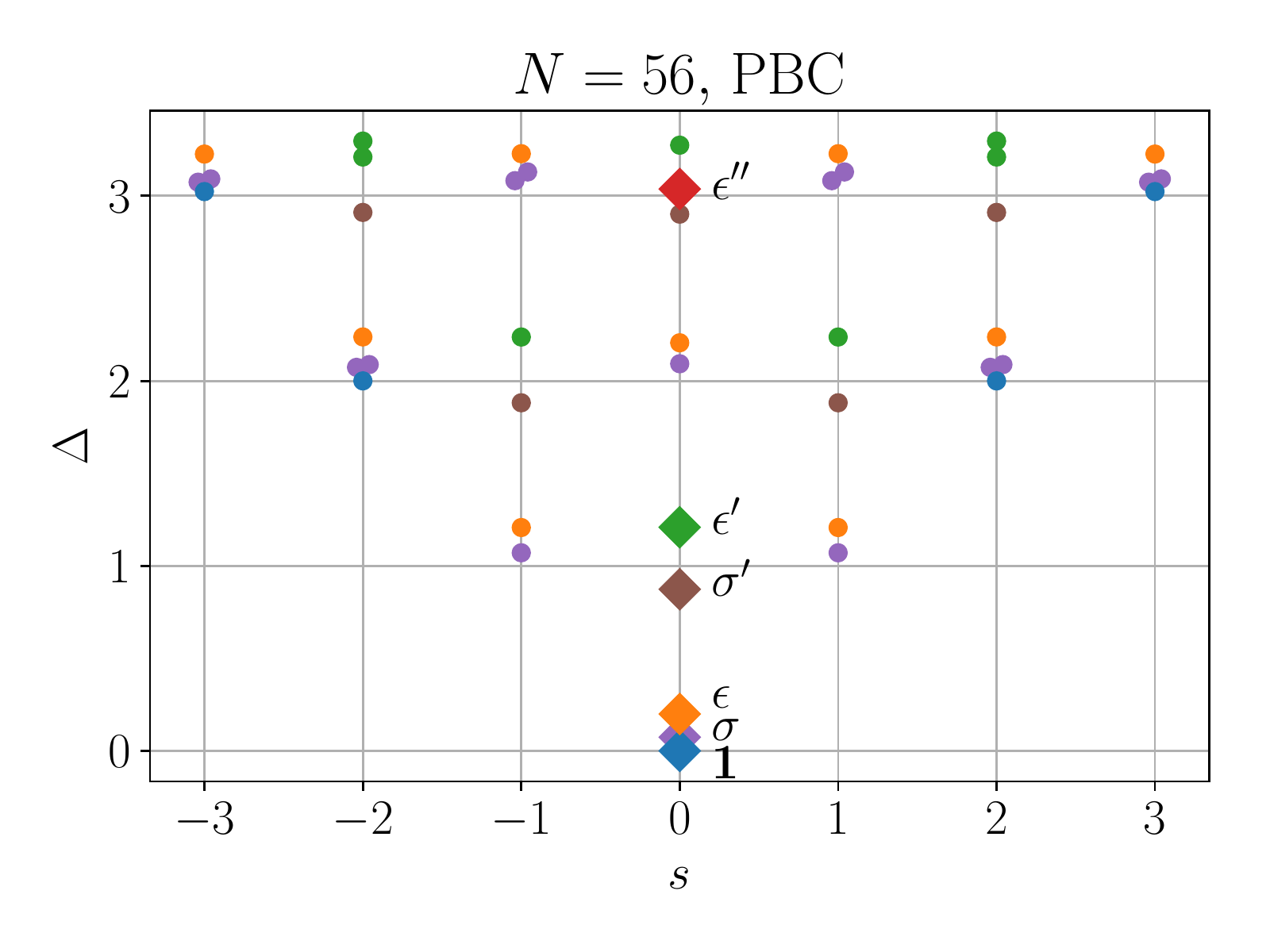} 
\includegraphics[width=0.49\linewidth]{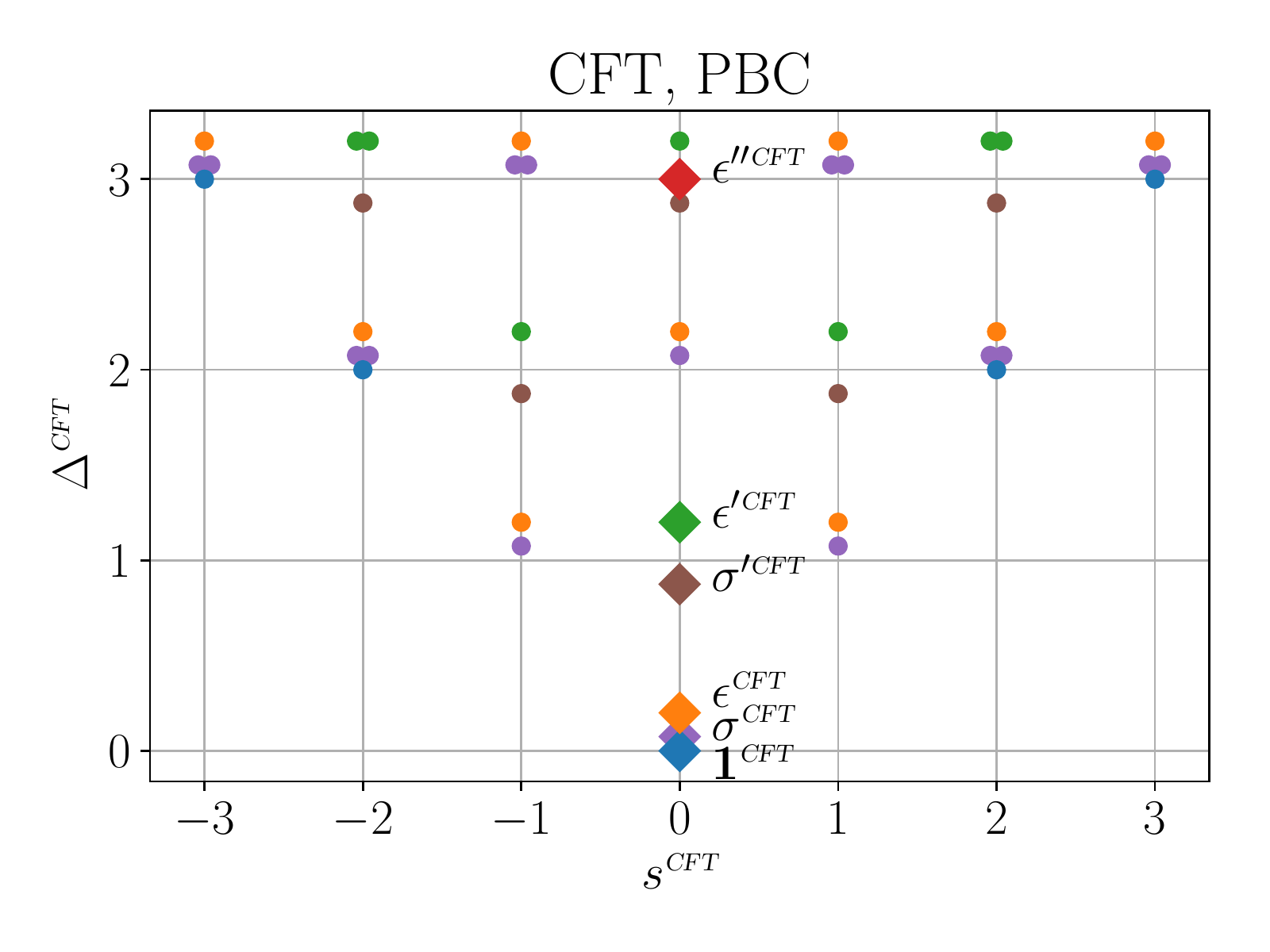} 
\includegraphics[width=0.49\linewidth]{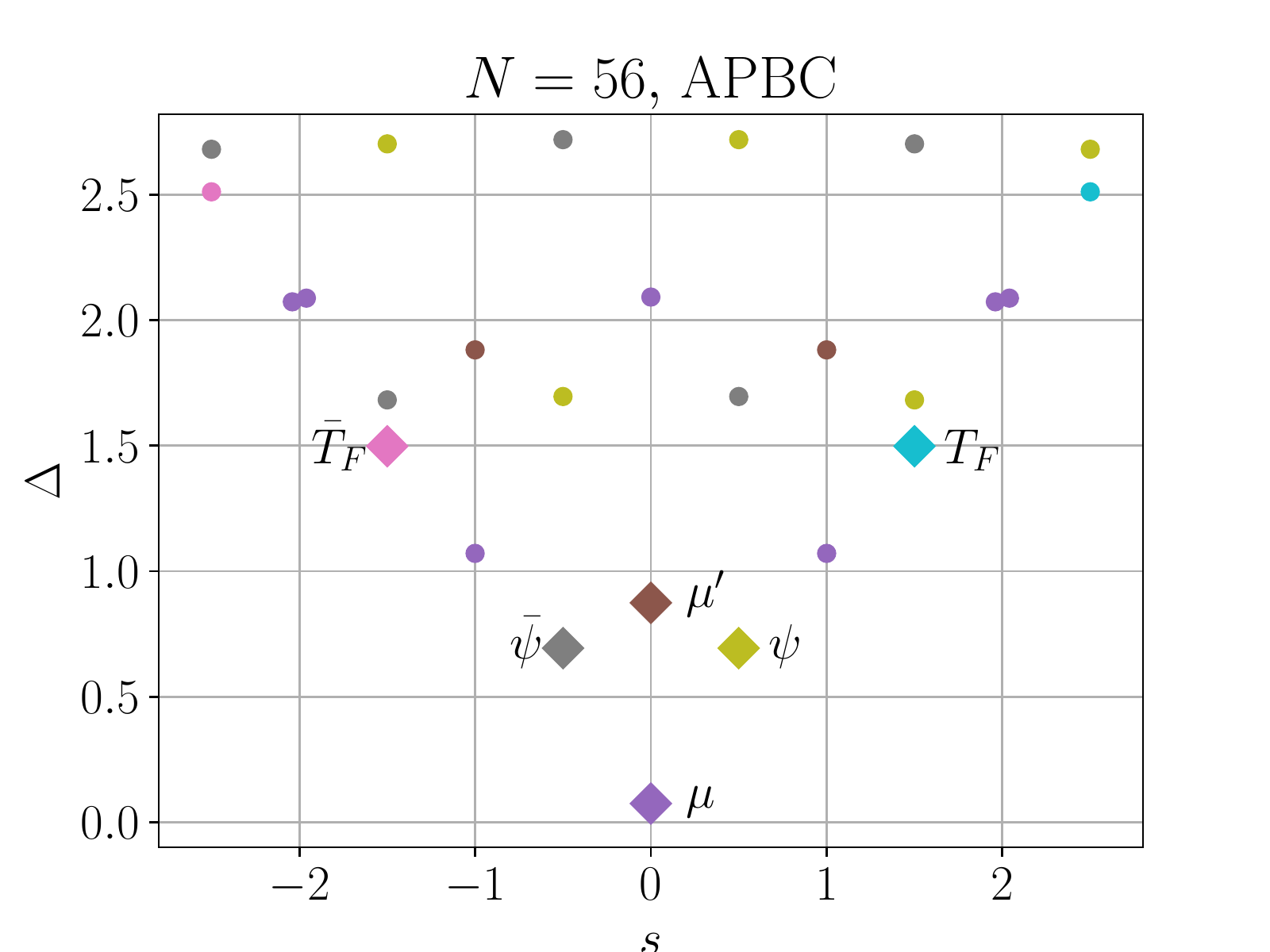} 
\includegraphics[width=0.49\linewidth]{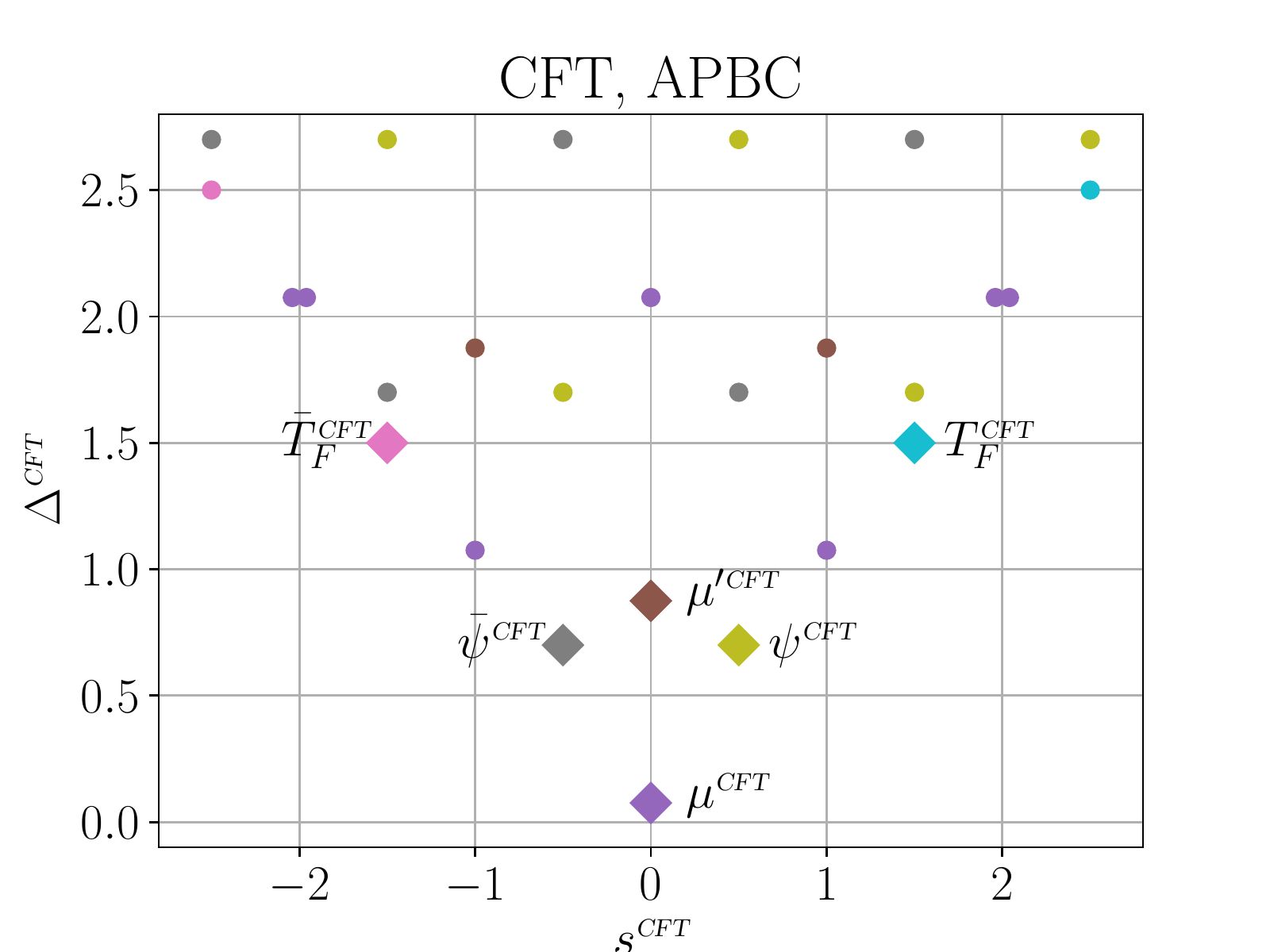} 
\caption{\label{Fig:TCIspec} (Left) Low-energy spectrum of the TCI model with PBC and APBC at $N=56$, diagonalized using puMPS with bond dimension $D=36$. Different colors indicate different conformal towers, with diamonds labeling the primary states. The exceptions are that the $\sigma(\sigma')$ and $\mu(\mu')$ towers are plotted with the same color. (Right) The spectrum of the TCI CFT up to $\Delta^{\CFT}\leq 3.2$ for PBC and $\Delta^{\CFT}\leq 2.7$ for APBC.}
\end{figure}

We use puMPS with bond dimension $20\leq D\leq 44$ to diagonalize the low-energy spectrum of the TCI model with both boundary conditions for $20\leq N\leq 80$. Similar to the Ising model, we plot the low-energy eigenstates up to scaling dimension $\Delta^{\CFT}\leq 3.2$ for the PBC and $\Delta^{\CFT}\leq 2.7$ for the APBC. The result is shown in Fig. (\ref{Fig:TCIspec}). We also see that all eigenstates corresponding to the CFT scaling operators in this range are captured by the puMPS ansatz.

Again, scaling dimensions and the central charge can be extracted with an extrapolation to the thermodynamic limit. The result is shown in Table \ref{table:TCIspec}. For simplicity we only show the scaling dimensions of the primary states. We see the error occurs after 3 significant digits, which roughly agrees with the accuracy of the tricritical point $\lambda^{*}$ in Eq.~\eqref{hTCI}.
\begin{table}[htbp]
\begin{tabular}{|c|c|c|c|}
\hline
        & Exact  & puMPS & error \\ \hline
  $c$ & $0.7$ & 0.6987 & $1.3\times 10^{-3}$  \\ \hline
  $\Delta_\epsilon$ & 0.2 & 0.2002 & $2\times 10^{-4}$   \\ \hline
  $\Delta_{\epsilon'}$ & 1.2 & 1.203 & $3\times 10^{-3}$  \\ \hline
  $\Delta_{\epsilon''}$ & 3 & 3.006 & $6.0\times 10^{-3}$  \\ \hline
  $\Delta_{\sigma}$ & 0.075 & 0.07493 & $7\times 10^{-5}$  \\ \hline
  $\Delta_{\sigma'}$ & 0.875 & 0.8748 & $2\times 10^{-4}$  \\ \hline
  $\Delta_\psi$ & 0.7 & 0.6979 &  $2.1\times 10^{-3}$ \\ \hline
  $\Delta_{\bar{\psi}}$ & 0.7 & 0.6979 & $2.1\times 10^{-3}$  \\ \hline
  $\Delta_{T_F}$ & 1.5 & 1.500 &  $1\times 10^{-4}$ \\ \hline
  $\Delta_{\bar{T}_F}$ &1.5 & 1.500 & $1\times 10^{-4}$  \\ \hline
  $\Delta_{\mu}$ & 0.075 & 0.07493 & $7\times 10^{-5}$   \\ \hline
  $\Delta_{\mu'}$ & 0.875 & 0.8748 & $2\times 10^{-4}$   \\ \hline
\end{tabular}
\caption{Scaling dimensions from the TCI model. All numerical values are extrapolated using $40\leq N \leq 80$, except $\Delta_{\epsilon''}$ where we use $20\leq N \leq 56$. All numerical values are kept to 4 significant digits.}
\label{table:TCIspec}
\end{table}

\subsection{OPE coefficients from the TCI model}
Following the general prescription in Section II and IV, we first need to construct lattice operators corresponding to CFT primary fields variationally. Here we choose to compute lattice representations of $\epsilon,\epsilon',\sigma,\sigma',\mu,\mu',\psi,\bar{\psi},T_F,\bar{T}_F$. These operators correspond to $\Delta^{\CFT}_\alpha\leq 1.2$ in the PBC sector and $\Delta^{\CFT}_\alpha\leq 1.5$ in the APBC sector. Notice that we will not construct the lattice operator for $\epsilon''$ with $\Delta^{\CFT}_{\epsilon''}=3$ because it is numerically difficult. The reason is that it requires a linear combination of many lattice operators with fine-tuned coefficients such that all contributions with lower scaling dimensions vanish. With an extrapolation of the finite-size $a_\alpha$'s, we obtain the results listed in Table \ref{table:OBFops}.
\begin{table}[htbp]
\begin{tabular}{|c|c|}
\hline
      Lattice  & CFT \\ \hline
       $X$     &  $0.7808\sigma^{\CFT}-0.1866\sigma'^{\CFT}$      \\ \hline 
       $XZ+ZX$     &  $0.3551\sigma^{\CFT}-0.9060\sigma'^{\CFT}$      \\ \hline
       $Z$     &  $0.5821\mathbf{1}^{\CFT}-0.4753\epsilon^{\CFT}-0.1760\epsilon'^{\CFT}$      \\ \hline 
       $XX$   &  $0.5821\mathbf{1}^{\CFT}+0.4753\epsilon^{\CFT}-0.1760\epsilon'^{\CFT}$      \\ \hline 
       $\mathcal{S}_I$     &  $0.7808\mu^{\CFT}-0.1866\mu'^{\CFT}$      \\ \hline 
       $\mathcal{S}_{XX}-\mathcal{S}_{YY}$     &  $0.3551\mu^{\CFT}-0.9060\mu'^{\CFT}$      \\ \hline 
       $\mathcal{S}_X$     &  $0.4042(\psi^{\CFT}+\bar{\psi}^{\CFT})+0.2462(T^{\CFT}_F+\bar{T}^{\CFT}_F)$      \\ \hline
       $\mathcal{S}_Y$     &  $-0.4042(\psi^{\CFT}-\bar{\psi}^{\CFT})+0.2462(T^{\CFT}_F-\bar{T}^{\CFT}_F)$      \\ \hline
       $\mathcal{S}_{IX}$     &  $0.4695(\psi^{\CFT}+\bar{\psi}^{\CFT})+0.0829(T^{\CFT}_F+\bar{T}^{\CFT}_F)$      \\ \hline
       $\mathcal{S}_{YZ}$     &  $-0.4695(\psi^{\CFT}-\bar{\psi}^{\CFT})+0.0829(T^{\CFT}_F-\bar{T}^{\CFT}_F)$      \\ \hline
\end{tabular}
\\

\begin{tabular}{|c|c|}
\hline
      CFT  & Lattice \\ \hline
       $\sigma^{\CFT}$     &  $1.413X-0.2910(XZ+ZX)$      \\ \hline 
       $\sigma'^{\CFT}$     &  $0.5539X-1.218(XZ+ZX)$      \\ \hline 
       $\epsilon^{\CFT}$     &  $1.052XX-1.052Z$      \\ \hline
       $\epsilon'^{\CFT}$     &  $3.307I-2.841XX-2.841Z$      \\ \hline  
       $\mu^{\CFT}$     &  $1.413 \mathcal{S}_I-0.2910(\mathcal{S}_{XX}-\mathcal{S}_{YY})$      \\ \hline
       $\mu'^{\CFT}$     &  $0.5539 \mathcal{S}_I-1.218(\mathcal{S}_{XX}-\mathcal{S}_{YY})$      \\ \hline  
       $\psi^{\CFT}$   &  $-0.5072(\mathcal{S}_X-\mathcal{S}_Y)+1.506(\mathcal{S}_{IX}-\mathcal{S}_{YZ})$      \\ \hline 
       $\bar{\psi}^{\CFT}$     &  $-0.5072(\mathcal{S}_X+\mathcal{S}_Y)+1.506(\mathcal{S}_{IX}+\mathcal{S}_{YZ})$      \\ \hline 
       $T^{\CFT}_F$   &  $2.860(\mathcal{S}_X+\mathcal{S}_Y)-2.462(\mathcal{S}_{IX}+\mathcal{S}_{YZ})$      \\ \hline 
       $\bar{T}^{\CFT}_F$     &  $2.860(\mathcal{S}_X-\mathcal{S}_Y)-2.462(\mathcal{S}_{IX}-\mathcal{S}_{YZ})$      \\ \hline 
\end{tabular}
\caption{Correspondence between lattice operators and CFT operators for the O'Brien-Fendley model. (Top) Correspondence between some lattice operators and a linear combination of CFT primary operators. (Bottom) Lattice operators that correspond to CFT primary operators. }
\label{table:OBFops}
\end{table}

\begin{table}
\begin{tabular}{|c|c|c|}
\hline
OPE coefficient & Exact & Numerical \\ \hline
$C_{\sigma\sigma\epsilon}$ & $0.9155$ & $0.9154$ \\ \hline
$C_{\sigma\sigma\epsilon'}$ & $0.1526$ & $0.1531$ \\ \hline
$C_{\sigma\sigma\epsilon''}$ & $0.0179$ & $0.0177$ \\ \hline
$C_{\sigma'\sigma\epsilon}$ & $0.5$ & $0.5007$ \\ \hline
$C_{\sigma'\sigma\epsilon'}$ & $0.75$ & $0.752$ \\ \hline
$C_{\sigma'\sigma'\epsilon''}$ & $0.875$ & $0.869$ \\ \hline
$C_{\epsilon\epsilon\epsilon'}$ & $0.610$ & $0.608$ \\ \hline
$C_{\epsilon'\epsilon\epsilon''}$ & $0.429$ & $0.437$ \\ \hline
$C_{\epsilon'\epsilon'\epsilon'}$ & $0.61$ & $0.58$ \\ \hline
\end{tabular}
\begin{tabular}{|c|c|c|}
\hline
OPE coefficient & Exact & Numerical \\ \hline
$C_{\psi T_F\epsilon}$ & $-0.655i$ & $-0.662i$ \\ \hline
$C_{\bar{\psi} T_F\epsilon'}$ & $0.655$ & $0.664$ \\ \hline
$C_{\bar{T}_F T_F\epsilon''}$ & $-i$ & $-0.03-1.02i$ \\ \hline
$C_{\bar{\psi} \bar{T}_F\epsilon}$ & $0.655i$ & $0.662i$ \\ \hline
$C_{\psi \bar{T}_F\epsilon'}$ & $0.655$ & $0.664$ \\ \hline
$C_{\psi \psi \epsilon'}$ & $-0.610$ & $-0.613$ \\ \hline
$C_{\bar{\psi} \bar{\psi} \epsilon'}$ & $-0.610$ & $-0.613$ \\ \hline
$C_{\bar{\psi} \psi \epsilon}$ & $0.610i$ & $0.612i$ \\ \hline
$C_{\bar{\psi} \psi \epsilon''}$ & $0.43i$ & $0.01+0.41i$ \\ \hline
\end{tabular}
\caption{OPE coefficients of the TCI CFT computed from the TCI model. The organization of the table follows the exact results listed before. All numerical results are kept to the last significant digits, and the exact results are shown with the same number of significant digit. All OPE coefficients involving $\epsilon''$ are extrapolated with data from $20\leq N \leq 56$, while those not involving $\epsilon''$ are extrapolated with data from $32\leq N \leq 72$.}
\label{table:TCIOPE}
\end{table}

\begin{table}
\begin{tabular}{|c|c|c|}
\hline
OPE coefficient & Exact & Numerical \\ \hline
$C_{\mu T_F \sigma}$ & $0.094-0.094i$ & $0.097-0.097i$ \\ \hline
$C_{\mu \bar{T}_F \sigma}$ & $0.094+0.094i$ & $0.097+0.097i$ \\ \hline
$C_{\mu' T_F \sigma'}$ & $-0.661+0.661i$ & $-0.669+0.669i$ \\ \hline
$C_{\mu' \bar{T}_F \sigma'}$ & $-0.661-0.661i$ & $-0.669-0.669i$ \\ \hline
$C_{\mu \psi \sigma}$ & $0.265-0.265i$ & $0.264-0.264i$ \\ \hline
$C_{\mu \psi \sigma'}$ & $-0.433+0.433i$ & $-0.434+0.434i$ \\ \hline
$C_{\mu' \psi \sigma}$ & $-0.433+0.433i$ & $-0.434+0.434i$ \\ \hline
$C_{\mu \bar{\psi} \sigma}$ & $0.264+0.264i$ & $0.264+0.264i$ \\ \hline
$C_{\mu' \bar{\psi} \sigma}$ & $-0.433-0.433i$ & $-0.434-0.434i$ \\ \hline
$C_{\mu' \bar{\psi} \sigma}$ & $-0.433-0.433i$ & $-0.434-0.434i$ \\ \hline
\end{tabular}

\begin{tabular}{|c|c|c|}
\hline
OPE coefficient & Exact & Numerical \\ \hline
$C_{\mu\mu\epsilon}$ & $-0.9154$ & $-0.9153$ \\ \hline
$C_{\mu\mu\epsilon'}$ & $0.1526$ & $0.1532$ \\ \hline

$C_{\mu\mu\epsilon''}$ & $-0.0179$ & $-0.0172$ \\ \hline

$C_{\mu'\mu\epsilon}$ & $-0.5$ & $-0.5008$ \\ \hline

$C_{\mu'\mu\epsilon'}$ & $0.75$ & $0.752$ \\ \hline

$C_{\mu'\mu'\epsilon''}$ & $-0.875$ & $-0.86$ \\ \hline
\end{tabular}
\caption{OPE coefficients of the TCI CFT computed from the TCI model, continued.}
\label{table:TCIOPE2}
\end{table}

We can then work out all OPE coefficients $C_{\alpha\beta\gamma}$ where $\phi_\beta\neq \epsilon''$ with Eqs.~\eqref{OPEPPP},\eqref{OPEAPBC}. For those OPE coefficients that are related by permuting indices, we will only show one particular order of indices for simplicity, e.g., we will compute $C_{\sigma\sigma\epsilon}$ but not $C_{\sigma\epsilon\sigma}$. With indices appropriately permuted, only the computation of $C_{\epsilon''\epsilon''\epsilon''}$ requires a lattice operator for $\epsilon''$, but we know $C^{\CFT}_{\epsilon''\epsilon''\epsilon''}$ vanishes because $\epsilon''$ is odd under Kramers-Wannier duality. Therefore in this case we can still extract a complete set of nonvanishing OPE coefficients, see Tables \ref{table:TCIOPE}, \ref{table:TCIOPE2}.

\section{Emergence of superconformal symmetry in the tricritical Ising model}
In this section we study the emergent superconformal symmetry in the TCI model. We first review the $\mathcal{N}=1$ superconformal algebra. We then show how to find the lattice operators that correspond to supervirasoro generators. We verify the action of the supervirasoro generators on low energy subspaces of the TCI model with both PBC and APBC. Some matrix elements are checked quantitatively with analytical results. In particular, a formula analogous to Eq.~\eqref{clat} for the central charge is proposed and checked numerically.
\subsection{$\mathcal{N}=1$ supersymmetry and the TCI model}
The $\mathcal{N}=1$ supersymmetry for $1+1$ dimensional quantum field theories is defined with two \textit{supercharges} $Q^{\QFT}$ and $\bar{Q}^{\QFT}$ which satisfy
\begin{eqnarray}
Q^{\QFT \dagger}&=&Q^{\QFT} \\
\bar{Q}^{\QFT \dagger}&=&\bar{Q}^{\QFT} \\
\label{Q1Q2}
~ [Q^{\QFT},\bar{Q}^{\QFT}]&=&0 \\
\label{HQ}
H^{\QFT}&=&(Q^{\QFT})^2+(\bar{Q}^{\QFT})^2.
\end{eqnarray}
The supercharges are fermionic operators. They map bosonic excitations into fermionic excitations and vice versa. 

 Each supercharge is associated with a \textit{supersymmetry current}, $T^{\QFT}_F,\bar{T}^{\QFT}_F$, which are the density of the supercharges,
\begin{equation}
Q^{\QFT}=\int dx \, T^{\QFT}_F(x), ~~\bar{Q}^{\QFT}=\int dx\, \bar{T}^{\QFT}_F(x).
\end{equation}

For a lattice model which flows into a supersymmetric quantum field theory, such as the TCI model Eq.~\eqref{hTCI}, the lattice version of Eqs.~\eqref{Q1Q2},\eqref{HQ} may not be exact. As pointed out by O'brien and Fendley \cite{obrien_lattice_2018}, the TCI Hamiltonian with density Eq.~\eqref{hTCI} can be expressed as
\begin{equation}
\label{TCIHsusy}
H_{\TCI} = Q^2+\bar{Q}^2+E_0,
\end{equation}
where $E_0$ is an energy shift, and
\begin{align}
Q&=\sum_{j} T_{F,j} \\
\label{Q1}
T_{F,j}&\propto(\gamma_{2j-1}+\gamma_{2j})+2i\lambda^{*}(\gamma_{2j-2}+\gamma_{2j+1})\gamma_{2j-1}\gamma_{2j}\\
\bar{Q}&=\sum_{j} \bar{T}_{F,j} \\
\label{Q2}
\bar{T}_{F,j}&\propto (\gamma_{2j-1}-\gamma_{2j})+2i\lambda^{*}(\gamma_{2j+1}-\gamma_{2j-2})\gamma_{2j-1}\gamma_{2j},
\end{align}
where $2\lambda^{*}\approx 0.856$. In terms of string operators, they are
\begin{align}
\label{TFstring}
T_{F}\propto(\mathcal{S}_{X}+\mathcal{S}_{Y})-2\lambda^{*}(\mathcal{S}_{YZ}+\mathcal{S}_{IX}), \\
\label{TFbarstring}
\bar{T}_{F}\propto (\mathcal{S}_{X}-\mathcal{S}_{Y})+2\lambda^{*}(\mathcal{S}_{YZ}-\mathcal{S}_{IX}).
\end{align}
It is simple to check that $Q$ and $\bar{Q}$ are Hermitian but $[Q,\bar{Q}]\neq 0$. Therefore, supersymmetry is not exact on the lattice. However, it has been shown numerically \cite{obrien_lattice_2018} that, under the RG flow, not only $Q$ and $\bar{Q}$ flow to the supercharges, but also $T_F$ and $\bar{T}_F$ flow to the supersymmetry currents. In the previous section, we have variationally found $T_{F,j}$ and $\bar{T}_{F,j}$ (Table \ref{table:OBFops}) without exploiting Eq.~\eqref{TCIHsusy},
\begin{align}
T_F=2.86[(\mathcal{S}_{X}+\mathcal{S}_{Y})-0.861(\mathcal{S}_{YZ}+\mathcal{S}_{IX})], \\
\bar{T}_F=2.86[(\mathcal{S}_{X}-\mathcal{S}_{Y})+0.861(\mathcal{S}_{YZ}-\mathcal{S}_{IX})].
\end{align}
They agree quantitatively with Eqs.~\eqref{TFstring},\eqref{TFbarstring} up to normalization, with the error in the third digit. 

We also note that in \cite{obrien_lattice_2018} the lattice operators corresponding to $\psi^{\CFT}$ and $\bar{\psi}^{\CFT}$ were also proposed,
\begin{align}
 \psi\propto(\mathcal{S}_{X}-\mathcal{S}_{Y})-2\lambda^{*}(\mathcal{S}_{YZ}-\mathcal{S}_{IX}), \\
 \bar{\psi}\propto (\mathcal{S}_{X}+\mathcal{S}_{Y})+2\lambda^{*}(\mathcal{S}_{YZ}+\mathcal{S}_{IX}).
 \end{align}
In this paper we obtain a very different coefficient between the two terms (see Table \ref{table:OBFops}). However, the two results do not contradict each other. They both correspond to $\psi^{\CFT} (\bar{\psi}^{\CFT})$ as the leading contribution, but our result also eliminates the contribution from $T^{\CFT}_F, \bar{T}^{\CFT}_F$. In this sense, we provide an improved lattice operator corresponding to $\psi^{\CFT}$ (as well as $\bar{\psi}^{\CFT}$). 

\subsection{The superconformal algebra}
If conformal symmetry is enhanced by the supersymmetry, the resulting quantum field theory is a \textit{superconformal} field theory (SCFT). In a SCFT,  the supercurrent $T^{\CFT}_F (\bar{T}^{\CFT}_F)$ are Virasoro primary fields with conformal dimensions $(3/2,0)$ and $(0,3/2)$, respectively. Expanding $T^{\CFT}_F$ in Fourier modes gives \textit{supervirasoro generators}
\begin{equation}
\label{Gn}
G^{\CFT}_{n}=\left(\frac{2\pi}{L}\right)^{-1/2}\int_0^L dx\, T^{\CFT}_F(x) e^{inx2\pi/L}.
\end{equation}
Together with $L^{\CFT}_n$, they satisfy the superconformal algebra
\begin{align}
\label{Svirasoro1}
[L^{\CFT}_n,L^{\CFT}_m]&=(n-m) L^{\CFT}_{n+m} + \frac{c^{\CFT}}{12}n (n^2-1) \delta_{n+m,0}  \\
\label{Svirasoro2}
~[L^{\CFT}_n,G^{\CFT}_m]&=\left(\frac{1}{2}n-m\right) G^{\CFT}_{n+m}  \\
\label{Svirasoro3}
~\{G^{\CFT}_n,G^{\CFT}_m\}&=2 L^{\CFT}_{n+m} + \frac{c^{\CFT}}{3} \left(n^2-\frac{1}{4}\right) \delta_{n+m,0},
\end{align}  
where the first identity is the Virasoro algebra, the second identity follows from the fact that $T^{\CFT}_F$ is a primary field with conformal dimensions $(3/2,0)$, and the third identity is the crucial feature of a supersymmetric theory that the anticommutator of two supersymmetry generators yields the generator of a spacetime transformation. Analogous to the Virasoro algebra, there is a copy of the same superconformal algebra for the anti-holomorphic generators $\bar{L}^{\CFT}_n,\bar{G}^{\CFT}_m$. Since $T^{\CFT}_F$ is a fermionic field, it follows that $m\in \mathbb{Z}+1/2$ for the NS boundary condition, and $m\in \mathbb{Z}$ for the R boundary condition. The corresponding superconformal algebras are called the NS algebra and the R algebra, respectively.

Let us analyze the action of the supervirasoro generators in more detail. First, let $n=0$ in Eq.~\eqref{Svirasoro2}. We obtain
\begin{equation}
[L^{\CFT}_0,G^{\CFT}_m]=-m G^{\CFT}_{m},
\end{equation}
which means that $G^{\CFT}_m$ changes the holomorphic dimension by $-m$. Therefore, $G^{\CFT}_m$ with $m<0$ is a raising operator, whereas with $m>0$ a lowering operator. A superconformal primary state $|\Phi^{\CFT}_\alpha\rangle$ is defined by
\begin{eqnarray}
L^{\CFT}_n|\Phi^{\CFT}_\alpha\rangle=0, ~~ G^{\CFT}_m|\Phi^{\CFT}_\alpha\rangle=0, ~(n,m>0) \\
\bar{L}^{\CFT}_n|\Phi^{\CFT}_\alpha\rangle=0, ~~ \bar{G}^{\CFT}_m|\Phi^{\CFT}_\alpha\rangle=0, ~(n,m>0).
\end{eqnarray}
By virtue of Eqs.~\eqref{Svirasoro1},\eqref{Svirasoro2}, the above equalities hold for all $n>0,m>0$ if they hold for $n=1,2$ and $m=1/2,3/2$ for the NS algebra (or $m=1$ for the R algebra).

In the $R$ algebra, $G^{\CFT}_0$ needs more attention. First, Eq.~\eqref{Svirasoro3} implies
\begin{equation}
\label{G0sq}
(G^{\CFT}_0)^2=L^{\CFT}_0-\frac{c^{\CFT}}{24}.
\end{equation}
Then, the Hamiltonian can be written as
\begin{equation}
H^{\CFT}=\frac{2\pi}{L}((G^{\CFT}_0)^2+(\bar{G}^{\CFT}_0)^2),
\end{equation}
which follows from Eq.~\eqref{HCFT}. We see that $G^{\CFT}_0$ and $\bar{G}^{\CFT}_0$ are proportional to supercharges, in accordance with Eqs.~\eqref{HQ},\eqref{Gn}. Second, $G^{\CFT}_0$ commutes with $L^{\CFT}_0$. This means that $G_0|\Phi^{\CFT}_\alpha\rangle$, if non-vanishing, has the same conformal dimensions as $|\Phi^{\CFT}_\alpha\rangle$. Eq.~\eqref{G0sq} implies that $G_0|\Phi^{\CFT}_\alpha\rangle$ is nonvanishing if
\begin{equation}
 h^{\CFT}_{\alpha}\neq\frac{c^{\CFT}}{24}.
\end{equation}
If this is true for some supervirasoro primary state in the R sector, the supervirasoro primary state is at least double degenerate. As a result, all descendant states will also be at least double degenerate.  In this case the supersymmetry is said to be spontaneously broken \cite{friedan_superconformal_1985}. We have seen that indeed there is a double degeneracy for each state in the R sector of the TCI model, one in the PBC of the spin chain and the other in the APBC of the spin chain.
\subsection{Supervirasoro primary states in the TCI CFT}
As noted in the previous section, there are 12 Virasoro primary states in the TCI CFT, where 8 of them are in the NS sector and 4 of them are in the R sector. In the R sector, all Virasoro primary states are also supervirasoro primary states. $\sigma^{\CFT}(\sigma'^{\CFT})$ is the superpartner of $\mu^{\CFT}(\mu'^{\CFT})$. They are related by the supercharge $Q^{\CFT}\propto G^{\CFT}_0$,
\begin{eqnarray}
\label{GnCFT1}
G^{\CFT}_0|\sigma^{\CFT}\rangle&=&a_{\sigma}|\mu^{\CFT}\rangle \\
\label{GnR2}
G^{\CFT}_0|\sigma'^{\CFT}\rangle&=&a_{\sigma'}|\mu'^{\CFT}\rangle,
\end{eqnarray}
where 
\begin{eqnarray}
a_{\alpha}=\sqrt{h^{\CFT}_\alpha-\frac{c^{\CFT}}{24}},
\end{eqnarray}
by virtue of Eq.~\eqref{G0sq}, where $h^{\CFT}_{\sigma}=3/80$, $h^{\CFT}_{\sigma'}=7/16$, $c^{\CFT}=7/10$.

\begin{figure}
\includegraphics[width=0.99\linewidth]{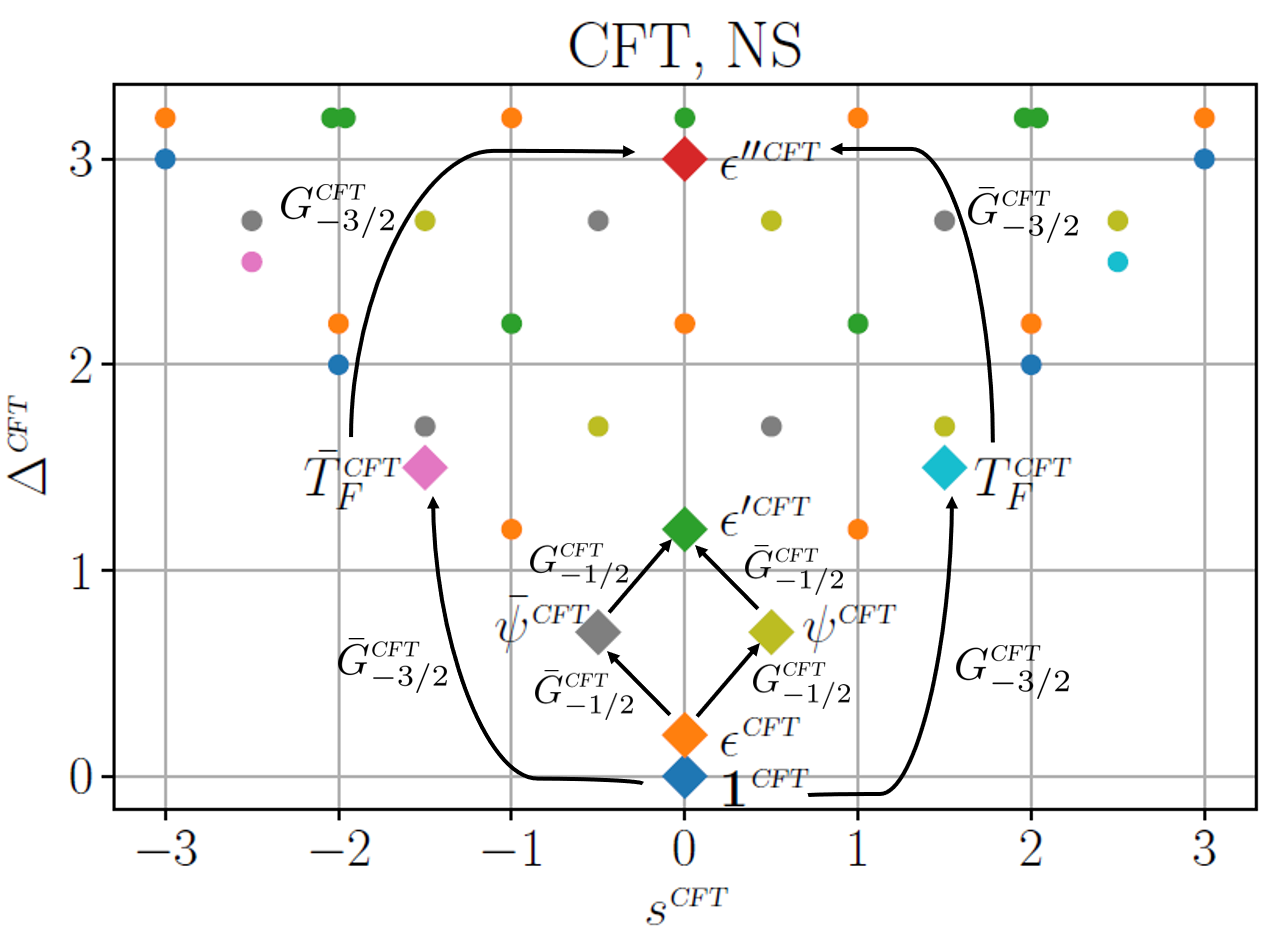} 
\caption{\label{Fig:TCINS} Spectrum of the TCI CFT in the NS sector. Primary states are labelled as diamonds. Arrows indicate that the primary states are related by the supervirasoro generators $G^{\CFT}_m$.}
\end{figure}

 In the NS sector, only $\mathbf{1}^{\CFT}$ and $\epsilon^{\CFT}$ are superconformal primary states. The rest of virasoro primary states are connected to the supervirasoro primary states by the $G^{\CFT}_m$, shown in Fig.(\ref{Fig:TCINS}). The matrix elements are
 \begin{eqnarray}
 \label{GnCFT3}
 \langle \psi^{\CFT}|G^{\CFT}_{-1/2}|\epsilon^{\CFT}\rangle&=&\langle \bar{\psi}^{\CFT}|\bar{G}^{\CFT}_{-1/2}|\epsilon^{\CFT}\rangle=a_{\epsilon} \\
 \label{GnCFT4}
 \langle \epsilon'^{\CFT}|G^{\CFT}_{-1/2}|\bar{\psi}^{\CFT}\rangle&=&\langle \epsilon'^{\CFT}|\bar{G}^{\CFT}_{-1/2}|\psi^{\CFT}\rangle=a_{\epsilon} \\
 \label{Gnc}
 \langle T^{\CFT}_F|G^{\CFT}_{-3/2}|0^{\CFT}\rangle&=&\langle \bar{T}^{\CFT}_F|\bar{G}^{\CFT}_{-3/2}|0^{\CFT}\rangle=a_{\mathbf{1}} \\
 \label{GnCFT2}
 \langle \epsilon''^{\CFT}|G^{\CFT}_{-3/2}|\bar{T}^{\CFT}_F\rangle&=&\langle \epsilon''^{\CFT}|\bar{G}^{\CFT}_{-3/2}|T^{\CFT}_F\rangle=a_{\mathbf{1}},
  \end{eqnarray}
  where $a_\epsilon=\sqrt{2h^{\CFT}_{\epsilon}}=1/\sqrt{5}$ and $a_{\mathbf{1}}=\sqrt{2c^{\CFT}/3}=\sqrt{7/15}$. They can be derived from the superconformal algebra (see Appendix). These matrix elements indicate that $|\psi^{\CFT}\rangle,|\bar{\psi}^{\CFT}\rangle,|\epsilon'^{\CFT}\rangle$ are supervirasoro descendants of $|\epsilon^{\CFT}\rangle$, and $|T^{\CFT}_F\rangle,|\bar{T}^{\CFT}_F\rangle,|\epsilon''^{\CFT}\rangle$ are supervirasoro descendants of $|\mathbf{1}^{\CFT}\rangle$. 

\subsection{Lattice supervirasoro generators}
In the CFT, the supervirasoro generators are Fourier modes of the fermionic stress tensors, Eq.~\eqref{Gn}. Therefore, we expect that on the lattice Fourier modes of the string operator $T_F$ also realize the supervirasoro operators at low energies. Recall the definition Eq.~\eqref{stringFourier},
\begin{equation}
\label{Gnlat}
G_{n}=\eta\sum_{j=1}^N \tilde{\mathcal{T}}^{j-1} T_{F,1} \mathcal{T}^{\dagger j-1} e^{inj 2\pi/N},
\end{equation}
where $\mathcal{T}$ and $\tilde{\mathcal{T}}$ are translation operators for the Hamiltonian with PBC and APBC, respectively, and $\eta$ is a normalization factor. The equation above applies to both the NS sector ($n\in 1/2+\mathbb{Z}$) and the R sector ($n\in \mathbb{Z}$). To fix the normalization factor $\eta$, we require that
\begin{equation}
\label{Gnnormalize}
\langle T_F|G_{1/2}|T\rangle=\sqrt{3}.
\end{equation} 
This comes from the CFT identity (derivation in Appendix)
\begin{equation}
\label{GnCFTnorm}
\langle T^{\CFT}_F|G^{\CFT}_{1/2}|T^{\CFT}\rangle=\sqrt{3}.
\end{equation}
Note that both $|T^{\CFT}\rangle$ and $|T^{\CFT}_F\rangle$ necessarily exist in a SCFT, such that the normalization condition Eq.~\eqref{Gnnormalize} is universally applicable.

\begin{figure}
\includegraphics[width=0.99\linewidth]{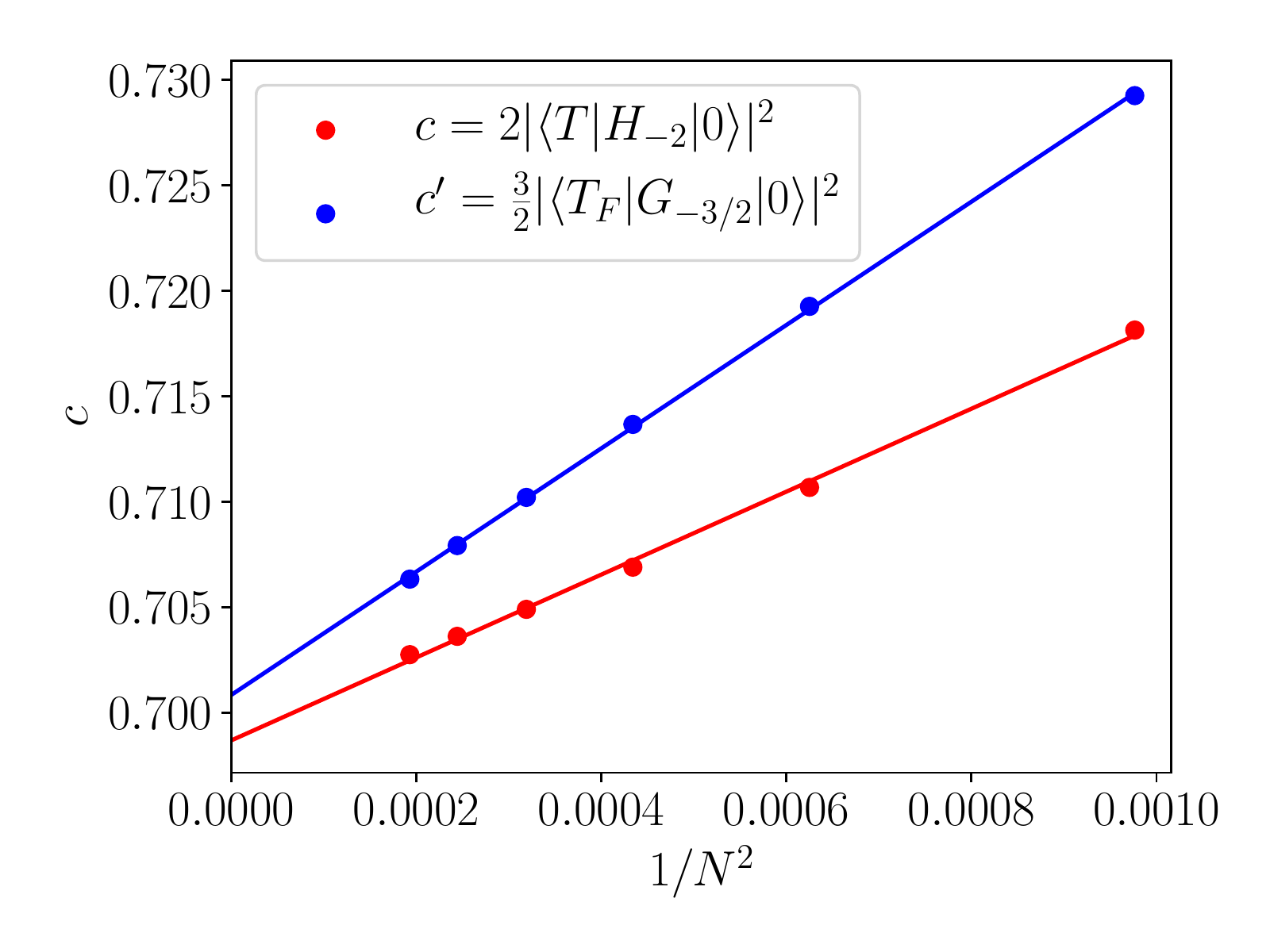}
\caption{\label{Fig:csusy} The central charge from Eqs.~\eqref{clat},\eqref{Gnc}.}
\end{figure}

\begin{figure}
\includegraphics[width=0.99\linewidth]{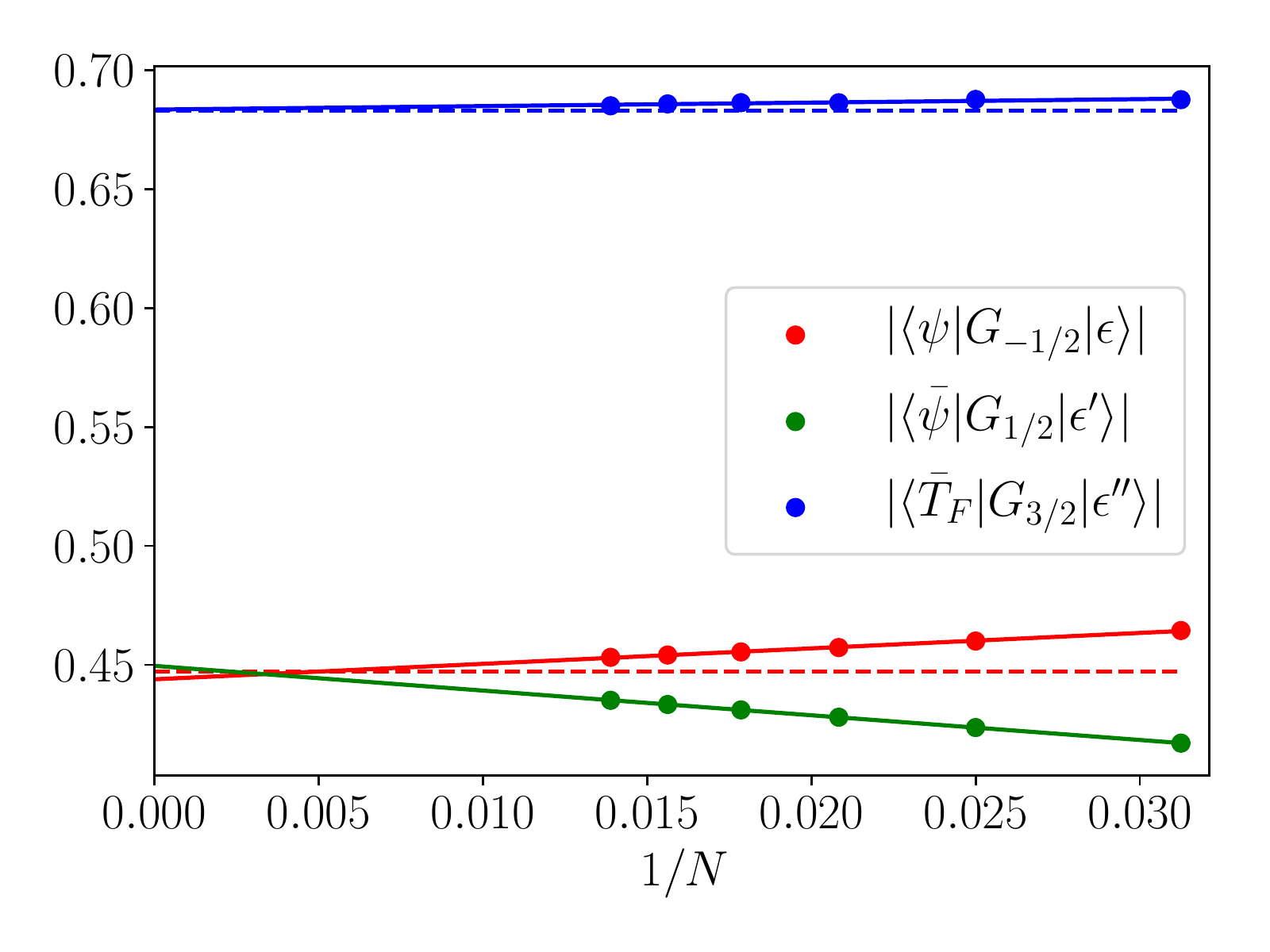}
\caption{\label{Fig:GnNS} Matrix elements of $G_n$ in the NS sector of the TCI model. The dashed lines represent the corresponding CFT matrix element Eqs.~\eqref{GnCFT3},\eqref{GnCFT4},\eqref{GnCFT2}. The CFT matrix elements Eqs.~\eqref{GnCFT3},\eqref{GnCFT4} have the same modulus, so we only show one of them in the figure. }
\end{figure}

Below we compare the matrix elements of $G_{n}$ with the CFT matrix elements Eqs.~\eqref{GnCFT1}-\eqref{GnCFT2}. In particular, Eq.~\eqref{Gnc} provides a way of verifying that central charge in Eqs.~\eqref{Svirasoro1} and \eqref{Svirasoro3} are the same,
\begin{equation}
\label{csusy}
c'=\frac{3}{2}|\langle T_F|G_{-3/2}|0\rangle|^2,
\end{equation}
which equals the central charge $c^{\CFT}$ in the thermodynamic limit. This equation can be viewed as the ''superpartner'' of Eq.~\eqref{clat}. The result is shown in Fig. (\ref{Fig:csusy}), where we also plot the result of Eq.~\eqref{clat} for comparison. We obtain $c'=0.701$ and $c=0.699$, with the errors on the same order. The other matrix elements in the NS sector are plotted in Fig.(\ref{Fig:GnNS}). We see that all matrix elements shown in the figure approximately converge to the nonzero CFT values Eqs.~\eqref{GnCFT3}, \eqref{GnCFT4}, and \eqref{GnCFT2} in the thermodynamic limit. Similarly, we can compute the matrix elements of $\bar{G}_n$, which are the Fourier modes of $\bar{T}_F$, and see that they agree with the CFT values in the thermodynamic limit. Therefore, we have verified in the TCI model that $|\psi\rangle,|\bar{\psi}\rangle,|\epsilon'\rangle$ are supervirasoro descendants of $|\epsilon\rangle$, and $|T_F\rangle,|\bar{T}_F\rangle,|\epsilon''\rangle$ are supervirasoro descendants of $|\mathbf{1}\rangle$. The only supervirasoro primaries in the NS sector, are therefore, $|\mathbf{1}\rangle$ and $|\epsilon\rangle$

\begin{figure}
\includegraphics[width=0.49\linewidth]{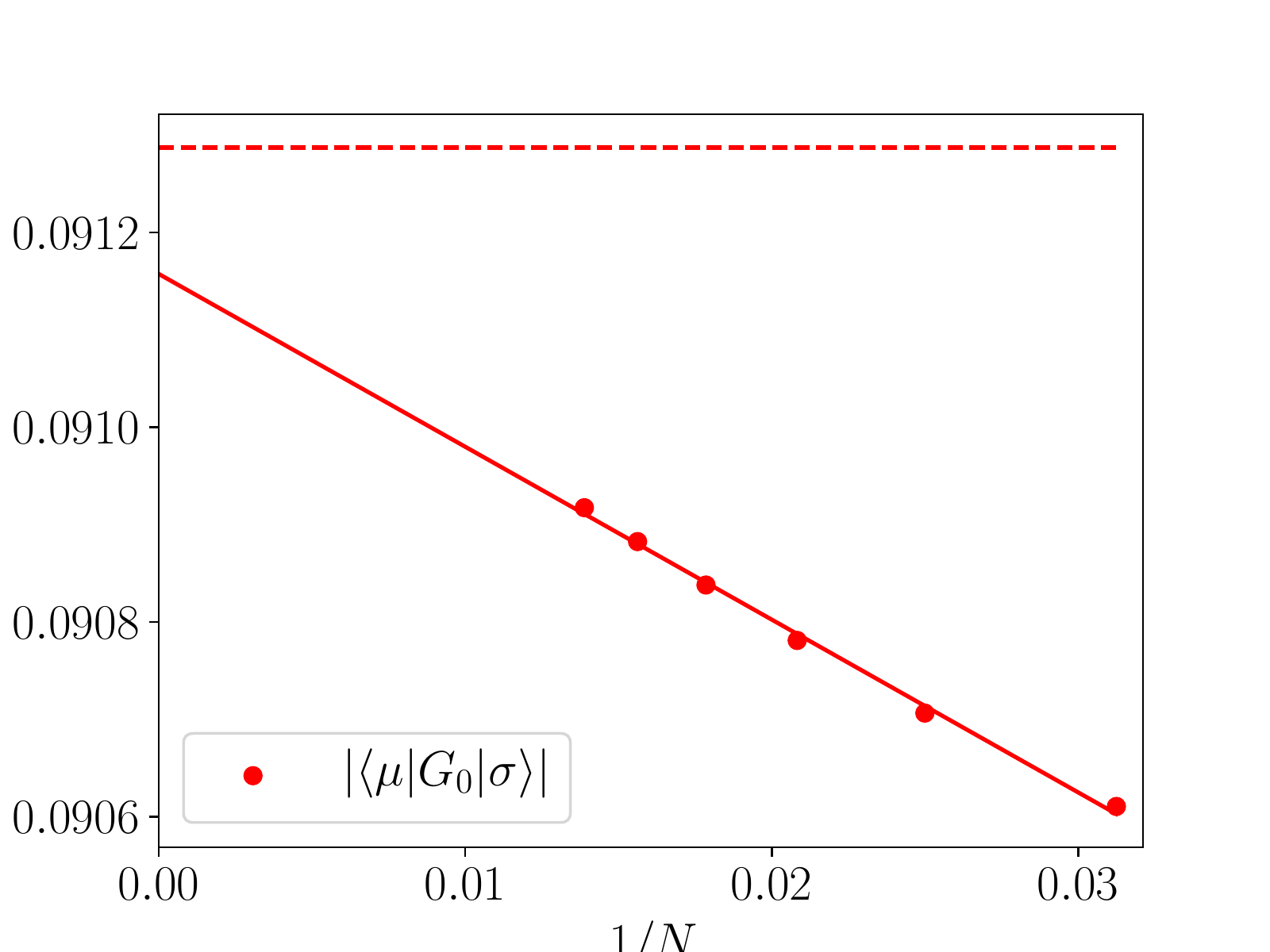}
\includegraphics[width=0.49\linewidth]{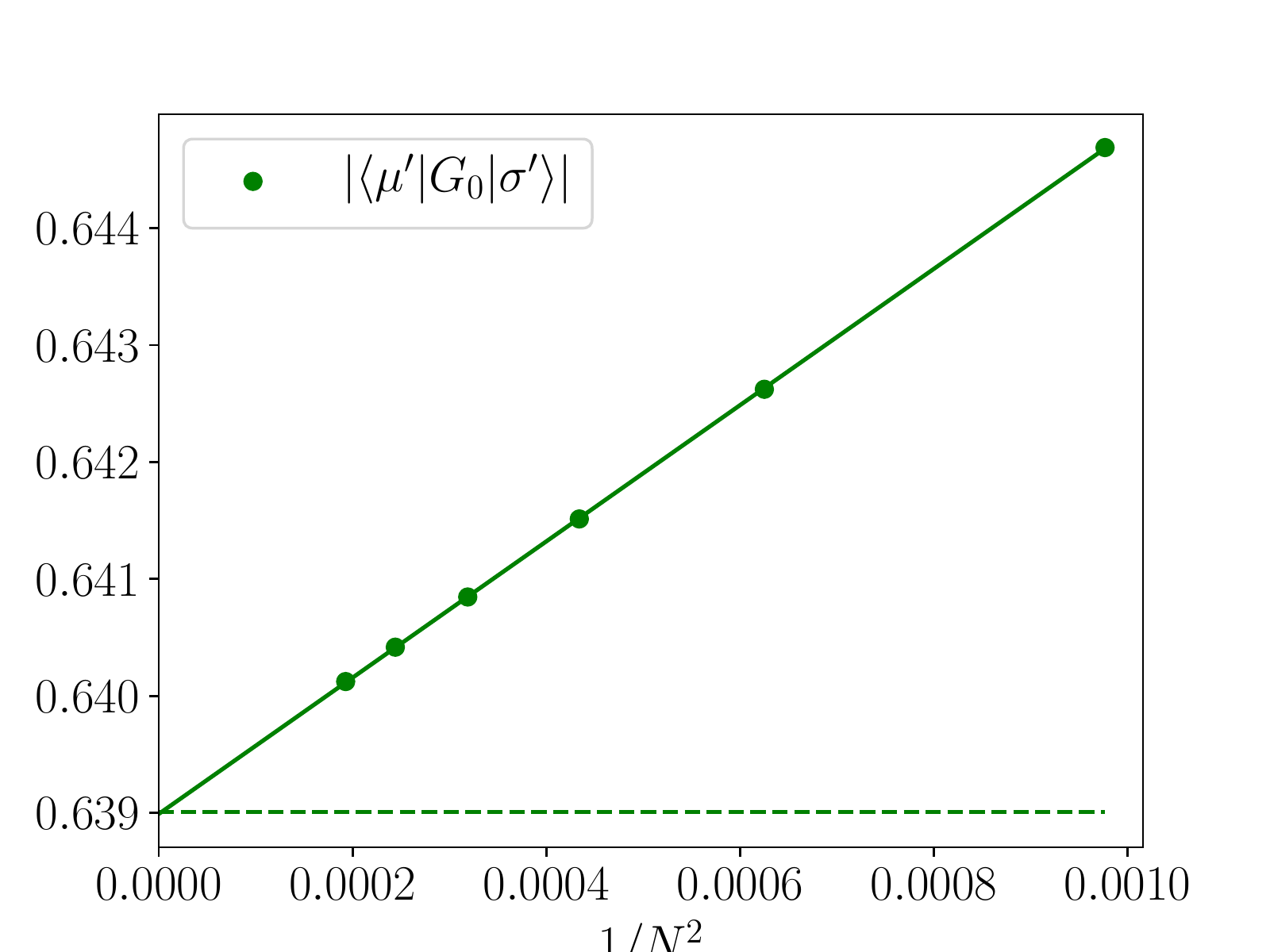}
\caption{\label{Fig:GnR} Matrix elements of $G_n$ in the R sector of the TCI model. The dashed lines represent the corresponding CFT matrix elements in Eqs.~\eqref{GnCFT1},\eqref{GnR2}.  }
\end{figure}

In the R sector, we can similarly compute the matrix elements of $G_n$ ($n\in \mathbb{Z}$). An important example is $G_0$ that relates superpartners, as in Eqs.~\eqref{GnCFT1} and \eqref{GnR2}. The  matrix elements can be computed on the lattice, shown in Fig. (\ref{Fig:GnR}). We also see that the numerical results agree with the CFT matrix elements.

To conclude, we have proposed the lattice supervirasoro generators $G_n$ as Fourier modes of the lattice string operator $T_F$, where the latter is found variationally. We have examined the action of $G_n$ in both NS and R sectors and seen that it agrees with the supervirasoro algebra. The whole construction only relies on the low-energy spectrum of the critical quantum spin chain. Therefore we expect that the above lattice construction of $G_n$ gives a generic method to identify supervirasoro primaries and superconformal towers.

\section{Conclusion}
In this paper, we have generalized the method \cite{zou_conformal_2018, zou_conformal_2019} of extracting conformal data from a critical quantum spin chains from PBC to APBC. 

Starting with the lattice Hamiltonian density, we first build the Hamiltonians $H^{\PBC}$ and $H^{\APBC}$ with PBC and APBC. We then diagonalize the low-energy eigenstates with both boundary conditions and various sizes $N$. To go beyond exact diagonalization, a tensor network method based on periodic uniform matrix product states can be used. The scaling dimensions $\Delta_\alpha$ and conformal spins $s_\alpha$ of scaling operators can be extracted from the energies $E_\alpha$ and momenta $P_\alpha$ of the eigenstates with both boundary conditions. Fourier modes $H_n$ of the Hamiltonian density with respect to the translation operator $\mathcal{T}$ ($\tilde{\mathcal{T}}$) of the (A)PBC Hamiltonian act as a linear combination of Virasoro generators on the low-energy subspace with (A)PBC. They allow us to identify each eigenstate with a CFT scaling operator. In particular, primary states and their conformal towers are identified. The central charge can be extracted with the matrix elements of $H_{-2}$.

We have shown that local operators correspond to PBC operators in the CFT, and that string operators correspond to APBC operators in the CFT. Given a lattice operator, we can associate it with a truncated linear combination of CFT scaling operators, whose coefficients are determined by minimizing a cost function. The lattice operators that correspond to CFT primary fields can be obtained by inverting the truncated linear expansion. OPE coeffcients can then be extracted from the matrix elements of the Fourier modes of the local or string operators that correspond to CFT primary fields. In the case of emergent superconformal symmetry, the fermionic stress tensor states $|T_F\rangle,|\bar{T}_F\rangle$ are always present in the low-energy spectrum of $H^{\APBC}$. The lattice supervirasoro generators $G_n,\bar{G}_n$ can be constructed as Fourier modes of the string operators that correspond to $T_F,\bar{T}_F$. They can be used to identify supervirasoro primary states and supervirasoro conformal towers. The matrix element of $G_{-3/2}$ gives another estimation of the central charge, which converges to the central charge of the SCFT in the thermodynamic limit. 

As an illustration of the general method, we have extracted complete conformal data of the Ising CFT and TCI CFT from the Ising spin chain and the TCI spin chain. We have correctly identified all primary states and their conformal towers. Scaling dimensions, conformal spins, the central charges and all OPE coefficients are obtained with high accuracy. For the TCI model, we have verified the action of the lattice supervirasoro generators on the low-energy eigenstates and showed that they agree with the expectation from the superconformal algebra. We stress that the only input of our method is the critical lattice Hamiltonian. It is interesting to apply our method to the cases where the underlying CFT has not been completely solved.

Apart from a complete set of conformal data, generators of extended symmetries (if any) can also be constructed on the lattice. In this paper we have investigated the superconformal symmetry, but other extended symmetry can be treated in the same way, such as the Kac-Moody algebra \cite{Wang_upcoming_2019}.

Our method can be generalized to other twisted boundary conditions that preserve emergent conformal symmetry. For an on-site symmetry defect, the generalization is straightforward. It is still an open question how to deal with more general conformal defects with our method. We would like to point out that for some topological conformal defects other methods such as tensor network renormalization \cite{hauru_topological_2016} and entanglement renormalization \cite{evenbly_nonlocal_2010} are available.   

\textit{Acknowlegments ---} We thank Cenke Xu, Andreas Ludwig, Zhenghan Wang, Ashley Milsted and Martin Ganahl for useful discussions and valuable comments. The authors acknowledge support from the Simons Foundation (Many Electron Collaboration) and Compute Canada. G. Vidal is a CIFAR fellow in the Quantum Information Science Program. Research at Perimeter Institute is supported by the Government of Canada through the Department of Innovation, Science and Economic Development Canada and by the Province of Ontario through the Ministry of Research, Innovation and Science. X is formerly known as Google[x] and is part of the Alphabet family of companies, which includes Google, Verily, Waymo, and others (www.x.company).  
 
 \bibliography{puMPS_CFT}
 
 \appendix
 
 \clearpage
 \section{puMPS algorithm for APBC}
 In this section we detail the puMPS algorithm for eigenstates with APBC, and how to compute matrix elements of local or string operators involving them. It is a straightforward generalization of the algorithm for eigenstates with PBC, which has been described in detail in \cite{zou_conformal_2018}.
 \subsection{Computing low-energy eigenstates}
 Recall the ansatz for APBC eigenstates with momentum $p$,
 \begin{equation}
 \label{phiAPBC}
|\Phi^{\APBC}_p(B;A)\rangle= \sum_{j=1}^{N}e^{-ipj} \tilde{\mathcal{T}}^j \sum_{\vec{s}=1}^d\mathrm{Tr}(B^{s_1}A^{s_2}\cdots A^{s_N})|\vec{s}\rangle,
\end{equation}
where $A$ is the puMPS tensor for the PBC ground state, satisfying
\begin{equation}
\label{symTA}
\sum_{s'}\mathcal{Z}_{ss'} A^{s'} = U_B(\mathcal{Z})A^s U^{\dagger}_B(\mathcal{Z}),
\end{equation}
and $B=B^{s}_{ab}$ contains the variational parameters to be computed. 

In \cite{zou_conformal_2018}, a reparametrization of the excitation ansatz has been shown useful for PBC eigenstates. Here we will use the same trick for APBC eigenstates. The trick consists of two steps. First, there is a gauge choice of the ground state puMPS tensor $A^s=A^{s}_C\lambda^{-1}$, where $A^s_C$ is a $D\times D$ matrix and $\lambda$ is $D \times D$ diagonal matrix, such that $A^s_L \equiv A^s=A^{s}_C\lambda^{-1}$ satisfies the left canonical condition and $A^s_R \equiv \lambda^{-1} A^{s}_C$ satisfies the right canonical condition. Second, we will reparameterize $B^s= B^s_C \lambda^{-1}$. The new parameterization consists of a $d \times D\times D$ tensor $B_C$ as variational parameters,
\begin{align}
&\,\,\,\,\,\,\,\,|\Phi^{\APBC}_p(B_C;A_L)\rangle \nonumber \\
&= \sum_{j=1}^{N}e^{-ipj} \tilde{\mathcal{T}}^j \sum_{\vec{s}=1}^d\mathrm{Tr}((B^{s_1}_C\lambda^{-1})A^{s_2}_L\cdots A^{s_N}_L)|\vec{s}\rangle.
\end{align}
Below we use $\mu=(s,a,b)$ to denote the combined index of the physical and bond indices.  
\begin{figure}
\includegraphics[width=\linewidth]{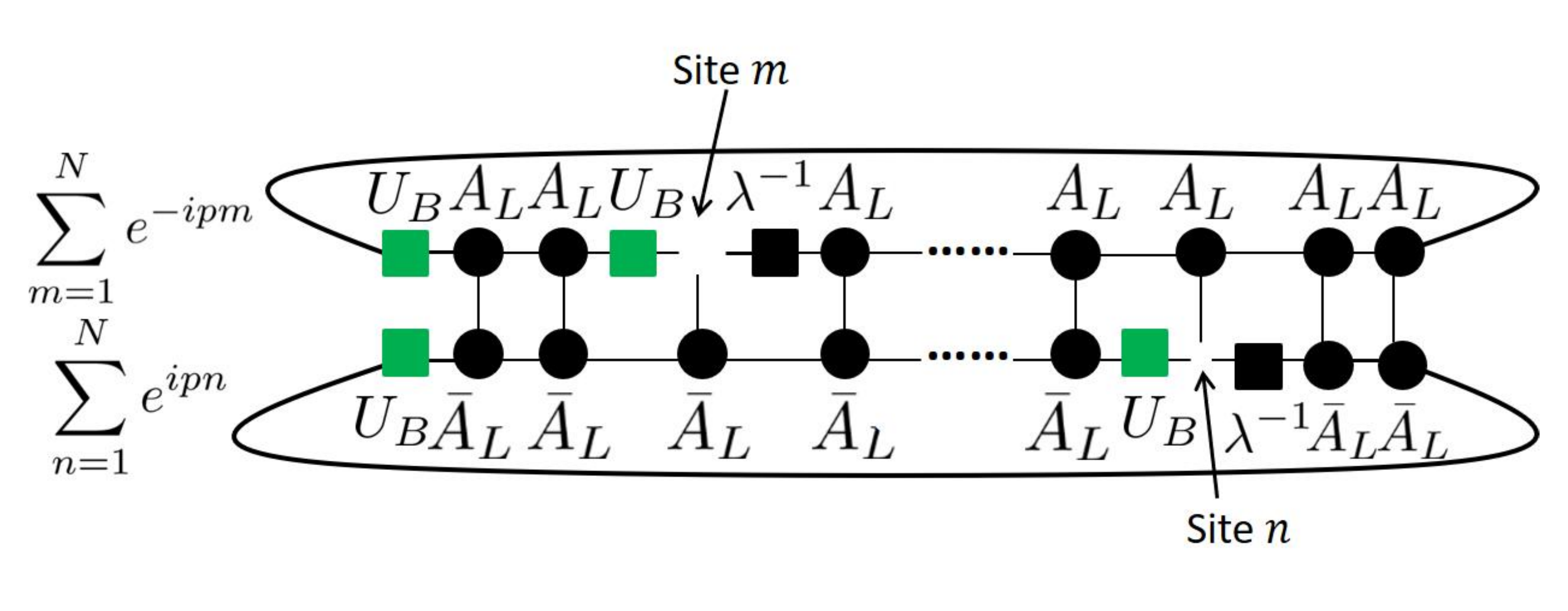}
\includegraphics[width=\linewidth]{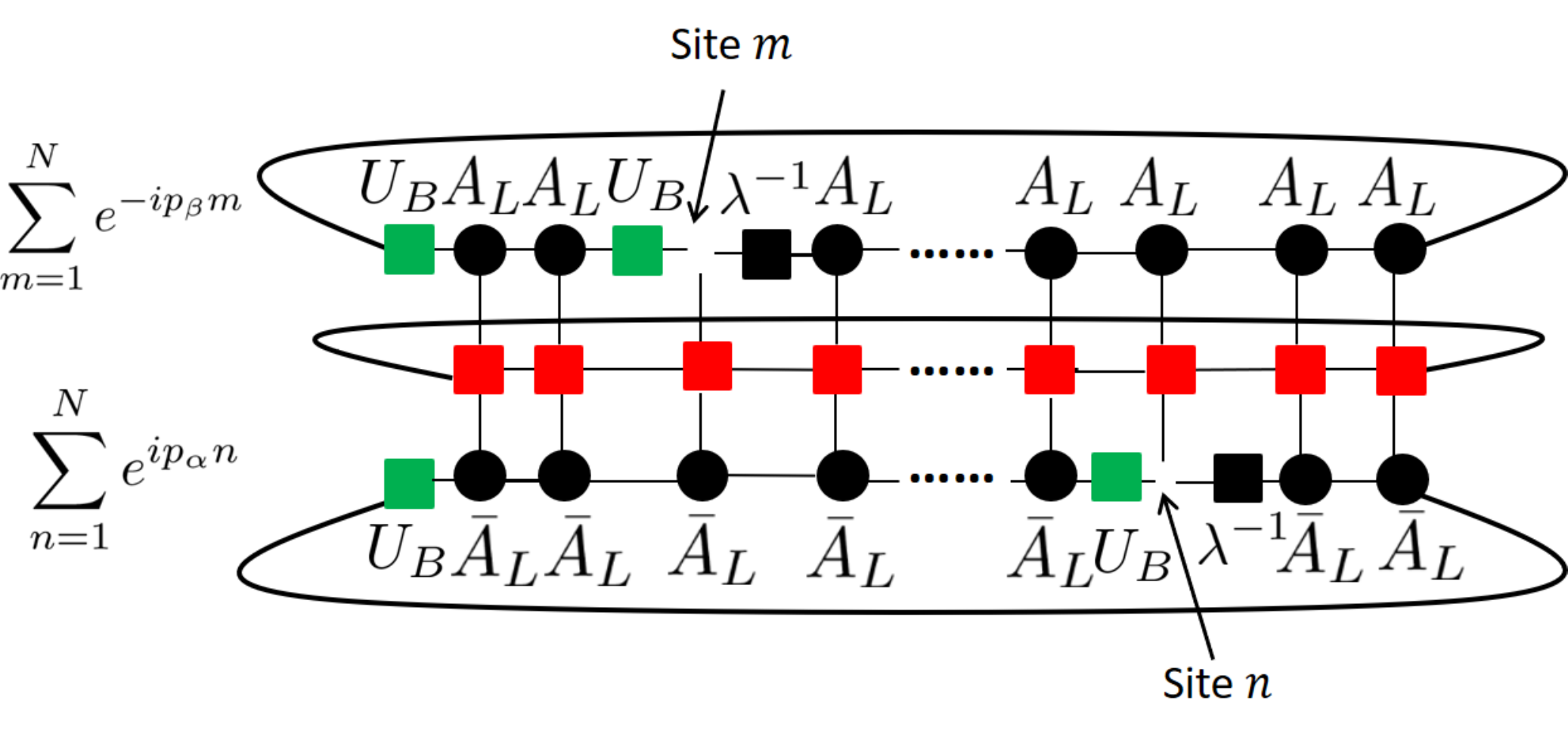}
\label{fig:NeffHeff}
\caption{(Top) The tensor network for $N^{\APBC}_{\mu\nu}(p)$ in Eq.~\eqref{Neff}. The green tensor $U_B \equiv U_B(\mathcal{Z})$ in Eq.~\eqref{UBZ}. (Bottom) The tensor network for $H^{\APBC}_{\mu\nu}(p)$ in Eq.~\eqref{Heff} if the red tensors form a matrix product operator (MPO) for the Hamiltonian $H$ and $p_\alpha=p_\beta=p$. It also represents $O_{\mu\nu}(p_\alpha,p_\beta) $ in Eq.~\eqref{Obilinear} if the red tensors form a MPO for $\tilde{O}^s$.}
\end{figure}
\begin{widetext}
Our approximation of the eigenstates of the Hamiltonian are obtained as the saddle point of the energy functional,
\begin{equation}
E_p(B_C,\bar{B}_C;A_L,\bar{A}_L)=\frac{\langle \Phi^{\APBC}_p(\bar{B}_C;\bar{A}_L)|H|\Phi^{\APBC}_p(B_C;A_L)\rangle}{\langle \Phi^{\APBC}_p(\bar{B}_C;\bar{A}_L)|\Phi^{\APBC}_p(B_C;A_L)\rangle}.
\end{equation}
The saddle point condition translates into a generalized eigenvalue equation for $B^{\mu}_C$,
\begin{equation}
\label{NBHB}
H^{\APBC}_{\mu\nu}(p)B^{\nu}_C=EN^{\APBC}_{\mu\nu}(p)B^{\nu}_C,
\end{equation}
where 
\begin{eqnarray}
\label{Neff}
N^{\APBC}_{\mu\nu}(p)&=&\left\langle \frac{\partial}{\partial \bar{B}^{\mu}_C}\Phi^{\APBC}_p(\bar{B}_C;\bar{A}_L)\right. \left| \frac{\partial}{\partial B^{\nu}_C}\Phi^{\APBC}_p(B_C;A_L)\right\rangle ~~~~~~ \\
\label{Heff}
H^{\APBC}_{\mu\nu}(p)&=&\left\langle \frac{\partial}{\partial \bar{B}^{\mu}_C}\Phi^{\APBC}_p(\bar{B}_C;\bar{A}_L)\right| H \left| \frac{\partial}{\partial B^{\nu}_C}\Phi^{\APBC}_p(B_C;A_L)\right\rangle. ~~~~~~
\end{eqnarray}
They are depicted as tensor networks in Fig.~\ref{fig:NeffHeff}. Note that the action of $\tilde{\mathcal{T}}=\mathcal{Z}_1 \mathcal{T}$ can be viewed as first acting with the ordinary translation operator and then acting with $\mathcal{Z}_1$, where the latter can be lifted to the bond indices by using Eq.~\eqref{symTA}. 
\end{widetext}
Contracting the tensor networks has the same leading cost $\mathcal{O}(ND^6)$ as the PBC case. The only difference from the case of PBC is the $U_B$ tensors appearing in the contraction. The periodic MPO for the Hamiltonian with PBC can be further decomposed into a MPO for the Hamiltonian with OBC and a local boundary term. This further lowers the cost compared to directly contracting the network with the periodic MPO. 

The generalized eigenvalue equation Eq.~\eqref{NBHB} can be translated into an ordinary eigenvalue problem by multiplying the pseudoinverse of $N^{\APBC}_{\mu\nu}$ on both sides,
\begin{equation}
\label{NiH}
\tilde{N}^{\APBC,\rho\mu}(p)H^{\APBC}_{\mu\nu}(p)B^{\nu}_C=E B^{\rho}_C,
\end{equation}
where $\tilde{N}^{\APBC}(p)$ is the pseudoinverse of $N^{\APBC}(p)$. Then Eq.~\eqref{NiH} can be solved in each momentum sector with a sparse eigenvalue solver such as the Arnoldi method. 
 \subsection{Computing matrix elements of local operators}
 Matrix elements of a Fourier mode of a local operator in the low-energy basis of $H^{\APBC}$ are bilinear functions of the $B$ tensors in Eq.~\eqref{phiAPBC},
 \begin{align}
 \label{Obilinear}
 &\,\,\,\,\,\,\,\, \langle \Phi^{\APBC}_{p_\alpha}(\bar{B}_{C,\alpha};\bar{A}_L)|\tilde{\mathcal{O}}^s|\Phi^{\APBC}_{p_\beta}(B_{C,\beta};A_L)\rangle \nonumber \\
 &= \bar{B}^{\mu}_{C,\alpha}\mathcal{O}_{\mu\nu}(p_\alpha,p_\beta)B^{\nu}_{C,\beta} . 
 \end{align}
This expression is nonzero only if momentum is conserved,
 \begin{equation}
 p_\alpha=\frac{2\pi}{N} s+p_\beta.
 \end{equation}
 The matrix $\mathcal{O}_{\mu\nu}(p_\alpha,p_\beta)$ is plotted at the bottom of Fig.~\ref{fig:NeffHeff}. The contraction also costs $\mathcal{O}(ND^6)$.
 \subsection{Computing matrix elements of string operators}
 Matrix elements of a Fourier mode of a string operator are also bilinear functions of the $B$ tensors.
  \begin{align}
 \label{Sbilinear}
&\,\,\,\,\,\,\,\,\, \langle \Phi^{\APBC}_{p_\alpha}(\bar{B}_{C,\alpha};\bar{A}_L)|\mathcal{S}^s_{\mathcal{O}}|\Phi^{\PBC}_{p_\beta}(B_{C,\beta};A_L)\rangle \nonumber \\
 &= \bar{B}^{\mu}_{C,\alpha}\mathcal{S}_{\mathcal{O},\mu\nu}(p_\alpha,p_\beta)B^{\nu}_{C,\beta}. 
 \end{align}
 Note that the ket is now a PBC low-energy eigenstate, and the bra is a APBC low-energy eigenstate. The conservation of momentum is necessary for the matrix element to be nonzero,
 \begin{equation}
 p_\alpha=\frac{2\pi}{N} s+p_\beta.
 \end{equation}
  Although string operators are nonlocal, they can be represented efficiently with a MPO. This means that $\mathcal{S}_{\mathcal{O},\mu\nu}(p_\alpha,p_\beta)$ can be represented as the same network as the bottom network of Fig.~\ref{fig:NeffHeff}, except that the $U_B$ tensors in the upper layer are removed. Therefore the computation of $\mathcal{S}_{\mathcal{O},\mu\nu}(p_\alpha,p_{\beta})$ has the same leading cost $\mathcal{O}(ND^6)$ as the computation of $\mathcal{O}_{\mu\nu}(p_\alpha,p_\beta)$.
  
If $\mathcal{O}$ is a one-site operator, then $\mathcal{S}^s_{\mathcal{O}}$ can be encoded in a MPO with bond dimension $2$,
\begin{equation}
\begin{bmatrix}
1 & 0
\end{bmatrix}
\prod_{j=1}^N
\begin{bmatrix}
Z_j & \mathcal{O}_j e^{-isj} \\
0 & I_j
\end{bmatrix}
\begin{bmatrix}
0 \\
1
\end{bmatrix}.
\end{equation}
If $\mathcal{O}$ is supported on $n$ sites, then an additional boundary term appears. In this case, $\mathcal{S}^s_{\mathcal{O}}$ can be decomposed into a MPO with bond dimension $n+1$ with open boundary conditions and a boundary term. For example, consider 
  \begin{equation}
  \mathcal{S}^s_{YZ}=\sum_{j=1}^{N-1}e^{-isj 2\pi/N} \left(\prod_{l=1}^{j-1} Z_l\right)Y_{j}Z_{j+1}+\mathcal{B}^{s}_{YZ},
  \end{equation}
 where the sum can be encoded in a MPO with open boundary conditions and bond dimenson $3$, and the boundary term is
 \begin{equation}
 \mathcal{B}^{s}_{YZ}=e^{-is2\pi} I_1 \left(\prod_{l=2}^{N-1} Z_{l}\right) Y_N.
\end{equation}
 If we act with $\mathcal{B}^{s}_{YZ}$ on a PBC eigenstate with parity $\mathcal{Z}_\alpha$, then the action  can be further simplified,
 \begin{equation}
 \mathcal{B}^{s}_{YZ}|\psi^{\PBC}_\alpha\rangle= e^{-is2\pi} (i X_NZ_1) \mathcal{Z}_\alpha|\psi^{\PBC}_\alpha\rangle.
 \end{equation}
 We see that the net effect of the boundary term of $\mathcal{S}^s_{\mathcal{O}}$ acting on a PBC eigenstate is equivalent to a local boundary term. Again we decompose $\mathcal{S}^s_{\mathcal{O}}$ into a MPO with OBC and a \textit{local} boundary term. This lowers the computational cost compared to directly contracting the tensor network with a periodic MPO. 
 
  \section{Fourier mode of multi-site operators and string operators}
  Recall that the Fourier mode of a local operator $\mathcal{O}$ is defined by
 \begin{equation}
\mathcal{O}^s= \frac{1}{N}\sum_{j=1}^N e^{-is x_j 2\pi/N} \mathcal{O}_j,
\end{equation}
where $x_j$ is the position of the operator $\mathcal{O}_j$ with support starting at site $j$. For a one-site operator, $x_j=j$. (In the main text we have use $j$ instead of $x_j$ for simplicity of notation.) However, this choice of $x_j$ is ambiguous for a multi-site operator, since it can be  anywhere inside the support. It has been shown \cite{zou_conformal_2019} that the corresponding CFT operator $\mathcal{O}^{\CFT}$ can change by a total spatial derivative if the choice of $x_j$ is changed. Similarly, a string operator 
\begin{equation}
\mathcal{S}_{\mathcal{O},j}=\left(\prod_{l=1}^{j-1} Z_l\right) \mathcal{O}_j.
\end{equation}
can be assigned some position $x_j$. The choice of $x_j$ affects the corresponding CFT operator up to a total spatial derivative.

 However, in the main text we have only considered the coefficients $a_\alpha$ of primary operators $\phi^{\CFT}_\alpha$ in the truncated expansion of $\mathcal{O}^{\CFT}$. The $a_\alpha$'s in the thermodynamic limit do not depend on the choice of $x_j$ since primary operators are not total spatial derivatives. Nevertheless, there is a preferred choice \cite{zou_conformal_2019} of $x_j$ which makes the finite-size scaling of the $a_\alpha$'s better, because they forbid some of the derivative descendants in the expansion of $\mathcal{O}^{\CFT}$ by symmetry. In this appendix, we list our choices of $x_j$ for both local and string operators that have appeared in the main text.
 
For a local operator $\mathcal{O}_j$, if it is $\mathbb{Z}_2$ odd and its support ranges from site $j$ to $j+n$, then we follow the ''middle point rule'', i.e., $x_j=j+n/2$. If the operator is $\mathbb{Z}_2$ even, we first rewrite it as a local product of Majorana operators. The Majorana operator $\gamma_{j'}$ is assigned position $x_{j'}=j'/2+1/4$. If the product of Majorana operators has support from $\gamma_{j'}$ to $\gamma_{j'+n'}$, then we follow the ''middle point rule'' $x_j=x_{j'}+n'/4=(2j'+n'+1)/4$. (Notice that two adjacent Majorana modes have a distance $1/2$ rather than $1$.)

If a string operator is $\mathbb{Z}_2$ odd, then it can be rewritten as a local product of Majorana operators. We can then use the ''middle point rule''. If the string operator is $\mathbb{Z}_2$ even, then we first use a Jordan-Wigner transformation to obtain a $\mathbb{Z}_2$ odd local operator, then the position of the string operator is assigned to the position of the local operator minus $1/2$. In summary, we list the result of $x_j$ for the operators that have appeared in the main text in Table \ref{table:positions}.

\begin{table}
\begin{tabular}{|c|c|c|c|}
\hline
$\mathcal{O}_j$ & $x_j$ & $\mathcal{S}_{\mathcal{O},j}$ & $x_j$  \\ \hline
$X_j$ & $j$ & $\mathcal{S}_{I,j}$ & $j-\frac{1}{2}$\\ \hline
$Y_j$ & $j$ & $\mathcal{S}_{X,j}$ & $j-\frac{1}{4}$ \\  \hline
$Z_j$ & $j$ & $\mathcal{S}_{Y,j}$ &  $j+\frac{1}{4}$\\  \hline
$X_jX_{j+1}$ & $j+\frac{1}{2}$ & $\mathcal{S}_{YZ,j}$& $j+\frac{3}{4}$ \\   \hline
$X_jZ_{j+1}$ & $j+\frac{1}{2}$ & $\mathcal{S}_{IX,j}$& $j+\frac{1}{4}$ \\   \hline
$Z_jX_{j+1}$ & $j+\frac{1}{2}$ & $\mathcal{S}_{XX,j}$& $j$ \\   \hline
$X_jX_{j+1}Z_{j+2}$ & $j+\frac{5}{4}$ & $\mathcal{S}_{YY,j}$& $j+1$ \\   \hline
$Z_jX_{j+1}X_{j+2}$ & $j+\frac{3}{4}$ &  & \\   \hline
\end{tabular}
\caption{Position assignment $x_j$ of local operators $\mathcal{O}_j$ and string operators $\mathcal{S}_{\mathcal{O},j}$.}
\label{table:positions}
\end{table}

 \section{CFT matrix elements with the supervirasoro algebra}
 In this appendix we derive the CFT matrix elements for the TCI CFT used in the main text. We will omit the $\CFT$ superscript because we will only compute quantities in the CFT. First, the central charge Eq.~\eqref{csusy} can be obtained by
 \begin{align}
 & \,\,\,\,\,\,\,\, |\langle T_F|G_{-3/2}|0\rangle|^2 \\
 &=\langle 0|G_{3/2}G_{-3/2}|0\rangle \\
 &=\langle 0|\{G_{3/2},G_{-3/2}\}|0\rangle \\
 &=\frac{2c}{3},
\end{align}
 where in the second line we use the fact that $G_{-3/2}$ acting on the ground state only gives the $|T_F\rangle$ state, in the third line we use the fact that $G_{3/2}$ annihilates the ground state, and in the last line we use the supervirasoro algebra and that $L_0$ annihilates the ground state.
  We can choose the phase of $G_n$ such that 
 \begin{equation}
 \label{TFC}
 \langle T_F|G_{-3/2}|0\rangle=\sqrt{\frac{2c}{3}}.
 \end{equation}
 
Similar calculation can be performed on other matrix elements. For example, in the NS sector we can compute
 \begin{align}
 & \,\,\,\,\,\,\,\, |\langle \psi|G_{-1/2}|\epsilon\rangle|^2 \\
 &=\langle \epsilon|G_{1/2}G_{-1/2}|\epsilon\rangle \\
 &=\langle \epsilon|\{G_{1/2},G_{-1/2}\}|\epsilon\rangle \\
 &=\langle \epsilon|2L_0|\epsilon\rangle \\
 &=\Delta_\epsilon+s_{\epsilon} \\
 &=0.2
\end{align}
and
 \begin{align}
 & \,\,\,\,\,\,\,\, |\langle \epsilon'|\bar{G}_{-1/2}|\psi\rangle|^2 \\
 &=\langle \psi|\bar{G}_{1/2}\bar{G}_{-1/2}|\psi\rangle \\
 &=\langle \psi|2\bar{L}_0|\psi\rangle \\
 &=\Delta_\psi-s_\psi \\
 &=0.2.
\end{align}
In the R sector we can compute
 \begin{align}
 & \,\,\,\,\,\,\,\, |\langle \mu|G_0|\sigma\rangle|^2 \\
 &=\langle \sigma|G^2_0|\sigma\rangle \\
 &=\langle \sigma|L_0-\frac{c}{24}|\sigma\rangle \\
 &=h_\sigma-\frac{c}{24}.
\end{align}
Similarly $|\langle \mu'|G_0|\sigma'\rangle|^2$ can be computed.

Finally, let us verify the normalization condition of $G_n$, Eq.~\eqref{GnCFTnorm}.
 \begin{align}
& \,\,\,\,\,\,\,\,  \langle T_F|G_{1/2}|T \rangle \\
&  = \sqrt{\frac{3}{2c}}\langle  0| G_{3/2}G_{1/2}|T\rangle \\
&  = \sqrt{\frac{3}{2c}}\langle  0| 2L_2 |T\rangle \\
&  = \sqrt{\frac{3}{2c}} 2\sqrt{\frac{c}{2}} \\
&  = \sqrt{3},
\end{align}
 where in the second line we use Eq.~\eqref{TFC}, in the third line we use the supervirasoro algebra and the fact that $G_{3/2}$ annihilates $|T\rangle$, and in the fourth line we use Eq.~\eqref{clat}.
 \end{document}